\PassOptionsToPackage{table}{xcolor}

\documentclass[sigconf]{acmart}

\def\BibTeX{{\rm B\kern-.05em{\sc i\kern-.025em b}\kern-.08emT\kern-.1667em\lower.7ex\hbox{E}\kern-.125emX}}

\usepackage{listings}   
\usepackage{pifont}
\newcommand{\cmark}{\ding{51}}%
\newcommand{\xmark}{\ding{55}}%

\usepackage{footmisc}
\usepackage{nicefrac}
\usepackage{siunitx}
\usepackage{array,framed}
\usepackage{color,soul}
\usepackage{booktabs}
\usepackage{
  color,
  float,
  epsfig,
  wrapfig,
  graphics,
  graphicx
}

\usepackage{placeins}
\usepackage{enumerate}
\usepackage{booktabs}
\usepackage{multirow}
\usepackage{graphicx}
\usepackage{subfigure}
\usepackage{enumitem}
\usepackage[table]{xcolor}
\usepackage{adjustbox}

\usepackage{textcomp}
\usepackage{setspace}
\usepackage{latexsym,fancyhdr,url}
\usepackage{enumerate}
\usepackage{algorithm2e}
\usepackage{algpseudocode}
\usepackage{graphics}
\usepackage{xparse} 
\usepackage{xspace}
\usepackage{multirow}
\usepackage{csvsimple}
\usepackage{balance}

\usepackage{
  tikz,
  pgfplots,
  pgfplotstable
}
\usepackage{hyperref}
\usepackage{breakurl}
\usetikzlibrary{
  shapes.geometric,
  arrows,
  external,
  pgfplots.groupplots,
  matrix
}

\pgfplotsset{compat=1.9}

\newtheorem{insight}{Insight}

\usepackage{mathtools,}

\DeclareMathAlphabet{\mathcal}{OMS}{cmsy}{m}{n}

\DeclareGraphicsExtensions{%
    .png,.PNG,%
    .pdf,.PDF,%
    .jpg,.mps,.jpeg,.jbig2,.jb2,.JPG,.JPEG,.JBIG2,.JB2}

\usepackage{xparse}
\newcommand{\bnm}{\begin{newmath}}
\newcommand{\enm}{\end{newmath}}

\newcommand{\bea}{\begin{eqnarray*}}%
\newcommand{\eea}{\end{eqnarray*}}%

\newcommand{\bne}{\begin{newequation}}
\newcommand{\ene}{\end{newequation}}

\newcommand{\bal}{\begin{newalign}}
\newcommand{\eal}{\end{newalign}}

\newenvironment{newalign}{\begin{align}%
\setlength{\abovedisplayskip}{4pt}%
\setlength{\belowdisplayskip}{4pt}%
\setlength{\abovedisplayshortskip}{6pt}%
\setlength{\belowdisplayshortskip}{6pt} }{\end{align}}

\newenvironment{newmath}{\begin{displaymath}%
\setlength{\abovedisplayskip}{4pt}%
\setlength{\belowdisplayskip}{4pt}%
\setlength{\abovedisplayshortskip}{6pt}%
\setlength{\belowdisplayshortskip}{6pt} }{\end{displaymath}}

\newenvironment{newequation}{\begin{equation}%
\setlength{\abovedisplayskip}{4pt}%
\setlength{\belowdisplayskip}{4pt}%
\setlength{\abovedisplayshortskip}{6pt}%
\setlength{\belowdisplayshortskip}{6pt} }{\end{equation}}

\newcounter{ctr}

\newcounter{mytable}
\def\mytable{\begin{centering}\refstepcounter{mytable}}
\def\endmytable{\end{centering}}

\newcounter{myfig}
\def\myfig{\begin{centering}\refstepcounter{myfig}}
\def\endmyfig{\end{centering}}

\newlength{\saveparindent}
\newlength{\saveparskip}
\newcommand{\E}{{\rm I\kern-.3em E}}

\renewcommand{\eqref}[1]{\mbox{Equation~(\ref{#1})}}

\def \part {part}

\renewcommand{\paragraph}[1]{\vspace*{6pt}\noindent\textbf{#1}\;}

\def \blackslug{\hbox{\hskip 1pt \vrule width 4pt height 8pt
    depth 1.5pt \hskip 1pt}}
\def \qed{\quad\blackslug\lower 8.5pt\null\par}

\newcounter{mynote}[section]

\newcommand\ignore[1]{}

\newcounter{rcnote}[section]

\newcounter{mrnote}[section]

\newcounter{fknote}[section]

\newcounter{anote}[section]

\DeclareMathSymbol{\mlq}{\mathord}{operators}{``}
\DeclareMathSymbol{\mrq}{\mathord}{operators}{`'}

\newcommand{\rhf}[2]{R_{f, \gamma}}

\DeclareDocumentCommand{\edist}{o o}{
  \ensuremath{
    \IfNoValueTF{#1}{{d}}{{\sf d}(#1,#2)}
  }
}

\newcommand{\olrk}[1]{\ifx\nursymbol#1\else\!\!\mskip4.5mu plus 0.5mu\left(\mskip0.5mu plus0.5mu #1\mskip1.5mu plus0.5mu \right)\fi}

\NewDocumentCommand{\indseq}{ O{1} O{r} }{{#1}\ldots {#2}}

\setcopyright{none}
\renewcommand\footnotetextcopyrightpermission[1]{} 
\begin{document}
\fancyhead{}
\def\thetitle{Transformer-Based Language Models for Software Vulnerability Detection}
\title{\thetitle}

\author{Chandra Thapa}
\affiliation{%
  \institution{CSIRO Data61, Sydney, Australia}}
\email{chandra.thapa@data61.csiro.au}

\author{Seung Ick Jang}
\affiliation{%
  \institution{CSIRO Data61, Sydney, Australia}}
\email{seung.jang@data61.csiro.au}

\author{Muhammad Ejaz Ahmed}
\affiliation{%
  \institution{CSIRO Data61, Sydney, Australia}}
\email{ejaz.ahmed@data61.csiro.au}

\author{Seyit Camtepe}
\affiliation{%
  \institution{CSIRO Data61, Sydney, Australia}}
\email{seyit.camtepe@data61.csiro.au}

\author{Josef Pieprzyk}
\affiliation{%
  \institution{CSIRO Data61, Sydney, Australia \& Institute of Computer Science, Polish Academy of Sciences, Warsaw, Poland}}
\email{josef.pieprzyk@data61.csiro.au}

\author{Surya Nepal}
\affiliation{%
  \institution{CSIRO Data61, Sydney, Australia}}
\email{surya.nepal@data61.csiro.au}



\date{}

\begin{abstract}

The large transformer-based language models demonstrate excellent performance in natural language processing. By considering the transferability of the knowledge gained by these models in one domain to other related domains, and the closeness of natural languages to high-level programming languages, such as C/C++, this work studies how to leverage (large) transformer-based language models in detecting software vulnerabilities and how good are these models for vulnerability detection tasks. In this regard, firstly, a systematic (cohesive) framework that details source code translation, model preparation, and inference is presented. Then, an empirical analysis is performed with software vulnerability datasets with C/C++ source codes having multiple vulnerabilities corresponding to the library function call, pointer usage, array usage, and arithmetic expression.   
Our empirical results demonstrate the good performance of the language models in vulnerability detection. Moreover, these language models have better performance metrics, such as F1-score, than the contemporary models, namely bidirectional long short term memory and bidirectional gated recurrent unit.
%
%
%
Experimenting with the language models is always challenging due to the requirement of computing resources, platforms, libraries, and dependencies. Thus, this paper also analyses the popular platforms to efficiently fine-tune these models and present recommendations while choosing the platforms.
\end{abstract}

\settopmatter{printfolios=true}
\maketitle


\section{Introduction}
\label{sec:introduction}

In Natural Language Processing (NLP), transformer-based models outperform existing models, including recurrent neural network (RNN) based architectures~\cite{bert,gpt,shoeybi2020megatronlm,roberta}. Furthermore, transformer-based language models are attractive and promising over RNN because, unlike RNN, it allows parallelization in the model's computation for faster processing. This is essential to reduce the model training/testing time if the model's size is large, which is a usual case for transformer-based models. Besides, their ability to remodel from natural language processing tasks to related tasks, through the process formally known as \emph{transfer learning}, enables us to extend their usage in other domains. 
Thus, it is prudent to effectively leverage these models beyond NLP, such as in software vulnerability detection, where most studies are limited to RNN-based models~\cite{main_paper,sysevr}.

As software, including operating systems, is an integral part of most computing devices, vulnerability detection at its source-code level is a must-have mechanism
for both proprietary and open-source software to ensure protection from adversaries. They can exploit the software vulnerabilities/weaknesses, allowing them not only to control its execution but also to steal or modify its data. For example, a simple buffer overflow bug in software such as NVIDIA SHIELD TV (a popular streaming media device)~\cite{nvidia}, macOS Catalina (an Apple operating system)~\cite{macos} and WhatsApp (a popular instant messaging application)~\cite{whatsapp} could lead to their exploit.

There are additional benefits of using the transformer-based models in software vulnerability detection. The benefits include the following: (i) it automates the detection that is not possible with the static analysis tools, which use heuristic methods to find the code constructions from the known vulnerabilities requiring extensive manual operations~\cite{flawfinder}, and (ii) it removes the need of extensive feature engineering requirements like in (classical) machine learning.

Although Bidirectional Encoder Representations from Transformers (BERT)~\cite{bert}, a transformer-based language model, is used recently in vulnerability detection~\cite{Bertforsecurity}, it is still unclear how the other models such as Generative Pre-trained Transformer (GPT)~\cite{gpt} perform. Moreover, training/fine-tuning these models is always challenging due to the computational requirements, libraries, and dependencies.
Thus, this paper aims to answer the following:
\begin{itemize}
    \item [\textbf{RQ1:}] \emph{What can be a systematic framework to leverage transformer-based language models for software vulnerability detection?}
    \item [\textbf{RQ2:}] \emph{How well existing transformer-based language models perform in detecting software vulnerabilities compared to other contemporary RNN-based models?}
    \item [\textbf{RQ3:}] \emph{Which platform is efficient to run these models?}
\end{itemize}

In our studies, we choose software source codes written in a high-level programming language, specifically C/C++, because of its popularity~\cite{useofC}, and it shares many characteristics with natural languages. Moreover, it inherits natural language grammar~\cite{vuldeelocator,main_paper,codebert}. Besides, both have a well-defined structure (or syntax) and contextual meaning (or semantics). Natural languages allow building long sentences from words and shorter phrases. Likewise, programming languages include a collection of instructions that can be used to write complex programs. The semantics of natural languages defines the meaning of sentences depending on the order and choice of words and their context. For programming languages, semantics refers to expected actions, their order, and results. Overall, these similarities are additional factors to motivate and enable us to leverage the transformer-based language models in software vulnerabilities detection efficiently.  


\textbf{Transformer-based language models for software vulnerability detection:} By considering (i) transformer-based (large) models of various architectures and sizes, including BERT~\cite{bert}, DistilBERT~\cite{distilbert}, CodeBERT~\cite{codebert}, GPT-2~\cite{gpt}, and Megatron language model variants~\cite{shoeybi2020megatronlm}, and (ii) RNN-based models, namely bidirectional long short term memory (BiLSTM)~\cite{main_paper}, and bidirectional gated recurrent unit (BiGRU)~\cite{gru}, this work contributes the following:

\begin{itemize}[leftmargin=3mm\relax]
    \item \textbf{Systematic framework:}
    To answer \textbf{RQ1}, we present a systematic framework for software vulnerability detection. The framework details the translation of the source codes to vectorized inputs for the models, description of the models, models' preparation, and inference. 
    
    \item \textbf{Comparative performances of the models on software vulnerability detection in C/C++ source code databases:}
    To answer \textbf {RQ2}, firstly, comparative performances under binary and multi-classification tasks are carried out to evaluate the models with a vulnerability dataset having \emph{Buffer Error} and \emph{Resource Management Error}~\cite{vuldeepecker_data}. Also, we demonstrate the need for data cleaning in these tasks. Besides, we provide the fine-tuning time to present an overall time cost of the models. Secondly, we further extend the performance analysis on multiple C/C++ vulnerabilities corresponding to the library function call, pointer usage, array usage, and arithmetic expression that are related to more than 341 Common Weakness Enumeration (CWE) IDs.  
    
    
    \item \textbf{Platform analysis:} 
    It is always confusing and challenging to handle a large model with billions of model parameters ({\em e.g.,} GPT-2 with 1.5B parameters). These models can easily exceed the capacity of available Graphics Processing Units (GPU) internal RAM ({\em e.g.,} 16GB) and usually require parallelisms, such as data parallelism and model parallelism. Finding a suitable platform to fine-tune and test these models effectively is as important as the main problem, {\em i.e.,} software vulnerability detection. Thus, as an answer to \textbf{RQ3}, we provide a platform analysis of four popular platforms, namely Horovod~\cite{horovod}, Megatron framework~\cite{shoeybi2020megatronlm}, DeepSpeed~\cite{deepspeed}, and HuggingFace~\cite{huggingface}, along with empirical analysis and our recommendations.
\end{itemize}


The remainder of this paper is structured as follows: 
Section~\ref{sec:systematic_framework} presents the details of our framework, including source code data translation, a brief introduction of the models under investigation, and the overall system flow to leverage transformer-based language models for vulnerability detection.  
All our results, including the performance of the models, are presented and discussed in Section~\ref{sec:results}. Section~\ref{sec:platformsurvey} analyzes the popular platforms and presents the challenges and recommendations for choosing the right platform to tune the models. 
Section~\ref{sec:background} presents the related works on machine learning-based vulnerability detection.
Finally, Section~\ref{sec:conclusion} concludes the paper.    

\section{Systematic framework}
\label{sec:systematic_framework}

\subsection{Data translation}
\label{sec:Data}

The models require formatted data to capture important features related to the vulnerabilities. Moreover, the model, including the transformer-based language model, inputs the vectorized form of the formatted data. The conversion of the data (\emph{e.g.}, C/C++ source codes) to an appropriate format, which is further transformed into vectorized inputs, is called data translation. In this process, the first step is to change the source code into code gadgets. 

\subsubsection{Code Gadgets and its extraction}
\begin{figure}[t]
\centering
   \includegraphics[width=1\linewidth]{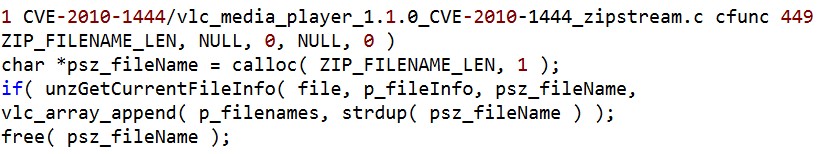}
   \caption{An example of a code gadget of a non-vulnerable library/API function. The first line of the gadget is a header, and the rest is its body.} 
   \vskip10pt
\label{fig:code_gadget-example}
\end{figure}

Code gadgets in software vulnerability are first proposed by Li et al.~\cite{main_paper}. It is generated as follows: 
\begin{itemize}[leftmargin=3mm]
\item Load all C/C++ files for analysis of relations between classes. 
\item Normalize source codes by applying regular expressions. This includes removing comments and non-ASCII characters.
\item Extract all function and variable definitions together with their usages.
\item Work through all source codes and if there is a library/API function call, perform a back-track as follows:
\begin{itemize}
    \item Extract all variable names from the function call.
    \item Stack up all lines which have relationships with the variables remaining within the scope of the library/API function. 
    \item If any variables are passed from a caller, perform another back-track for the caller.
\end{itemize}
\end{itemize}
Overall, each code gadget can be seen as an assembled semantically related statement slices having data dependency or control dependencies with each other. It can be associated either with a vulnerability or without any vulnerability. In this work, we consider the code gadgets formed based on data dependencies and labeled \emph{``1''} if they are vulnerable and \emph{``0''} otherwise. For an example of a code gadget, refer to Figure~\ref{fig:code_gadget-example}.
\begin{table}[t]
\vskip12pt
\footnotesize
\centering
\caption{Division of VulDeePecker dataset based on its type.}
\label{table:data}
\begin{adjustbox}{max width=0.5\textwidth}
\begin{tabular}{|c|c|c|c|c|c|}
\hline
\rowcolor[HTML]{FFFFFF} 
Dataset & \textbf{Type} & \multicolumn{1}{l|}{\cellcolor[HTML]{FFFFFF}\textbf{Original}} & \multicolumn{1}{l|}{\cellcolor[HTML]{FFFFFF}\textbf{Cleaned}} & \textbf{Train} & \textbf{Test} \\ \hline
\rowcolor[HTML]{FFFFFF} 
\cellcolor[HTML]{FFFFFF} & Buffer Error (BE) & 10440 & 7649 & 6161 & 1488 \\ \cline{2-6} 
\rowcolor[HTML]{FFFFFF} 
\multirow{-2}{*}{\cellcolor[HTML]{FFFFFF}Group 1} & Non-vulnerable & 29313 & 12262 & 9768 & 2494 \\ \hline
\rowcolor[HTML]{FFFFFF} 
\cellcolor[HTML]{FFFFFF} & \begin{tabular}[c]{@{}c@{}}Resource\\ Management Error (RME)\end{tabular} & 7285 & 2757 & 2214 & 543 \\ \cline{2-6} 
\rowcolor[HTML]{FFFFFF} 
\multirow{-2}{*}{\cellcolor[HTML]{FFFFFF}Group 2} & Non-vulnerable & 14600 & 5010 & 4000 & 1010 \\ \hline
\rowcolor[HTML]{FFFFFF} 
\multicolumn{1}{|l|}{\cellcolor[HTML]{FFFFFF}} & BE+ RME & 17725 & 10395 & 8368 & 2027 \\ \cline{2-6} 
\rowcolor[HTML]{FFFFFF} 
\multicolumn{1}{|l|}{\multirow{-2}{*}{\cellcolor[HTML]{FFFFFF}Group 3}} & Combined Non-vulnerable & 43913 & 17197 & 13704 & 3491 \\ \hline
\end{tabular}%
\end{adjustbox}
\end{table}

\subsubsection{Data preparation}
Code gadgets are processed through multiple stages before inputting into the model. 

\paragraph{Data cleaning:}
As the code gadgets are extracted from multiple sources, the dataset can have the following: (i) duplicate code gadgets with the same label and (ii) duplicate code gadgets with different labels ({\em i.e.,} label conflict). For example, we discover these two issues on the \emph{VulDeePecker dataset}~\cite{main_paper} (refer to Appendix~\ref{appendix:vul_data} for the details of this dataset).
Duplicate gadgets with the same label can leak data to the test set. On the other hand, duplicate gadgets with different labels have a negative impact on model training/testing. Thus, we have to clean the dataset. In this regard, firstly, we find the duplicate code gadget by mapping all gadgets into hash values using the SHA256 hashing algorithm provided by python \emph{hashlib} library. We choose hashing method for finding the duplicates because it is much faster than Regex or naive string comparison methods. For code gadgets with conflicting labels, we remove all such code gadgets, and for same code gadgets with the same labels, we removed their copies from the dataset. Refer to Table~\ref{table:data} for the number of samples in cleaned and uncleaned ({\em i.e.,} original) VulDeePecker dataset.        

\paragraph{Data pre-processing:}
Firstly, if there are any comments in the code gadget, those are removed. Secondly, user-defined names are replaced by their symbolic equivalents. This is done by replacing (i) user-defined function name by ``FUNC'' (or using consecutive natural numbers as postfix to ``FUNC'', like ``FUNC\_1'' and ``FUNC\_2'', if multiple functions), and (ii) user assigned variable name by ``VAR'' (or using consecutive natural numbers as postfix to ``VAR'', like ``VAR\_1'' and ``VAR\_2'', if multiple variables). This way, we normalize the code gadget. 
%
Thirdly, we create subsets of data based on the available vulnerabilities. For example, we make two sets of data from the VulDeePecker dataset; one with Buffer Error (BE) and its non-vulnerable versions, and the other with Resource Management Error (RME) and its non-vulnerable versions. As we perform both the binary classification and multi-class classification; we assign the labels in the following ways:
\begin{itemize}[leftmargin=3mm]
    \item For binary classification labeling, we perform experiments separately for each of the vulnerabilities. For example, BE and RME datasets of the VulDeePecker dataset. If code gadget has vulnerability its label is \emph{``1''}, and \emph{``0''} otherwise.  
    \item For multi-class classification labeling, we perform experiments on the union of the vulnerabilities, and we provide the label \emph{``0''} for the clean data and \emph{``1''} onward in an increasing order based on the available vulnerability types in the data. For example, the code gadget with BE, RME and non-vulnerable are labelled \emph{``1''}, \emph{``2''} and \emph{``0''}, respectively, in VulDeePecker dataset.
\end{itemize}

\paragraph{Dataset partitioning:}
Following our data pre-processing step, we divide the dataset into multiple groups for the experiments. For example, the VulDeePecker dataset is divided into three groups; Group~1 with BE and its non-vulnerable code gadgets, Group~2 with RME and its non-vulnerable code gadgets, and Group~3 with combined BE and RME and their non-vulnerable code gadgets.
Dataset of Group~1 and Group~2 are used while performing binary classification separately, and Group~3 is used while performing three-class classification. 
The dataset is split into a train and test set at a ratio of 80:20 (\emph{e.g.}, the number of samples in train and test in the VulDeePecker dataset are as shown in Table~\ref{table:data}). We perform three-fold cross-validation and present the overall results for the testing set.

\paragraph{Word embeddings:}
\begin{table}[]
\centering
\caption{The list of the word embedding methods used for the models considered in this paper. Refer to Huggingface's site~\cite{huggingface_token} for details on the implementation of these word embedding methods and functions.}
\resizebox{\columnwidth}{!}{%
\begin{tabular}{|l|c|l|}
\hline
\multicolumn{1}{|c|}{\textbf{Word Embeddings}} & \textbf{\begin{tabular}[c]{@{}c@{}}Embedding\\ size\end{tabular}} & \multicolumn{1}{c|}{\textbf{Models}} \\ \hline
Tokenizer (our own)  and Word2Vec & 512 & BiLSTM, BiGRU \\ \hline
\begin{tabular}[c]{@{}l@{}}Huggingface's  Bert Tokenizer \\ (Tokenize based on WordPiece)\end{tabular} & 512 & \begin{tabular}[c]{@{}l@{}}BERT (BERTBase, MegatronBERT)\end{tabular} \\ \hline
\begin{tabular}[c]{@{}l@{}}Huggingface's DistilBert Tokenizer\\ (Runs end-to-end tokenization based \\ on punctuation splitting and WordPiece)\end{tabular} & 512 & DistilBERT \\ \hline
\begin{tabular}[c]{@{}l@{}}Huggingface's Roberta Tokenizer\\ (Derived from GPT-2 tokenizer using\\ byte-level Byte-Pair-Encoding)\end{tabular} & 514 & RoBERTa, CodeBERT \\ \hline
\begin{tabular}[c]{@{}l@{}}Huggingface's GPT-2 Tokenizer\\ (Based on byte-level Byte-Pair-Encoding)\end{tabular} & 1024 & \begin{tabular}[c]{@{}l@{}}GPT-2 (GPT-2 Base, GPT-2 Large, \\ GPT-2 XL, MegatronGPT-2), GPT-J\end{tabular} \\ \hline
\begin{tabular}[c]{@{}l@{}}Huggingface's GPT-2 Tokenizer\\ (Based on byte-level Byte-Pair-Encoding)\end{tabular} & 2048 & GPT-J \\
\hline
\end{tabular}%
}
\label{tab:word_emb}
\vskip-5pt
\end{table}
The learned representation for text data, such as source code, where words are mapped to numeric data vectors in a predefined vector space that encodes the word's meaning, is called word embedding. This process mainly has two associated operations: tokenization and embedding. Tokenization converts the input sentence into characters, words, or sub-words, which are semantically-viable segments, and these smaller segments are called tokens. In embedding, these tokens are mapped to relevant vectors using various trained embedding methods, capturing the context around the token in the sentence. In this regard, we use Word2Vec for the BiLSTM and BiGRU models, a WordPiece-based (subword-based) tokenizer for BERT models, and a  byte-level Byte-Pair-Encoding based tokenizer for GPT-2 models. The embedding functions corresponding to the models investigated in this paper are listed in Table~\ref{tab:word_emb}.

\subsection{Models}
Our studies investigate a recurrent neural network-based model and various transformer-based models for vulnerability detection. These models are listed in Table~\ref{tab:models}. We briefly discuss these models in the following (refer to Appendix~\ref{appendix:models} for details). 

Under recurrent neural networks, we consider BiLSTM~\cite{lstm} and BiGRU~\cite{gru}. We use BiLSTM and BiGRU models that are close to those used in software vulnerability detection as presented in~\cite{main_paper} (the model is not publicly available) and \cite{sysevr}, respectively. These models have two issues: (i) they cannot perform well when the input sequence is too long, and (ii) hard to parallelize their operations at sequence level~\cite{attention}. These shortcomings are removed in transformers. 

Under transformer-based models, we consider BERT~\cite{bert}, DistilBERT~\cite{distilbert}, Robustly optimized BERT approach (RoBERTa)~\cite{roberta}, CodeBERT~\cite{codebert}, GPT-2~\cite{gpt}, and Megatron-Language Model variants~\cite{shoeybi2020megatronlm}. In this paper, we consider the BERT model, called BERTBase, having 110M parameters with \emph{12 layers, 768 hidden size, and 12 attention heads}. DistilBERT is a smaller, faster, cheaper, and lighter version of BERT, and it has 66M parameters with \emph{6 layers, 768 hidden size, and 12 attention heads}. RoBERTa has the BERTBase architecture and is pre-trained towards better optimization, performance, and robustness across NLP tasks. CodeBERT is a bimodal pre-trained transformer model for both programming and natural languages. It is trained on bimodal data (code \& documents)~\cite{codeserachnet}, with codes from Python, Java, JavaScript, Ruby, Go, and PHP. Its architecture follows BERT and RoBERTa.  

For GPT-based models, their architecture is based on decoder blocks of transformer and masked self-attention~\cite{gpt}. GPT outperforms available models that are based on recursive neural networks, convolutional neural networks, and LSTMs~\cite{gpt}. In this paper, we consider GPT-2 models of various sizes: (1) GPT-2 Base, which has 117M parameters with \emph{12 layers, 768 hidden size, and 12 attention heads}, (2) GPT-2 Large, which has 774M parameters with \emph{36 layers, 1280 hidden size, and 20 attention heads}, (3) GPT-2 XL, which has 1.5B parameters with \emph{48 layers, 1600 hidden size, and 25 attention heads}, and (4) GPT-J, which has 6B parameters with \emph{28 layers, 4096 hidden size, and 16 attention heads}, pre-trained on the dataset called Pile having a diverse text data~\cite{pile}.        

Megatron-LMs are transformer-based models developed by NVIDIA. They are generated by enabling intra-layer model-parallelism on the architecture level of the existing language models, such as BERT and GPT-2~\cite{shoeybi2020megatronlm}. In this paper, we consider Megatron versions of BERT and GPT-2 models provided by Nvidia; (1) MegatraonBERT having 345M parameters with \emph{24 layers, 1024 hidden size, and 16 attention heads}, and (2) MegatronGPT-2 having 345M parameters with \emph{24 layers, 1024 hidden size, and 16 attention heads}.

\begin{table}[]
\centering
\caption{Models considered in our studies and their sizes.}
\label{tab:models}
\begin{adjustbox}{max width=0.4\textwidth}
\begin{tabular}{|l|c|c|c|}
\hline
\multicolumn{1}{|c|}{\textbf{Provider}} & \textbf{Language Model} & \textbf{Size} & \textbf{\#Parameters} \\ \hline
\rowcolor[HTML]{FFFFFF} 
\cellcolor[HTML]{FFFFFF} & MegatronBERT & Standard & 345M \\ \cline{2-4} 
\rowcolor[HTML]{FFFFFF} 
\multirow{-2}{*}{\cellcolor[HTML]{FFFFFF}Nvidia} & MegatronGPT-2 & Standard & 345M \\ \hline
Hugging Face & BERT & Base Model & 110M \\ \hline
\rowcolor[HTML]{FFFFFF} 
\cellcolor[HTML]{FFFFFF} & \cellcolor[HTML]{FFFFFF} & Base Model & 117M \\ \cline{3-4} 
\rowcolor[HTML]{FFFFFF} 
\cellcolor[HTML]{FFFFFF} & \cellcolor[HTML]{FFFFFF} & Large Model & 774M \\ \cline{3-4} 
\rowcolor[HTML]{FFFFFF} 
\multirow{-3}{*}{\cellcolor[HTML]{FFFFFF}OpenAI} & \multirow{-3}{*}{\cellcolor[HTML]{FFFFFF}GPT-2} & XL Model & 1.5B \\ \hline
EleutherAI & GPT-J & Standard & 6B \\ \hline
\rowcolor[HTML]{FFFFFF} 
Hugging Face & DistilBERT & Standard & 66M \\ \hline
Microsoft & CodeBERT & Standard & 125M \\ \hline
\rowcolor[HTML]{FFFFFF} 
Hugging Face & RoBERTa & Standard & 125M \\ \hline
VulDeePecker & BiLSTM & Standard & 1.2M \\ \hline
SySeVR & BiGRU & Standard & 1.6M \\ \hline
\end{tabular}%
\end{adjustbox}
\vskip-10pt
\end{table}

\subsection{System flow}
\label{sec:systemflow}
\begin{figure}[t]
\vskip-5pt
\centering
   \includegraphics[trim=5cm 0.5cm 5cm 0.5cm, clip=true, width=\linewidth]{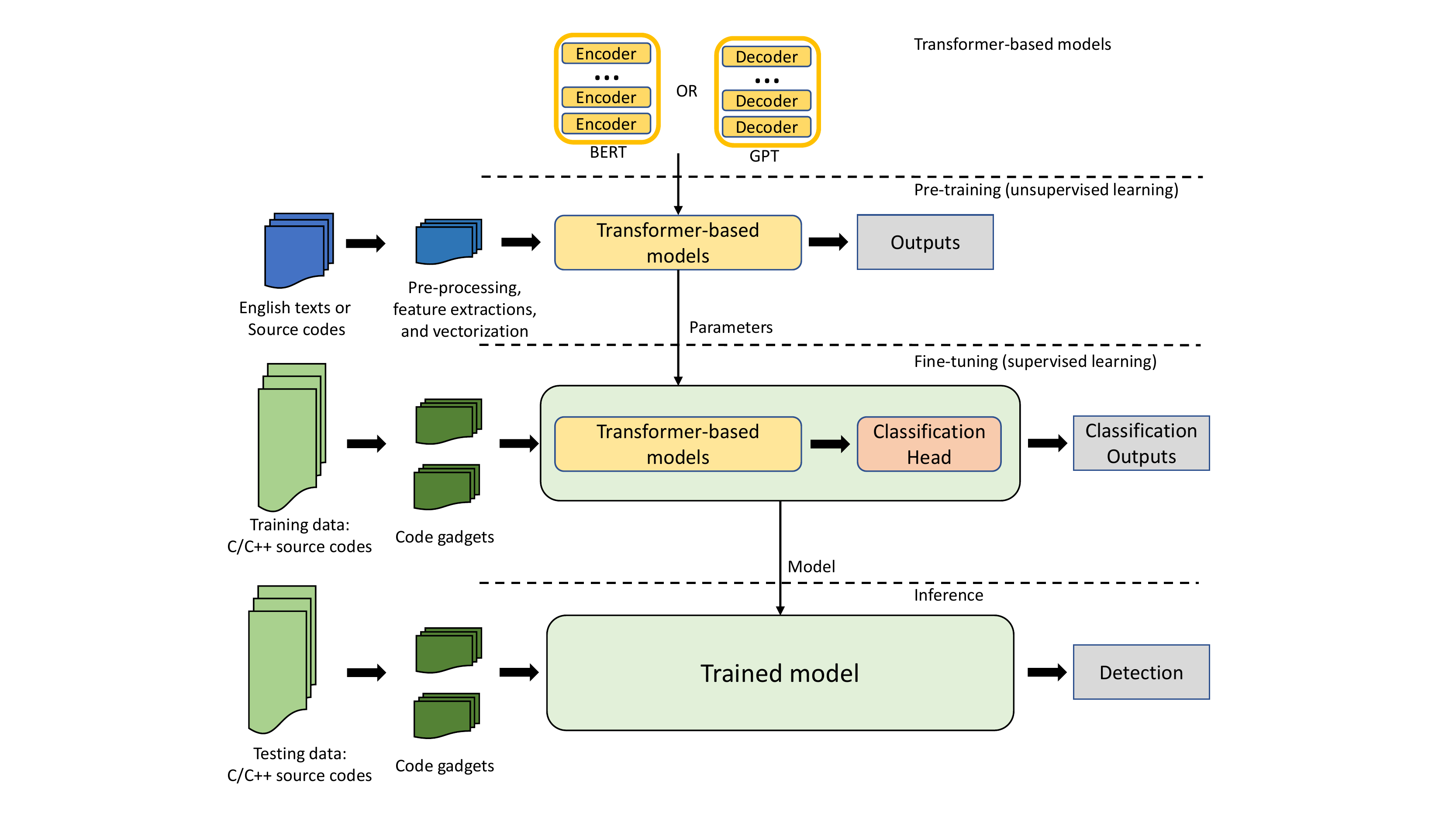}
   \caption{Our systematic framework for software vulnerability detection: pre-training, fine-tuning and inference.}
\label{fig:systemflow}
\vskip10pt
\end{figure}

In this paper, we consider the pre-trained model approach of transfer learning, where we first pick a pre-trained transformer-based model, then a classification head is attached at the top of the final layer of the model, and the resulting model is fine-tuned through the software vulnerability dataset consisting of C/C++ source codes. The overall systematic framework is illustrated in Figure~\ref{fig:systemflow}, where the process is divided into three main steps: pre-training, fine-tuning, and inference.

\subsubsection{Pre-training}
In this work, we choose the transformer-based models trained on a large corpus of English texts, except CodeBERT, which is trained on source codes of various programming languages. During the pre-training step, the models are trained in an unsupervised fashion to understand the context of the English words and sentences, including their syntax and semantics. 

Except for CodeBERT, all BERT-based models, including RoBERTa, and DistilBERT, are pre-trained using \emph{masked language modeling} that masked 15\% of input token positions randomly, then those masked tokens are predicted, and the model is optimized based on the original masked tokens and the predicted tokens. Moreover, while training, the masked tokens are replaced (1) by token ``[MASK]" for 80\%, (2) by a random token different than the replaced one for 10\%, and (3) by leaving the masked token as it is for 10\%. The models learn a bidirectional representation of the sentence through masked language modeling. On the other hand, the CodeBERT model uses two objectives; masked language modeling for the generator blocks and \emph{replaced token detection} for the discriminator block~\cite{codebert}. The replaced token detection enables the discriminator block to learn effectively about the real and fake output tokens from the generators instead of predicting masked tokens like in masked language modeling. 
Unlike BERT, GPT-based models, including MegatronGPT-2, use \emph{casual language modeling} objective in which the next word is predicted by providing all previous words of an input sentence~\cite{gpt2}. Thus, the learning in the GPT model is unidirectional in nature; hence it is also called an autoregressive model.

\subsubsection{Fine-tuning and inference}
Initially, all of our models, except CodeBERT, are pre-trained on the natural language data. Now they are specialized in the software vulnerability detection task, which is a classification task, through fine-tuning. Usually, this is performed with a small dataset than the pre-training dataset and carried under supervised learning that requires the knowledge of the data labels. For this task, a classification head is added to the top of the pre-trained model. In this regard, we use the architecture of the classification heads as depicted in Table~\ref{tab:classificaitonhead} for the models. For CodeBERT, only the discriminator block is used for fine-tuning and inference.    

\begin{table}[t]
\centering
\caption{Architecture of the classification heads for our models' fine tuning.}
\label{tab:classificaitonhead}
\resizebox{0.98\columnwidth}{!}{%
\begin{tabular}{|l|l|}
\hline
\rowcolor[HTML]{FFFFFF} 
\textbf{Language Model} & \textbf{Classification Head} \\ \hline
\rowcolor[HTML]{FFFFFF} 
BERT, MegatronBERT & Dropout Layer + Linear Layer (size = 3072) \\ \hline
\rowcolor[HTML]{FFFFFF} 
DistilBERT & \begin{tabular}[l]{@{}l@{}}Linear Layer (size = 3072) + ReLU + Dropout Layer + \\ Linear Layer (size = 3072)\end{tabular} \\ \hline
\rowcolor[HTML]{FFFFFF} 
RoBERTa, CodeBERT & \begin{tabular}[l]{@{}l@{}}Dropout Layer + Linear Layer (size = 3072) + tanh + \\ Dropout Layer + Linear Layer (size = 3072)\end{tabular} \\ \hline
\rowcolor[HTML]{FFFFFF} 
GPT-2, MegatronGPT-2 & Dropout Layer + Linear Layer (size = 1024) \\ \hline
\rowcolor[HTML]{FFFFFF} 
GPT-J & Linear Layer (size = 2048) \\ \hline
\end{tabular}%
}
\vskip-10pt
\end{table}

For all the transformer-based models in this paper, we allow the entire model architecture to update during the fine-tuning step, which is performed with a low learning rate. This allows the model's pre-trained weights to be adjusted based on our software vulnerability dataset (C/C++ source codes). Besides, all the transformer-based models are fine-tuned for 10 epochs (whereas BiLSTM and BiGRU are trained for 100 and 20 epochs, respectively, following the existing literature). Then, the resulting models are tested on the test dataset in the inference step, and various evaluation metrics, False Positive Rate (FPR) and False Negative Rate (FNR) (refer to Appendix~\ref{appendix:performance_metrics} for details), are calculated.


\section{Experiments and Results}
\label{sec:results}
This section presents our empirical results and observations under various experiments.
All the transformer-based models in this section are fine-tuned with the training dataset for 10 epochs with {\small \tt learning\_rate = 1.0e-05, weight\_decay = 0.06}, and {\small \tt warmup\_steps = 500}. The weight decay and the warmup steps control the learning rate with the iterations while training. 
The total number of iterations is calculated as follows:
\begin{equation*}
    \text{Total iterations} = \frac{\text{Total number of samples} * \text{Training epochs}}{\text{Batch size}}.
\end{equation*}
In all our experiments, the batch size equals 16, and a linear learning rate scheduler is used.
Now, when we set {\small \tt warmup\_steps = 500}, and {\small \tt weight\_decay = 0.06}, then every 500 iterations (steps), our learning rate will be decayed by 6\%. 
\subsection{Need for clean data and testing our BiLSTM}
\label{sec:bilstm}

In this paper, we consider a BiLSTM model similar to the one reported by Li et al.~\cite{main_paper}. We call the previously reported model \emph{VulDeePecker Original} model. Considering the VulDeePecker dataset under binary classification, we experiment with our BiLSTM model and BERTBase model on the clean and the original datasets as depicted in Table~\ref{table:data}. In binary classification, the model considers only one vulnerability; for other vulnerabilities, the model is trained separately and independently. The BiLSTM model is trained for 100 epochs, and the BERT Base model is fine-tuned for 10 epochs.  
Our results are depicted in Table~\ref{tab:cleanvsunclean_biLSTM}\footnote{\label{fold}These results are average results of three folds of the test sets. Let a test set is represented by $T_i$, for each fold $i\in \{1,2,3\}$. If $ D$ represents the clean VulDeePecker dataset, then the corresponding fine-tuning train set is $D \setminus T_i$. The results for BiLSTM are obtained at the end of the 100 epoch, whereas for all other models, it is obtained at the end of the 10 epoch.\label{data_footnote}}, and they demonstrate the following:
\begin{itemize}[leftmargin=3mm\relax]
    \item The results of BiLSTM with the original data are better than with the clean data for both Group~1 and Group~2 datasets. In the original dataset, considering the high number of redundant samples in the original dataset (see Table~\ref{tab:data_redundancy}), we confirm a high possibility of data leakage in the test dataset. Consequently, it leads to apparently better performance. Thus, for fair results, clean data is required for our experiments. We get the same inference from the performance of BERTBase on the Group~2 dataset; however, there are improvements for the Buffer Error.  
    \item The performance of our BiLSTM on original data and VulDeePecker Original is similar for the Group~2 dataset but different for the Group~1 dataset, specifically, FPR and FNR. Although we cannot confirm the closeness of the original and our BiLSTM model, we keep our BiLSTM model and its result for comparison. 
\end{itemize}

\begin{table}[t]
\vskip12pt
\centering
\caption{Number of samples (code gadgets) having confliction and redundancy in the uncleaned VulDeePecker dataset.}
\resizebox{0.48\textwidth}{!}{%
\begin{tabular}{|l|c|c|c|}
\hline
 & Confliction & Redundancy & Both Confliction \& Redundancy \\ \hline
CWE119 - Buffer & 645 & 18,989 & 208 \\ \hline
CWE399 - Resource & 86 & 13,992 & 40 \\ \hline
Sub-total & 731 & 32,981 & 248 \\ \hline
Merged & 741 & 33,050 & 257 \\ \hline
\end{tabular}%
}
\label{tab:data_redundancy}
\end{table}

\begin{table}[t]
\vskip3pt
\caption{Test performance of BiLSTM and BERTBase for the binary classification on the original and clean dataset.}
\centering
\resizebox{1\linewidth}{!}{%
\begin{tabular}{|c|c|c|cll|cll|l|cll|cll|}
\cline{1-9} \cline{11-16}
\textbf{\begin{tabular}[c]{@{}c@{}}Dataset and \\ Vulnerability\end{tabular}}                       & \textbf{Metrics} & \textbf{\begin{tabular}[c]{@{}c@{}}VulDeePecker\\      Original~\cite{main_paper}\end{tabular}} & \multicolumn{3}{c|}{\textbf{\begin{tabular}[c]{@{}c@{}}BiLSTM\\ (Original Data)\end{tabular}}} & \multicolumn{3}{c|}{\textbf{\begin{tabular}[c]{@{}c@{}}BiLSTM\\ (Clean Data)\end{tabular}}} &                          & \multicolumn{3}{c|}{\textbf{\begin{tabular}[c]{@{}c@{}}BERTBase\\ (Original Data)\end{tabular}}} & \multicolumn{3}{c|}{\textbf{\begin{tabular}[c]{@{}c@{}}BERTBase \\ (Clean Data)\end{tabular}}} \\ \cline{1-9} \cline{11-16} 
                                                                                                    & FPR              & \cellcolor[HTML]{9AFF99}2.90\%                                                & \multicolumn{3}{c|}{24.99\%}                                                                   & \multicolumn{3}{c|}{33.86\%}                                                                & \cellcolor[HTML]{FFFFFF} & \multicolumn{3}{c|}{\cellcolor[HTML]{9AFF99}2.49\%}                                              & \multicolumn{3}{c|}{4.05\%}                                                                    \\ \cline{2-9} \cline{11-16} 
                                                                                                    & FNR              & 18.00\%                                                                       & \multicolumn{3}{c|}{\cellcolor[HTML]{9AFF99}8.46\%}                                            & \multicolumn{3}{c|}{15.27\%}                                                                & \cellcolor[HTML]{FFFFFF} & \multicolumn{3}{c|}{10.21\%}                                                                     & \multicolumn{3}{c|}{\cellcolor[HTML]{9AFF99}6.52\%}                                            \\ \cline{2-9} \cline{11-16} 
                                                                                                    & Precision        & \cellcolor[HTML]{9AFF99}82.00\%                                               & \multicolumn{3}{c|}{78.57\%}                                                                   & \multicolumn{3}{c|}{71.46\%}                                                                & \cellcolor[HTML]{FFFFFF} & \multicolumn{3}{c|}{92.76\%}                                                                     & \multicolumn{3}{c|}{\cellcolor[HTML]{9AFF99}93.56\%}                                           \\ \cline{2-9} \cline{11-16} 
                                                                                                    & Recall           & \cellcolor[HTML]{9AFF99}91.70\%                                               & \multicolumn{3}{c|}{91.54\%}                                                                   & \multicolumn{3}{c|}{84.73\%}                                                                & \cellcolor[HTML]{FFFFFF} & \multicolumn{3}{c|}{89.79\%}                                                                     & \multicolumn{3}{c|}{\cellcolor[HTML]{9AFF99}93.48\%}                                           \\ \cline{2-9} \cline{11-16} 
\multirow{-5}{*}{\begin{tabular}[c]{@{}c@{}}Group 1,\\ Buffer Error\end{tabular}}                   & F1-score         & \cellcolor[HTML]{9AFF99}86.60\%                                               & \multicolumn{3}{c|}{84.55\%}                                                                   & \multicolumn{3}{c|}{77.50\%}                                                                & \cellcolor[HTML]{FFFFFF} & \multicolumn{3}{c|}{91.25\%}                                                                     & \multicolumn{3}{c|}{\cellcolor[HTML]{9AFF99}93.52\%}                                           \\ \cline{1-9} \cline{11-16} 
                                                                                                    & FPR              & \cellcolor[HTML]{9AFF99}2.80\%                                                & \multicolumn{3}{c|}{5.93\%}                                                                    & \multicolumn{3}{c|}{16.10\%}                                                                & \cellcolor[HTML]{FFFFFF} & \multicolumn{3}{c|}{\cellcolor[HTML]{9AFF99}1.03\%}                                              & \multicolumn{3}{c|}{3.32\%}                                                                    \\ \cline{2-9} \cline{11-16} 
                                                                                                    & FNR              & \cellcolor[HTML]{9AFF99}4.70\%                                                & \multicolumn{3}{c|}{\cellcolor[HTML]{FFFFFF}5.76\%}                                            & \multicolumn{3}{c|}{\cellcolor[HTML]{FFFFFF}12.63\%}                                        & \cellcolor[HTML]{FFFFFF} & \multicolumn{3}{c|}{\cellcolor[HTML]{9AFF99}3.48\%}                                              & \multicolumn{3}{c|}{5.82\%}                                                                    \\ \cline{2-9} \cline{11-16} 
                                                                                                    & Precision        & \cellcolor[HTML]{9AFF99}95.30\%                                               & \multicolumn{3}{c|}{\cellcolor[HTML]{FFFFFF}94.09\%}                                           & \multicolumn{3}{c|}{\cellcolor[HTML]{FFFFFF}84.50\%}                                        & \cellcolor[HTML]{FFFFFF} & \multicolumn{3}{c|}{\cellcolor[HTML]{9AFF99}97.92\%}                                             & \multicolumn{3}{c|}{93.79\%}                                                                   \\ \cline{2-9} \cline{11-16} 
                                                                                                    & Recall           & \cellcolor[HTML]{9AFF99}94.60\%                                               & \multicolumn{3}{c|}{\cellcolor[HTML]{FFFFFF}94.24\%}                                           & \multicolumn{3}{c|}{\cellcolor[HTML]{FFFFFF}87.37\%}                                        & \cellcolor[HTML]{FFFFFF} & \multicolumn{3}{c|}{\cellcolor[HTML]{9AFF99}96.52\%}                                             & \multicolumn{3}{c|}{94.18\%}                                                                   \\ \cline{2-9} \cline{11-16} 
\multirow{-5}{*}{\begin{tabular}[c]{@{}c@{}}Group 2,\\ Resource \\ Management\\ Error\end{tabular}} & F1-score         & \cellcolor[HTML]{9AFF99}95.00\%                                               & \multicolumn{3}{c|}{\cellcolor[HTML]{FFFFFF}94.16\%}                                           & \multicolumn{3}{c|}{\cellcolor[HTML]{FFFFFF}85.86\%}                                        & \cellcolor[HTML]{FFFFFF} & \multicolumn{3}{c|}{\cellcolor[HTML]{9AFF99}97.21\%}                                             & \multicolumn{3}{c|}{93.98\%}                                                                   \\ \cline{1-9} \cline{11-16} 
\end{tabular}
}
\label{tab:cleanvsunclean_biLSTM}
\vskip-10pt
\end{table}
\subsection{Performance of the transformer-based models on VulDeePecker dataset}
\label{sec:all_performance}

In this paper, we perform experiments considering binary and multi-class classifications separately. Moreover, binary classification enables us to know the performance of our models on each specific vulnerability independently. In contrast, multi-class classification allows us to see the model's ability to deal with multiple vulnerabilities jointly. In our experiments, we find the differences in the results, and they are presented in the following sections. 

\subsubsection{Binary classification}
\label{sec:binary_classification}
\begin{table*}[t]
\centering
\caption[]{Average test performance\footref{fold} of various models for the binary classification on clean VulDeePecker dataset Group 1 and Group 2.}
\label{tab:binaryclassfication_all_results}
\resizebox{\textwidth}{!}{%
\begin{tabular}{|c|c|c|c|c|c|c|c|c|c|c|c|c|c|c|}
\hline
\textbf{\begin{tabular}[c]{@{}c@{}}Dataset and \\ Vulnerability\end{tabular}}                              & \textbf{Metrics} & \textbf{\begin{tabular}[c]{@{}c@{}}VulDeePecker\\ Original~\cite{main_paper}\end{tabular}} & \textbf{BiLSTM} & \textbf{BiGRU} & \textbf{BERTBase} & \textbf{GPT-2 Base} & \textbf{CodeBERT} & \textbf{DistilBERT} & \textbf{RoBERTa} & \textbf{GPT-2 Large}            & \textbf{GPT-2 XL}               & \textbf{MegatronBERT} & \textbf{MegatronGPT-2}          & \textbf{GPT-J} \\ \hline
                                                                                                           & FPR              & 2.90\%                                                                   & 33.86\%         & 15.19\%        & 4.05\%            & 4.20\%              & 2.97\%            & 3.85\%              & 4.48\%           & 2.67\%                          & \cellcolor[HTML]{9AFF99}2.66\%  & 3.25\%                & 2.81\%                          & 2.74\%         \\ \cline{2-15} 
                                                                                                           & FNR              & 18.00\%                                                                  & 15.27\%         & 35.49\%        & 6.52\%            & 6.44\%              & 4.85\%            & 6.75\%              & 6.56\%           & \cellcolor[HTML]{9AFF99}4.72\%  & 4.94\%                          & 5.24\%                & 5.61\%                          & 5.76\%         \\ \cline{2-15} 
                                                                                                           & Precision        & 82.00\%                                                                  & 71.46\%         & 73.04\%        & 93.56\%           & 93.35\%             & 95.27\%           & 93.86\%             & 92.95\%          & 95.74\%                         & \cellcolor[HTML]{9AFF99}95.75\% & 94.84\%               & 95.49\%                         & 95.61\%        \\ \cline{2-15} 
                                                                                                           & Recall           & 91.70\%                                                                  & 84.73\%         & 64.51\%        & 93.48\%           & 93.56\%             & 95.15\%           & 93.25\%             & 93.44\%          & \cellcolor[HTML]{9AFF99}95.28\% & 95.06\%                         & 94.76\%               & 94.39\%                         & 94.24\%        \\ \cline{2-15} 
\multirow{-5}{*}{\begin{tabular}[c]{@{}c@{}}Group 1,\\ Buffer Error\\ (BE)\end{tabular}}                   & F1-score         & 86.60\%                                                                  & 77.50\%         & 68.37\%        & 93.52\%           & 93.45\%             & 95.21\%           & 93.55\%             & 93.19\%          & \cellcolor[HTML]{9AFF99}95.51\% & 95.40\%                         & 94.80\%               & 94.94\%                         & 94.90\%        \\ \hline
                                                                                                           & FPR              & 2.80\%                                                                   & 16.10\%         & 4.40\%         & 3.32\%            & 3.81\%              & 3.09\%            & 4.40\%              & 2.92\%           & \cellcolor[HTML]{9AFF99}1.71\%  & 1.77\%                          & 2.40\%                & 2.50\%                          & 2.17\%         \\ \cline{2-15} 
                                                                                                           & FNR              & 4.70\%                                                                   & 12.63\%         & 10.34\%        & 5.82\%            & 5.01\%              & 4.71\%            & 7.12\%              & 5.20\%           & 3.10\%                          & 3.28\%                          & 3.53\%                & \cellcolor[HTML]{9AFF99}3.03\%  & 3.96\%         \\ \cline{2-15} 
                                                                                                           & Precision        & 95.30\%                                                                  & 84.50\%         & 91.58\%        & 93.79\%           & 92.97\%             & 94.25\%           & 91.82\%             & 94.51\%          & \cellcolor[HTML]{9AFF99}96.79\% & 96.66\%                         & 95.54\%               & 95.38\%                         & 95.96\%        \\ \cline{2-15} 
                                                                                                           & Recall           & 94.60\%                                                                  & 87.37\%         & 89.66\%        & 94.18\%           & 94.99\%             & 95.29\%           & 92.88\%             & 94.80\%          & 96.90\%                         & 96.72\%                         & 96.47\%               & \cellcolor[HTML]{9AFF99}96.97\% & 96.04\%        \\ \cline{2-15} 
\multirow{-5}{*}{\begin{tabular}[c]{@{}c@{}}Group 2,\\ Resource\\ Management\\ Error\\ (RME)\end{tabular}} & F1-score         & 95.00\%                                                                  & 85.86\%         & 90.59\%        & 93.98\%           & 93.96\%             & 94.76\%           & 92.34\%             & 94.65\%          & \cellcolor[HTML]{9AFF99}96.84\% & 96.69\%                         & 96.00\%               & 96.16\%                         & 95.98\%        \\ \hline
\end{tabular}
}
\end{table*}

\begin{table*}[t]
\vskip10pt
\centering
\caption[]{Average test performance\footref{fold} of various models for the multi-class classification on clean VulDeePecker dataset Group 3.}
\resizebox{\textwidth}{!}{%
\begin{tabular}{|c|c|c|c|c|c|c|c|c|c|c|c|c|c|}
\hline
\textbf{\begin{tabular}[c]{@{}c@{}}Dataset and \\ Vulnerability\end{tabular}}                                 & \textbf{Metrics} & \textbf{BiLSTM} & \textbf{BiGRU} & \textbf{BERTBase} & \textbf{GPT-2 Base} & \textbf{CodeBERT} & \textbf{DistilBERT} & \textbf{RoBERTa} & \textbf{GPT-2 Large}             & \textbf{GPT-2 XL}                & \textbf{MegatronBERT}           & \textbf{MegatronGPT-2} & \textbf{GPT-J}                  \\ \hline
                                                                                                              & FPR              & 21.29\%         & 7.03\%         & 2.37\%            & 2.95\%              & 1.91\%            & 2.42\%              & 2.41\%           & 1.60\%                          & 1.47\%                          & 1.83\%                          & 2.03\%                 & \cellcolor[HTML]{9AFF99}1.44\%  \\ \cline{2-14} 
                                                                                                              & FNR              & 13.95\%         & 38.64\%        & 4.85\%            & 5.41\%              & 4.83\%            & 5.83\%              & 5.68\%           & 4.96\%                          & 4.77\%                          & \cellcolor[HTML]{9AFF99}4.61\%  & 5.08\%                 & 5.22\%                          \\ \cline{2-14} 
                                                                                                              & Precision        & 61.00\%         & 78.41\%        & 93.95\%           & 92.55\%             & 95.08\%           & 93.78\%             & 93.82\%          & 95.84\%                         & 96.17\%                         & 95.28\%                         & 94.78\%                & \cellcolor[HTML]{9AFF99}96.22\% \\ \cline{2-14} 
                                                                                                              & Recall           & 86.05\%         & 61.36\%        & 95.15\%           & 94.59\%             & 95.17\%           & 94.17\%             & 94.32\%          & 95.04\%                         & 95.23\%                         & \cellcolor[HTML]{9AFF99}95.39\% & 94.92\%                & 94.78\%                         \\ \cline{2-14} 
\multirow{-5}{*}{\begin{tabular}[c]{@{}c@{}}Group 3, \\ Buffer Error\\ (BE)\end{tabular}}                     & F1-score         & 71.39\%         & 68.03\%        & 94.55\%           & 93.56\%             & 95.12\%           & 93.97\%             & 94.07\%          & 95.43\%                         & \cellcolor[HTML]{9AFF99}95.70\% & 95.34\%                         & 94.85\%                & 95.49\%                         \\ \hline
                                                                                                              & FPR              & 3.40\%          & 0.66\%         & 0.66\%            & 1.02\%              & 0.64\%            & 0.73\%              & 0.75\%           & 0.42\%                          & 0.42\%                          & 0.50\%                          & 0.56\%                 & \cellcolor[HTML]{9AFF99}0.25\%  \\ \cline{2-14} 
                                                                                                              & FNR              & 10.68\%         & 7.49\%         & 4.07\%            & 4.67\%              & 4.12\%            & 4.55\%              & 4.07\%           & \cellcolor[HTML]{9AFF99}2.85\%  & 3.10\%                          & 3.64\%                          & 3.27\%                 & 5.88\%                          \\ \cline{2-14} 
                                                                                                              & Precision        & 74.48\%         & 93.99\%        & 94.12\%           & 91.20\%             & 94.34\%           & 93.54\%             & 93.40\%          & 96.23\%                         & 96.27\%                         & 95.52\%                         & 95.02\%                & \cellcolor[HTML]{9AFF99}97.68\% \\ \cline{2-14} 
                                                                                                              & Recall           & 89.32\%         & 92.51\%        & 95.93\%           & 95.33\%             & 95.88\%           & 95.45\%             & 95.93\%          & \cellcolor[HTML]{9AFF99}97.15\% & 96.90\%                         & 96.36\%                         & 96.73\%                & 94.12\%                         \\ \cline{2-14} 
\multirow{-5}{*}{\begin{tabular}[c]{@{}c@{}}Group 3, \\ Resource \\ Management \\ Error\\ (RME)\end{tabular}} & F1-score         & 81.20\%         & 93.23\%        & 95.02\%           & 93.21\%             & 95.10\%           & 94.48\%             & 94.65\%          & \cellcolor[HTML]{9AFF99}96.68\% & 96.58\%                         & 95.93\%                         & 95.86\%                & 95.87\%                         \\ \hline
                                                                                                              & Precision        & 64.14\%         & 83.08\%        & 93.99\%           & 92.19\%             & 94.88\%           & 93.72\%             & 93.71\%          & 95.94\%                         & 96.20\%                         & 95.34\%                         & 94.83\%                & \cellcolor[HTML]{9AFF99}96.60\% \\ \cline{2-14} 
                                                                                                              & Recall           & 86.92\%         & 69.57\%        & 95.36\%           & 94.79\%             & 95.36\%           & 94.51\%             & 94.74\%          & 95.59\%                         & \cellcolor[HTML]{9AFF99}95.68\% & 95.65\%                         & 95.39\%                & 94.61\%                         \\ \cline{2-14} 
\multirow{-3}{*}{\begin{tabular}[c]{@{}c@{}}Group 3, \\ BE + RME \\ (Global Avg.)\end{tabular}}               & F1-score         & 73.80\%         & 75.25\%        & 94.67\%           & 93.47\%             & 95.12\%           & 94.11\%             & 94.22\%          & 95.76\%                         & \cellcolor[HTML]{9AFF99}95.94\% & 95.49\%                         & 95.11\%                & 95.59\%                         \\ \hline
                                                                                                              & Precision        & 67.74\%         & 86.20\%        & 94.04\%           & 91.88\%             & 94.71\%           & 93.66\%             & 93.61\%          & 96.03\%                         & 96.22\%                         & 95.40\%                         & 94.90\%                & \cellcolor[HTML]{9AFF99}96.95\% \\ \cline{2-14} 
                                                                                                              & Recall           & 87.69\%         & 76.93\%        & 95.54\%           & 94.96\%             & 95.53\%           & 94.81\%             & 95.12\%          & \cellcolor[HTML]{9AFF99}96.09\% & 96.07\%                         & 95.87\%                         & 95.82\%                & 94.45\%                         \\ \cline{2-14} 
\multirow{-3}{*}{\begin{tabular}[c]{@{}c@{}}Group 3, \\ BE+ RME \\ (Macro Avg.)\end{tabular}}                 & F1-score         & 76.42\%         & 81.08\%        & 94.78\%           & 93.39\%             & 95.11\%           & 94.23\%             & 94.36\%          & 96.06\%                         & \cellcolor[HTML]{9AFF99}96.15\% & 95.63\%                         & 95.36\%                & 95.68\%                         \\ \hline
\end{tabular}
}
\label{tab:multiclass_classification}
\end{table*}

\begin{figure}[t]
	\centering
		\subfigure[Binary Classification.]{
			\includegraphics[width=0.9\columnwidth]{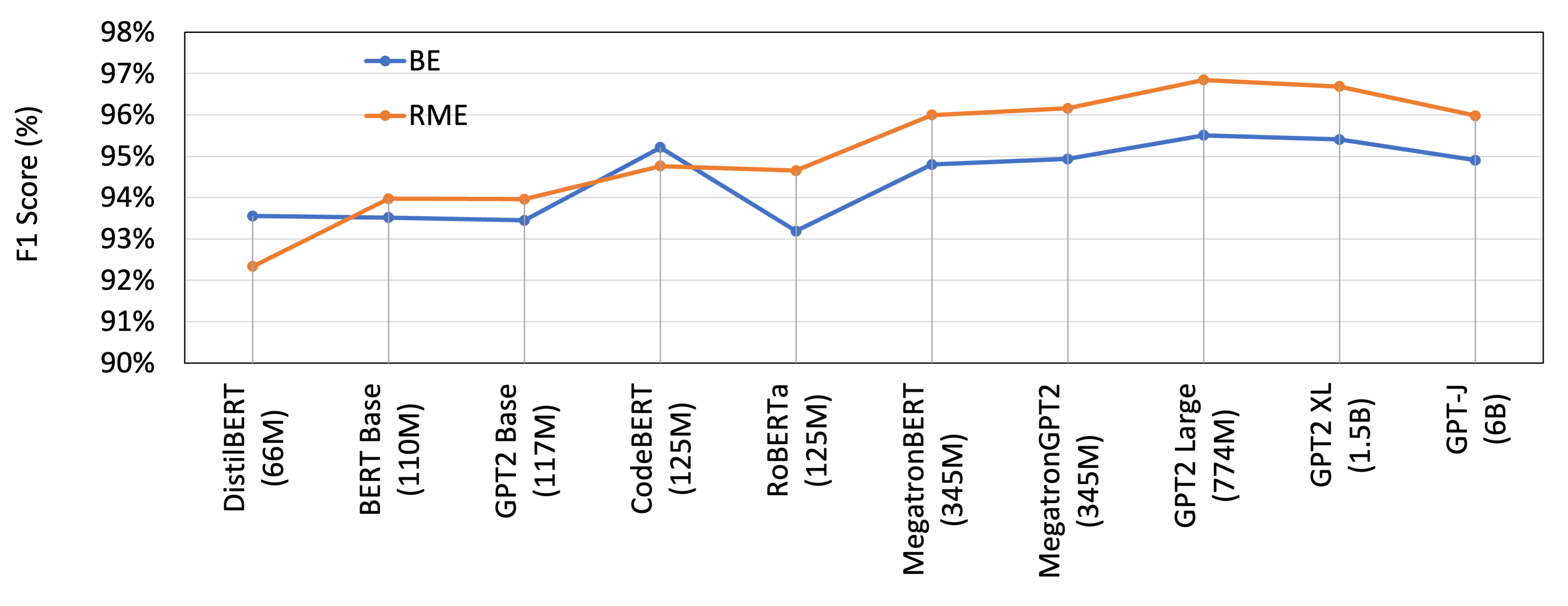}
			\label{fig:f1_overall_trend_binary}
		}
		\vskip-5pt
		\subfigure[Multi-class Classification.]{
			\includegraphics[width=0.9\columnwidth]{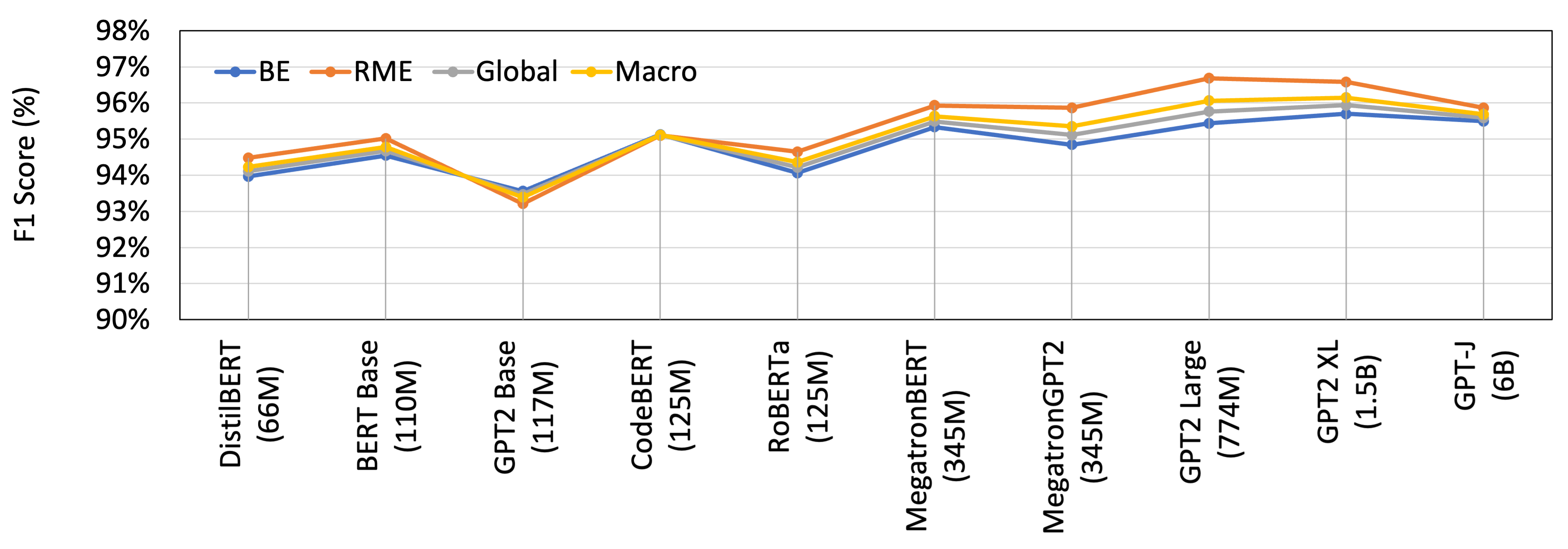}
			\label{fig:f1_overall_trend_multi}
		}
		\caption{The overall trend of F1-score with the increasing size of the Transformer-based models.}
		\label{fig:f1_overall_trend}
\end{figure}

The test results for the various models in the binary classification task are presented in Table~\ref{tab:binaryclassfication_all_results}. 
Separate experiments were performed for Group~1 and Group~2 datasets (see Table~\ref{table:data}), and all the models consider two output labels, viz., 0 (non-vulnerable), 1 (vulnerable: BE if Group~1 dataset and RME if Group~2 dataset).
We observe an overall improvement in results for the transformer-based models over BiLSTM, including VulDeePecker Original model and BiGRU: 
\begin{itemize}[leftmargin=3mm\relax]
    \item For Buffer Error, FPR is slightly improved; however, FNR reduction is significant. For example, GPT-2 XL has only 4.72\% compared to 18\% reported originally for BiLSTM. Besides, improvements in Precision, Recall, and F1-scores are also significant. For example, GPT-2 Large has an F1-score of 95.51\% compared to 86.6\% reported originally for BiLSTM.
    \item For Resource Management Error, considering the results reported originally for BiLSTM, GPT-2 Large, GPT-2 XL, MegatronBERT, MegatraonGPT-2, and GPT-J models have an improvement in all metrics. 
\end{itemize}
Among transformer-based models, GPT-2 Large and GPT-2 XL show better vulnerability detection. Most interestingly, GPT-J, the largest GPT-2-based model, is not better in detection than its smaller counterparts, such as GPT-2 Large. The possible reason for this is that (1) our dataset size might not be sufficient to fine-tune GPT-J, and (2) GPT-J might need adjustment to its fine-tuning hyperparameters such as learning rate and warm-up steps. We left further exploration in this direction as future work. 
Now, analyzing the overall trend in the improvements, the performance is in an increasing trend (having some slight fall and rise) with the increase in the model's size. Refer to Figure~\ref{fig:f1_overall_trend_binary} for an illustration. 

\subsubsection{Multi-class classification}

The test results for the various models in the multi-class classification are presented in Table~\ref{tab:multiclass_classification}\footref{data_footnote}. All the experiments were performed using the Group~3 dataset (see Table~\ref{table:data}), and all the models consider three output labels, viz., 0 (non-vulnerable), 1 (BE), and 2 (RME).
Like in binary classification, we observe an overall improvement in results for the transformer-based models over BiLSTM and BiGRU.
Among transformer-based models, considering the global average results, {GPT-2~XL} shows better Recall and F1-score; whereas GPT-J delivers better precision.    
Unlike binary classification results, GPT-J, the largest GPT-2-based model considered in this paper, has shown some better results than its smaller counterparts for FPR and Precision. 
Now, analyzing the overall trend in the improvements, the performance is in an increasing trend (having some slight fall and rise) with the increase in the model's size. Refer to Figure~\ref{fig:f1_overall_trend_multi} for an illustration. 

\subsubsection{Fine-tuning time}

\begin{figure}[t]
\begin{center}
   \includegraphics[width=0.9\linewidth]{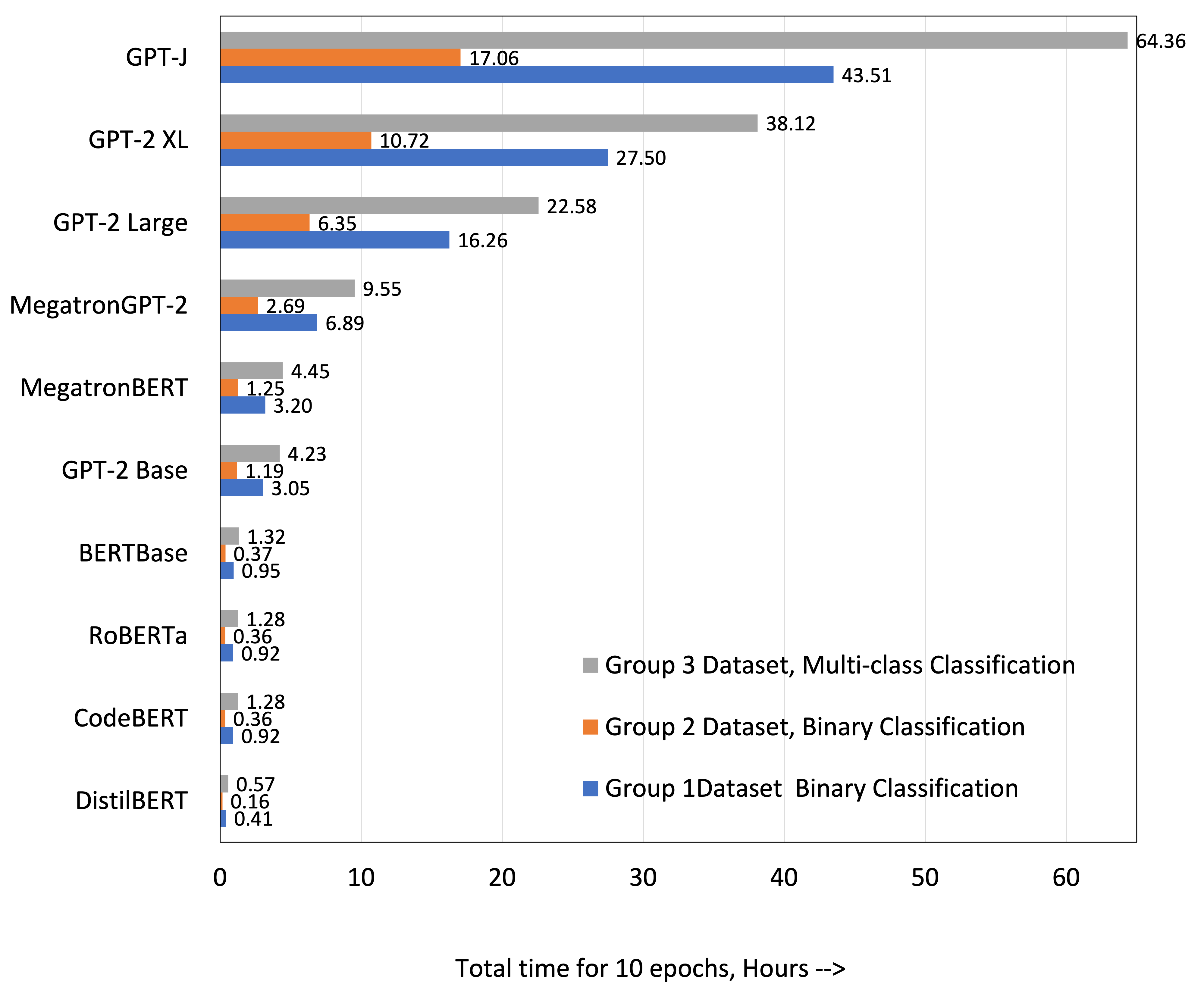}
\end{center}
   \caption{Total time is taken in hours to fine-tune our models for 10 epochs using one NVIDIA GPU RTX A6000 with 48GB GDDR6 GPU memory.}
\label{fig:time}
\end{figure}

To give an idea of how long the fine-tuning of these large models will take for the software vulnerability detection, we present the fine-tuning time taken by our models for 10 epochs with binary and multi-classification tasks. We ran our test on our system with NVIDIA GPU RTX A6000 with 48GB GDDR6 GPU memory. All the models were fine-tuned using only one GPU for the time measurement\footnote{We fine-tuned all the models with HuggingFace, and we applied DeepSpeed ZeRO Stage 2 on GPT-J only to resolve \emph{CUDA Error: out of memory} issue.}. Our result is depicted in Figure~\ref{fig:time}. As per expectation for multi-class classification tasks, among our models, GPT-J being the largest model with 6B model parameters, took the highest time, around 65 hours, to run for 10 epochs, and DistilBERT, the smallest model with 66M parameters, took about 35 minutes to run for 10 epochs. Also, we evaluated the models under the test dataset at the end of each epoch in these runs. The pattern is similar for the binary classification tasks with Group 1 and Group 2 datasets. For a model, the fine-tuning time is different for different datasets because of the different sizes of the fine-tuning datasets (see Table~\ref{table:data} where the training dataset is used for fine-tuning these models).

\begin{insight}
While choosing the model for the vulnerability detection, if there is no time constraint for the model's fine-tuning, then we can pick one of the best performing models, \emph{e.g.}, GPT-2 Large for the Group 1 dataset and F1-score. If there is a time constraint, then we need to pick the model that has the best trade-off between the performance and fine-tuning time, \emph{e.g.}, CodeBERT for Group 1 dataset and F1-score.   
\end{insight}


\subsection{Performance of the transformer-based models on the SeVC dataset}
\label{sec:multiple_vul}

For the studies with more than two vulnerabilities, we consider the Semantics-based Vulnerability Candidate (SeVC) dataset~\cite{sysevr} having 126 different vulnerabilities (refer to Appendix~\ref{appendix:sysevr_data} for details). Broadly SeVC is divided into four categories based on the cause of vulnerabilities; API Function Call, Arithmetic Expression, Array Usage, and Pointer Usage. Overall, we divided the dataset into 5 groups for our studies. Refer to Table~\ref{table:sysevr} for details.

\begin{table}[t]
\vskip7pt
\footnotesize
\centering
\caption{Division of SeVC dataset based on its categories. The number of vulnerable and non-vulnerable samples are indicated by `V' and `NV', respectively, in the table.}
\label{table:sysevr}
\begin{adjustbox}{max width=1\linewidth}
\begin{tabular}{|c|c|c|c|c|c|}
\hline
\textbf{Dataset} & \textbf{Categories}                                                  & \textbf{Original} & \textbf{Cleaned} & \textbf{Train}                                                           & \textbf{Test}                                                         \\ \hline
Group 4          & \begin{tabular}[c]{@{}c@{}}API Function\\ Call (AFC)\end{tabular}    & 64403             & 54181            & \begin{tabular}[c]{@{}c@{}}43344\\ (V: 10647, NV:32697)\end{tabular}     & \begin{tabular}[c]{@{}c@{}}10837\\ (V: 2611, NV: 8226)\end{tabular}   \\ \hline
Group 5          & \begin{tabular}[c]{@{}c@{}}Arithmetic\\ Expression (AE)\end{tabular} & 22154             & 14454            & \begin{tabular}[c]{@{}c@{}}11563\\ (V: 2642, NV: 8921)\end{tabular}      & \begin{tabular}[c]{@{}c@{}}2891\\ (V: 648, NV: 2243)\end{tabular}     \\ \hline
Group 6          & \begin{tabular}[c]{@{}c@{}}Array\\ Usage (AU)\end{tabular}           & 42229             & 34166            & \begin{tabular}[c]{@{}c@{}}27332\\ (V: 8237, NV: 19095)\end{tabular}     & \begin{tabular}[c]{@{}c@{}}6834\\ (V: 2145, NV: 4689)\end{tabular}    \\ \hline
Group 7          & \begin{tabular}[c]{@{}c@{}}Pointer\\ Usage (PU)\end{tabular}         & 291841            & 176883           & \begin{tabular}[c]{@{}c@{}}141506\\ (V: 21189, NV: 120317 )\end{tabular} & \begin{tabular}[c]{@{}c@{}}35377\\ (V: 5335, NV: 30042)\end{tabular}  \\ \hline
Group 8          & \begin{tabular}[c]{@{}c@{}}AFC + AE +\\ AU + PU\end{tabular}         & 420627            & 206376           & \begin{tabular}[c]{@{}c@{}}165100\\ (V: 274804, NV: 145823)\end{tabular} & \begin{tabular}[c]{@{}c@{}}41276\\ (V: 4769 , NV: 36507)\end{tabular} \\ \hline
\end{tabular}
\end{adjustbox}
\end{table}

\begin{table*}[t]
\footnotesize
\centering
\caption{Test performance of various models for the binary classification on clean SeVC dataset. `NA' refers to `Not Available'.}
\label{table:sevc_result_binary}
\begin{adjustbox}{max width=0.7\linewidth}
\begin{tabular}{|c|c|c|c|c|c|c|c|}
\hline
{\color[HTML]{000000} \textbf{\begin{tabular}[c]{@{}c@{}}Dataset and\\ Vulnerability\end{tabular}}}                           & {\color[HTML]{000000} \textbf{Metrics}} & {\color[HTML]{000000} \textbf{\begin{tabular}[c]{@{}c@{}}VulDeePecker\\ (BiLSTM~\cite{sysevr})\end{tabular}}} & {\color[HTML]{000000} \textbf{\begin{tabular}[c]{@{}c@{}}SySeVR\\ (BiLSTM~\cite{sysevr})\end{tabular}}} & {\color[HTML]{000000} \textbf{\begin{tabular}[c]{@{}c@{}}Our\\ BiLSTM\end{tabular}}} & {\color[HTML]{000000} \textbf{BiGRU}}                  & {\color[HTML]{000000} \textbf{BERTBase}}               & {\color[HTML]{000000} \textbf{GPT-2 Base}}             \\ \hline
{\color[HTML]{000000} }                                                                                                       & {\color[HTML]{000000} FPR}              & {\color[HTML]{000000} 5.50\%}                                                                   & \cellcolor[HTML]{9AFF99}{\color[HTML]{000000} 2.10\%}                                     & {\color[HTML]{000000} 21.08\%}                                                       & {\color[HTML]{000000} 15.50\%}                         & {\color[HTML]{000000} 3.63\%}                          & {\color[HTML]{000000} 3.28\%}                          \\ \cline{2-8} 
{\color[HTML]{000000} }                                                                                                       & {\color[HTML]{000000} FNR}              & {\color[HTML]{000000} 22.50\%}                                                                  & {\color[HTML]{000000} 17.50\%}                                                            & {\color[HTML]{000000} 8.91\%}                                                        & \cellcolor[HTML]{9AFF99}8.80\%                         & {\color[HTML]{000000} 9.06\%}                          & {\color[HTML]{000000} 11.01\%}                         \\ \cline{2-8} 
{\color[HTML]{000000} }                                                                                                       & {\color[HTML]{000000} Precision}        & {\color[HTML]{000000} 79.10\%}                                                                  & \cellcolor[HTML]{9AFF99}{\color[HTML]{000000} 91.50\%}                                    & {\color[HTML]{000000} 81.21\%}                                                       & {\color[HTML]{000000} 85.47\%}                         & {\color[HTML]{000000} 89.14\%}                         & {\color[HTML]{000000} 89.87\%}                         \\ \cline{2-8} 
{\color[HTML]{000000} }                                                                                                       & {\color[HTML]{000000} Recall}           & {\color[HTML]{000000} 77.52\%}                                                                  & {\color[HTML]{000000} 82.56\%}                                                            & {\color[HTML]{000000} 91.09\%}                                                       & \cellcolor[HTML]{9AFF99}{\color[HTML]{000000} 91.20\%} & {\color[HTML]{000000} 90.94\%}                         & {\color[HTML]{000000} 88.99\%} \\ \cline{2-8} 
\multirow{-5}{*}{{\color[HTML]{000000} \begin{tabular}[c]{@{}c@{}}Group 4, \\ API \\ Function \\ Call \\ (AFC)\end{tabular}}} & {\color[HTML]{000000} F1- score}        & {\color[HTML]{000000} 78.30\%}                                                                  & {\color[HTML]{000000} 86.80\%}                                                            & {\color[HTML]{000000} 85.87\%}                                                       & {\color[HTML]{000000} 88.24\%}                         & \cellcolor[HTML]{9AFF99}{\color[HTML]{000000} 90.03\%} & {\color[HTML]{000000} 89.43\%}                         \\ \hline
{\color[HTML]{000000} }                                                                                                       & {\color[HTML]{000000} FPR}              & {\color[HTML]{000000} NA}                                                                       & {\color[HTML]{000000} 3.80\%}                                                             & {\color[HTML]{000000} 15.43\%}                                                       & {\color[HTML]{000000} 18.34\%}                         & {\color[HTML]{000000} 3.09\%}                          & \cellcolor[HTML]{9AFF99}{\color[HTML]{000000} 2.32\%}  \\ \cline{2-8} 
{\color[HTML]{000000} }                                                                                                       & {\color[HTML]{000000} FNR}              & {\color[HTML]{000000} NA}                                                                       & {\color[HTML]{000000} 17.10\%}                                                            & {\color[HTML]{000000} 8.30\%}                                                        & \cellcolor[HTML]{9AFF99}{\color[HTML]{000000} 6.55\%}  & {\color[HTML]{000000} 9.32\%}                          & {\color[HTML]{000000} 11.21\%}                         \\ \cline{2-8} 
{\color[HTML]{000000} }                                                                                                       & {\color[HTML]{000000} Precision}        & {\color[HTML]{000000} NA}                                                                       & {\color[HTML]{000000} 88.30\%}                                                            & {\color[HTML]{000000} 85.60\%}                                                       & {\color[HTML]{000000} 83.59\%}                         & {\color[HTML]{000000} 90.16\%}                         & \cellcolor[HTML]{9AFF99}{\color[HTML]{000000} 92.28\%} \\ \cline{2-8} 
{\color[HTML]{000000} }                                                                                                       & {\color[HTML]{000000} Recall}           & {\color[HTML]{000000} NA}                                                                       & {\color[HTML]{000000} 82.87\%}                                                                 & {\color[HTML]{000000} 91.70\%}                                                       & \cellcolor[HTML]{9AFF99}{\color[HTML]{000000} 93.45\%} & {\color[HTML]{000000} 90.68\%}                         & {\color[HTML]{000000} 88.79\%}                         \\ \cline{2-8} 
\multirow{-5}{*}{{\color[HTML]{000000} \begin{tabular}[c]{@{}c@{}}Group 5, \\ Arithmetic \\ Expression \\ (AE)\end{tabular}}} & {\color[HTML]{000000} F1- score}        & {\color[HTML]{000000} NA}                                                                       & {\color[HTML]{000000} 85.50\%}                                                            & {\color[HTML]{000000} 88.55\%}                                                       & {\color[HTML]{000000} 88.25\%}                         & {\color[HTML]{000000} 90.42\%}                         & \cellcolor[HTML]{9AFF99}{\color[HTML]{000000} 90.50\%} \\ \hline
{\color[HTML]{000000} }                                                                                                       & {\color[HTML]{000000} FPR}              & {\color[HTML]{000000} NA}                                                                       & \cellcolor[HTML]{9AFF99}{\color[HTML]{000000} 1.50\%}                                     & {\color[HTML]{000000} 19.73\%}                                                       & {\color[HTML]{000000} 18.15\%}                         & {\color[HTML]{000000} 3.58\%}                          & {\color[HTML]{000000} 4.32\%}                          \\ \cline{2-8} 
{\color[HTML]{000000} }                                                                                                       & {\color[HTML]{000000} FNR}              & {\color[HTML]{000000} NA}                                                                       & {\color[HTML]{000000} 18.30\%}                                                            & {\color[HTML]{000000} 14.26\%}                                                       & {\color[HTML]{000000} 13.59\%}                         & {\color[HTML]{000000} 14.35\%}                         & \cellcolor[HTML]{9AFF99}{\color[HTML]{000000} 12.24\%} \\ \cline{2-8} 
{\color[HTML]{000000} }                                                                                                       & {\color[HTML]{000000} Precision}        & {\color[HTML]{000000} NA}                                                                       & {\color[HTML]{000000} 87.90\%}                                                            & {\color[HTML]{000000} 81.29\%}                                                       & {\color[HTML]{000000} 82.64\%}                         & \cellcolor[HTML]{9AFF99}{\color[HTML]{000000} 91.30\%} & {\color[HTML]{000000} 89.92\%}                         \\ \cline{2-8} 
{\color[HTML]{000000} }                                                                                                       & {\color[HTML]{000000} Recall}           & {\color[HTML]{000000} NA}                                                                       & {\color[HTML]{000000} 81.72\%}                                                                 & {\color[HTML]{000000} 85.74\%}                                                       & {\color[HTML]{000000} 86.41\%}                         & {\color[HTML]{000000} 85.65\%}                         & \cellcolor[HTML]{9AFF99}{\color[HTML]{000000} 87.76\%} \\ \cline{2-8} 
\multirow{-5}{*}{{\color[HTML]{000000} \begin{tabular}[c]{@{}c@{}}Group 6, \\ Array \\ Usage \\ (AU)\end{tabular}}}           & {\color[HTML]{000000} F1- score}        & {\color[HTML]{000000} NA}                                                                       & {\color[HTML]{000000} 84.70\%}                                                            & {\color[HTML]{000000} 83.46\%}                                                       & {\color[HTML]{000000} 84.49\%}                         & {\color[HTML]{000000} 88.38\%}                         & \cellcolor[HTML]{9AFF99}{\color[HTML]{000000} 88.82\%} \\ \hline
{\color[HTML]{000000} }                                                                                                       & {\color[HTML]{000000} FPR}              & {\color[HTML]{000000} NA}                                                                       & \cellcolor[HTML]{9AFF99}{\color[HTML]{000000} 1.30\%}                                     & {\color[HTML]{000000} 15.66\%}                                                       & {\color[HTML]{000000} 12.99\%}                         & {\color[HTML]{000000} 1.40\%}                          & {\color[HTML]{000000} 1.54\%}                          \\ \cline{2-8} 
{\color[HTML]{000000} }                                                                                                       & {\color[HTML]{000000} FNR}              & {\color[HTML]{000000} NA}                                                                       & {\color[HTML]{000000} 19.70\%}                                                            & {\color[HTML]{000000} 5.43\%}                                                        & \cellcolor[HTML]{9AFF99}{\color[HTML]{000000} 4.78\%}  & {\color[HTML]{000000} 7.96\%}                          & {\color[HTML]{000000} 8.25\%}                          \\ \cline{2-8} 
{\color[HTML]{000000} }                                                                                                       & {\color[HTML]{000000} Precision}        & {\color[HTML]{000000} NA}                                                                       & {\color[HTML]{000000} 87.30\%}                                                            & {\color[HTML]{000000} 85.80\%}                                                       & {\color[HTML]{000000} 87.99\%}                         & \cellcolor[HTML]{9AFF99}{\color[HTML]{000000} 92.02\%} & {\color[HTML]{000000} 91.29\%}                         \\ \cline{2-8} 
{\color[HTML]{000000} }                                                                                                       & {\color[HTML]{000000} Recall}           & {\color[HTML]{000000} NA}                                                                       & {\color[HTML]{000000} 80.39\%}                                                                 & {\color[HTML]{000000} 94.57\%}                                                       & \cellcolor[HTML]{9AFF99}{\color[HTML]{000000} 95.22\%} & {\color[HTML]{000000} 92.04\%}                         & {\color[HTML]{000000} 91.75\%}                         \\ \cline{2-8} 
\multirow{-5}{*}{{\color[HTML]{000000} \begin{tabular}[c]{@{}c@{}}Group 7, \\ Pointer \\ Usage \\ (PU)\end{tabular}}}         & {\color[HTML]{000000} F1- score}        & {\color[HTML]{000000} NA}                                                                       & {\color[HTML]{000000} 83.70\%}                                                            & {\color[HTML]{000000} 89.97\%}                                                       & {\color[HTML]{000000} 91.46\%}                         & \cellcolor[HTML]{9AFF99}{\color[HTML]{000000} 92.03\%} & {\color[HTML]{000000} 91.52\%}                         \\ \hline
\end{tabular}
\end{adjustbox}
\end{table*}


We have performed binary and multi-class classification tasks for the SeVC dataset considering BiLSTM, BiGRU, BERTBase, and GPT-2 Base models. For the binary classification task, BERTBase and {GPT-2} Base have an improvement over BiLSTM in all metrics except FPR and Precision for Group~4 and FPR for Group~6 \& Group~7. For example, the Group~4 dataset has an F1-score of 90.03\% with BERTBase compared to the previously reported results of 86.80\% and 78.30\% with BiLSTMs. Refer to Table~\ref{table:sevc_result_binary} for details. Besides, except for the Group 6 dataset, BiGRU has better FNR compared to other models. 
There is no result reported in the previous work for multi-class classification tasks, so we compare BERTBase and GPT-2 Base. Considering the global averaged results, BERTBase has performed better than GPT-2 Base. It has 88.34\% F1-score. Refer to Table~\ref{table:sevc_result_multiclass} for details. In contrast to binary classification, multi-class classification with SeVC has a low F1-score. This is understandable from the fact that SeVC has multiple vulnerabilities that increase the complexity of machine learning.

\begin{table}[t]
\vskip5pt
\footnotesize
\centering
\caption{Test performance of various models for the multi-class classification on clean SeVC dataset.}
\label{table:sevc_result_multiclass}
\begin{adjustbox}{max width=0.7\linewidth}
\begin{tabular}{|c|c|c|c|}
\hline
\textbf{\begin{tabular}[c]{@{}c@{}}Dataset and\\ Vulnerability\end{tabular}}                             & \textbf{Metrics} & \textbf{BERTBase}               & \textbf{GPT-2 Base}             \\ \hline
                                                                                                         & FPR              & 0.11\%                          & \cellcolor[HTML]{9AFF99}0.05\%  \\ \cline{2-4} 
                                                                                                         & FNR              & \cellcolor[HTML]{9AFF99}25.64\% & 33.33\%                         \\ \cline{2-4} 
                                                                                                         & Precision        & 83.45\%                         & \cellcolor[HTML]{9AFF99}90.83\% \\ \cline{2-4} 
                                                                                                         & Recall           & \cellcolor[HTML]{9AFF99}74.36\% & 66.67\%                         \\ \cline{2-4} 
\multirow{-5}{*}{\begin{tabular}[c]{@{}c@{}}Group 8, \\ API \\ Function \\ Call \\ (AFC)\end{tabular}}   & F1- score        & \cellcolor[HTML]{9AFF99}78.64\% & 76.89\%                         \\ \hline
                                                                                                         & FPR              & \cellcolor[HTML]{9AFF99}0.21\%  & 0.27\%                          \\ \cline{2-4} 
                                                                                                         & FNR              & 9.96\%                          & \cellcolor[HTML]{9AFF99}9.60\%  \\ \cline{2-4} 
                                                                                                         & Precision        & \cellcolor[HTML]{9AFF99}85.40\% & 81.94\%                         \\ \cline{2-4} 
                                                                                                         & Recall           & 90.04\%                         & \cellcolor[HTML]{9AFF99}90.40\% \\ \cline{2-4} 
\multirow{-5}{*}{\begin{tabular}[c]{@{}c@{}}Group 8, \\ Arithmetic \\ Expression \\ (AE)\end{tabular}}   & F1- score        & \cellcolor[HTML]{9AFF99}87.65\% & 85.96\%                         \\ \hline
                                                                                                         & FPR              & \cellcolor[HTML]{9AFF99}0.39\%  & 0.44\%                          \\ \cline{2-4} 
                                                                                                         & FNR              & 12.44\%                         & \cellcolor[HTML]{9AFF99}11.18\% \\ \cline{2-4} 
                                                                                                         & Precision        & \cellcolor[HTML]{9AFF99}85.16\% & 83.70\%                         \\ \cline{2-4} 
                                                                                                         & Recall           & 87.56\%                         & \cellcolor[HTML]{9AFF99}88.82\% \\ \cline{2-4} 
\multirow{-5}{*}{\begin{tabular}[c]{@{}c@{}}Group 8, \\ Array \\ Usage \\ (AU)\end{tabular}}             & F1- score        & \cellcolor[HTML]{9AFF99}86.34\% & 86.19\%                         \\ \hline
                                                                                                         & FPR              & \cellcolor[HTML]{9AFF99}0.79\%  & 0.87\%                          \\ \cline{2-4} 
                                                                                                         & FNR              & \cellcolor[HTML]{9AFF99}9.60\%  & 11.35\%                         \\ \cline{2-4} 
                                                                                                         & Precision        & \cellcolor[HTML]{9AFF99}89.85\% & 88.77\%                         \\ \cline{2-4} 
                                                                                                         & Recall           & \cellcolor[HTML]{9AFF99}90.40\% & 88.65\%                         \\ \cline{2-4} 
\multirow{-5}{*}{\begin{tabular}[c]{@{}c@{}}Group 8, \\ Pointer \\ Usage \\ (PU)\end{tabular}}           & F1- score        & \cellcolor[HTML]{9AFF99}90.12\% & 88.71\%                         \\ \hline
                                                                                                         & Precision        & \cellcolor[HTML]{9AFF99}87.95\% & 86.88\%                         \\ \cline{2-4} 
                                                                                                         & Recall           & \cellcolor[HTML]{9AFF99}88.73\% & 87.47\%                         \\ \cline{2-4} 
\multirow{-3}{*}{\begin{tabular}[c]{@{}c@{}}Group 8, \\ AFC + AE + AU + PU\\ (Global Avg.)\end{tabular}} & F1 score         & \cellcolor[HTML]{9AFF99}88.34\% & 87.18\%                         \\ \hline
                                                                                                         & Precision        & 85.97\%                         & \cellcolor[HTML]{9AFF99}86.31\% \\ \cline{2-4} 
                                                                                                         & Recall           & \cellcolor[HTML]{9AFF99}85.59\% & 83.63\%                         \\ \cline{2-4} 
\multirow{-3}{*}{\begin{tabular}[c]{@{}c@{}}Group 8, \\ AFC + AE + AU + PU\\ (Macro Avg.)\end{tabular}}  & F1 score         & \cellcolor[HTML]{9AFF99}85.78\% & 84.95\%                         \\ \hline
\end{tabular}
\end{adjustbox}
\end{table}

\section{Platform analysis}
\label{sec:platformsurvey}


There is a rising trend of releasing bigger models, usually transformer-based models. For instance, the BERTBase model has 110 million parameters, and the BERTLarge has 340 million parameters~\cite{attention,bert,how_bert_works}, which Google released in 2018. OpenAI released the GPT-2 language model with 1.5 billion parameters in 2019, and it was followed by the GPT-3 model, which has 175 billion parameters~\cite{gpt,gpt2}. These models are shown to be outperforming the existing models in terms of accuracy or precision.  
These models are used in various fields via the downstream task, where the model is further trained to adjust to the new targeted dataset. 
On the flip side, handling these large models requires substantial computational resources. Precisely, a single GPU is not enough to train extra-large models with large-scale data because these models have too many parameters to train,  and there are too many samples to process on a single GPU. Thus, usually end up with a ``\emph{CUDA out of memory}” error. For example, in our experiments, we could not run GPT-2 Large with 774M model parameters on an NVIDIA V100 GPU with 16GB internal memory, even with a batch size of 1.
%
\begin{figure}[t]
\centering
   \includegraphics[trim=6cm 4cm 6cm 4cm, clip=true, width=0.9\linewidth]{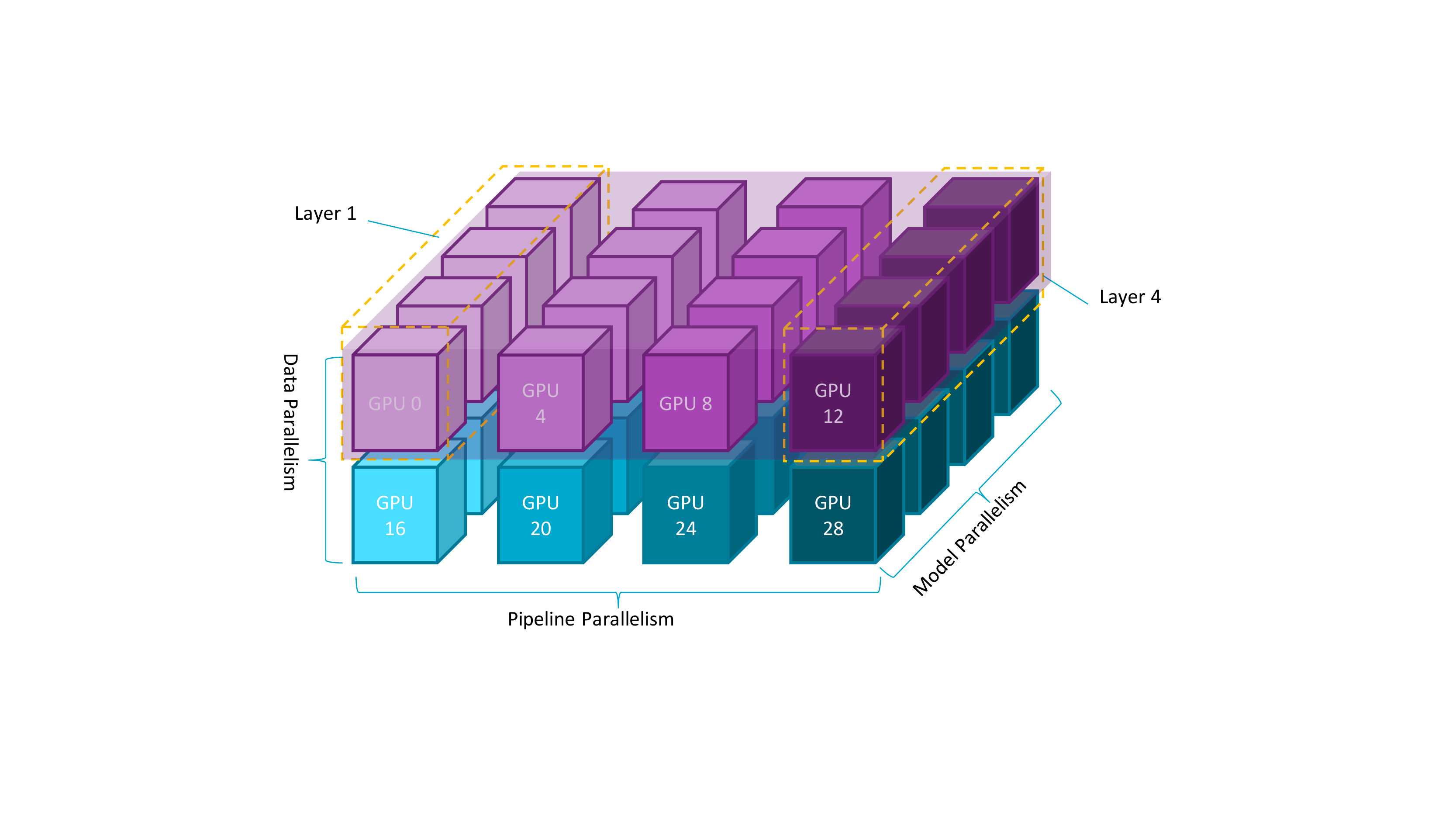}
   \caption{Deep learning model parallelism approaches.}
\label{fig:model_parallel}
\end{figure}
%
In this regard, we consider deep learning parallelism approaches (presented in ~\cite{gpt}) to resolve the challenge due to limited GPU memory. For an illustration of the approaches, refer to Figure~\ref{fig:model_parallel}. There are multiple types of parallelism methods that are defined based on how the data and models are computed collaboratively with multiple GPUs.
%
%
Usually, these methods are collectively called a 3D Parallelism and are described in the following:
\begin{itemize}[leftmargin=3mm\relax]
    \item \textbf{Data parallelism:} Data parallelism splits a large dataset into smaller batches, and each GPU (or GPU group) holds an identical copy of the model parameters. Then, each GPU (or GPU group) sends the computed gradients to a parameter server. The parameter server aggregates the computed gradients and calculates the updates. Afterward, it sends the updates to each GPU (or GPU group) for updating, and each GPU (or GPU group) processes the next batch. 
    \item \textbf{Pipeline parallelism (vertical parallelism):} Pipeline parallelism enables model parallelism. This approach shares neural network layers into stages with an equal number (desirably) of layers on each stage. Each stage is processed by one GPU (or GPU group), and the calculated output is forwarded to the next stage. For example, splitting a model with 24 layers across 4 stages would mean each stage gets 6 layers. Then, each GPU (or GPU group) processes the assigned stage and passes the output to the following GPU (or GPU group).
    \item \textbf{Model parallelism (tensor parallelism or horizontal parallelism):} Model parallelism splits the execution of a single layer over multiple GPUs, while Pipeline parallelism splits multiple layers across multiple GPUs. Each layer is split up into multiple chunks, and each piece belongs to a designated GPU. The processed results are synced at the end of the step.
\end{itemize}

\subsection{Platforms}
To leverage the parallelism while model training/testing in our studies, we have analyzed four popular open-sourced platforms. Table~\ref{tab:framework_comparison} summarizes these platforms, and details are presented in the following paragraphs.

\begin{table}[t]
\vskip10pt
\centering
\caption{Summary of popular open-sourced machine learning platforms.}
\label{tab:framework_comparison}
\resizebox{0.48\textwidth}{!}{%
\begin{tabular}{|l|c|c|c|c|}
\hline
 & \textbf{HuggingFace} & \textbf{Megatron} & \textbf{DeepSpeed} & \textbf{Horovod} \\ \hline
Description & \begin{tabular}[c]{@{}c@{}}A machine learning \\ framework for Jax, \\ Pytorch and TensorFlow\end{tabular} & \begin{tabular}[c]{@{}c@{}}An implementation \\ of Transformer\end{tabular} & \begin{tabular}[c]{@{}c@{}}A deep learning \\ optimization library for\\  distributed training\end{tabular} & \begin{tabular}[c]{@{}c@{}}A python library \\ for data parallelism\end{tabular} \\ \hline
Data Parallelism & \cmark & \cmark & \cmark & \cmark \\ \hline
Pipeline Parallelism & Partial (need customization) & \cmark & \cmark & \xmark \\ \hline
Tensor Parallelism & \xmark & \cmark & \cmark & \xmark \\ \hline
Memory efficiency & Normal & Normal & Excellent & Normal \\ \hline
Training speed & Normal & Good & Great & Normal \\ \hline
Type & \begin{tabular}[c]{@{}c@{}}Can use Megatron-LM, \\ and all models\end{tabular} & \begin{tabular}[c]{@{}c@{}}Dedicated only \\ for Megatron-LM\end{tabular} & \begin{tabular}[c]{@{}c@{}}Just a library, supplement\\  tool for memory \\ efficiency and speed\end{tabular} & \begin{tabular}[c]{@{}c@{}}Dedicated only for \\ Data Parallelism\end{tabular} \\ \hline
\end{tabular}
}
\vskip-10pt
\end{table} 

\paragraph{Horovod}
Horovod~\cite{horovod} is a stand-alone Python library for data parallelism. It uses an optimized ring-all reduce algorithm to improve both performance and usability. It supports TensorFlow, Keras, PyTorch, and Apache MXNet. It can achieve linear performance gain provided that the portion of parameters in the dense layers to all parameters is small~\cite{horovod1}. Horovod developers claimed that it achieved 90\% scaling efficiency for Inception V3 and ResNet-101 models and 68\% scaling efficiency for the VGG-16 model.
Model parallelism means models are split and can be evaluated concurrently. With this definition, Horovod supports model partitioning for workload division but does not support model or pipeline parallelism. Without modification of Horovod implementation, it can train only models that fit into a single GPU. Consequently, we do not consider Horovod as a development framework in this work.

\paragraph{Megatron Framework}
NVIDIA released Megatron language models and a PyTorch-based framework to train/test the models~\cite{shoeybi2020megatronlm}. The framework and the model support not only model parallelism (pipeline and tensor) but also data parallelism. NVIDIA trained MegatronGPT-2 (8.3 billion parameters) with 8-way model parallelism and 64-way data parallelism, trained on 512 GPUs (NVIDIA Tesla V100) using mixed precision. This is $24\times$ the size of BERT and $5.6\times$ the size of GPT-2 (the previous largest version). MegatronBERT has 3.9 billion parameters, which is $12\times$ the size of the BERTLarge model~\cite{gpt,gpt2}.
As the Megatron framework supports all three parallelism approaches, we utilised it as one of the development frameworks for this work. However, the Megatron framework is model-specific, and hard to utilise other pre-trained models within the framework. So, we only evaluated Megatron BERT 345M model and the GPT-2 345M model by implementing our own model providers, data providers, and metric function provider functions. 

\paragraph{DeepSpeed}
DeepSpeed~\cite{deepspeed} is an open-source deep-learning optimization library released by Microsoft for PyTorch. It delivers extreme-scale, extreme memory efficient, and extremely communication efficient model training by leveraging model parallelism on existing computer resources. DeepSpeed introduces Zero Redundancy Optimizer (ZeRO), and it enables $10\times$ larger models, $10\times$ faster training, and minimal code change. As a specific example, the memory consumption for training a 7.5B parameter model is about 120GB of GPU RAM (the calculation is based on 64 GPUs). The ZeRO Stage~1 partitions optimizer states across the GPUs and requires 31.4GB of GPU RAM (the calculation is based on 64 GPUs). The ZeRO Stage~2 reduces 32-bit gradients to 16-bit for updating the model weights, and it requires 16.6GB of GPU RAM. In ZeRO Stage~3, the 16-bit model parameters are partitioned across the GPU, and it requires only 1.9GB of GPU RAM.
Recently DeepSpeed introduced ZeRO-Infinity, which leverages the total memory capacity of a system, concurrently exploiting all heterogeneous memory (GPU, CPU, and NVMe). They claimed that with a single NVIDIA DGX-2 node, a 1000 billion parameter model could be trained. The key idea of ZeRO-Infinity is offloading data into other types of memory. It splits a model into multiple states and stores into CPU or NVMe device memory, which are usually much bigger than GPU memory, and GPU holds only a small portion of states for computing. Due to the overall features of DeepSpeed, we have used it to carry out our experiments in this work.

\paragraph{HuggingFace}
HuggingFace~\cite{huggingface} supports only data parallelism but did not officially implement model parallelism (neither pipeline nor tensor). However, HuggingFace integrates DeepSpeed, enabling users to enjoy the benefits of DeepSpeed, such as ZeRO stages. There are two options to integrate DeepSpeed:
\begin{itemize}[leftmargin=3mm\relax]
    \item Integration of the core DeepSpeed features via Trainer: Users need to use Trainer and supply the DeepSpeed config file. The rest of the things will be done automatically. We strongly recommend this integration method.
    \item Integrate DeepSpeed by ourself: Core functionality functions must be implemented, such as {\small \ttfamily{from\_pretrained} and \ttfamily{from\_config}}.
\end{itemize}

\subsection{Discussion on the platforms}
In this section, firstly, we present the challenges that we faced, then provide our recommendations.

\subsubsection{Challenges}

\noindent
\textbf{(1) No admin privilege:} 
Institutions have HPC clusters that consist of multiple nodes. For example, we have an HPC cluster with 114 nodes. Each node has two Xeon 14-core E5-2690 v4 CPUs and four NVIDIA Tesla P100 16 gigabytes GPUs. Due to security and maintenance reasons, the service owner does not provide admin privilege to HPC users, and this policy has created many issues. For instance, we cannot install system-dependent programs/libraries by ourselves, and due to the version of HPC OS (SUSE Linux 12.4), we could not use the latest versions of NVIDIA drivers and CUDA. We have used Anaconda~\cite{anaconda} to set up virtual environments, and within the virtual environments, we installed pre-requisites and dependencies, which were not supported by the HPC environment directly (in non-virtual environments). 

\noindent
\textbf{(2) Model Parallelism:}
Some models in the HuggingFace framework support a naive pipeline model parallelism, such as GPT-2 and T5. Users need to define a dictionary that maps attention layers to GPUs, and the embedding layer and LMHead are always automatically mapped to the first device. This is an experimental feature and has some GPU idling problems (it is hard to balance workload between GPUs). Megatron supports tensor model parallelism and pipeline model parallelism. However, NVIDIA does not provide parallelised pre-trained models, and users need to write a program to split/merge Megatron-LM models for model parallelism. 

\noindent
\textbf{(3) Small GPU RAM:}
Many internal users share the HPC cluster in an institution, and acquiring GPU computing resources is very competitive. Moreover, each GPU can have less RAM size. For example, our has only 16 gigabyte GPU RAM. These restrictions created issues, especially when we fine-tuned the large models. For instance, to fine-tune the GPT-2 Extra Large model for the dataset, we used 4 GPUs with the HuggingFace pipeline model parallelism. Due to the overhead of parallelism and the GPU idling problems, the fine-tuning task required a very long processing time, but we could only acquire some of the required computing resources, and it delayed our experiments.

\subsubsection{Our recommendations}

As a machine learning framework, HuggingFace provides thousands of pre-trained models to perform various tasks on text, image, and sound. The framework provides not only easy-to-use APIs but also affordable memory efficiency and training speed. However, it supports only data parallelism and is very hard to train large models due to the lack of model parallelism supports. Megatron framework can be an alternative platform to perform tasks with large models as it supports 3D parallelism. Nevertheless, the Megatron framework only provides a limited number of pre-trained language models.
Besides, the DeepSpeed framework supports 3D parallelism with very high memory efficiency and high training speed. Moreover, it can be easily integrated with HuggingFace models. We applied DeepSpeed ZeRO Stage~2 on HuggingFace models to resolve our engineering challenges. With a single 16GB RAM GPU, we could fine-tune the GPT-2 XL model. Even though it was not practical, we could barely fine-tune the GPT-J model, consisting of 6 billion parameters, only with a single GPU.

We evaluated the training performance of the DeepSpeed framework in terms of processing speed and memory efficiency. We ran our test on our system with two RTX A6000 48GB GPUs, and the performance comparison results are depicted in Table~\ref{tab:performance_comparison}. When we were fine-tuning the GPT-2 XL model with a single GPU, the average train runtime per epoch was $3\textup{H} : 49\textup{M} : 03\textup{S}$ without DeepSpeed. But with DeepSpeed, the average train runtime per epoch was $2\textup{H} : 49\textup{M} : 23\textup{S}$, 26.05\% of the performance gain. More interestingly, when we applied DeepSpeed on the fine-tuning task, GPU RAM usage was cut by 57.77\%, making a large model fit into a small GPU RAM. Similarly, we could get 12.45\% processing speed gain and 65.64\% memory gain from the 2-GPU comparison.

\begin{table}[]
\vskip8pt
\caption{Fine-tuning performance comparison results of GPT-2 XL model with/without DeepSpeed.}
\label{tab:performance_comparison}
\resizebox{0.48\textwidth}{!}{%
\begin{tabular}{|c|c|c|c|c|}
\hline
Model & GPT-2 XL & GPT-2 XL & GPT-2 XL & GPT-2 XL \\ \hline
Number of GPU & 1 & 2 & 1 & 2 \\ \hline
Applied DeepSpeed & No & No & Stage 2 & Stage 2 \\ \hline
Parallelization & - & Data & - & Data \\ \hline
Epoch & 2 & 2 & 2 & 2 \\ \hline
Batch Size / GPU & 16 & 16 & 16 & 16 \\ \hline
Train Samples & 22072 & 22072 & 22072 & 22072 \\ \hline
Train Runtime & 7\textup{H}:38\textup{M}:06\textup{S} & 3\textup{H}:52\textup{M}:09\textup{S} & 5\textup{H}:38\textup{M}:46\textup{S} & 3\textup{H}:23\textup{M}:16\textup{S} \\ \hline
Train Runtime / epoch & 3\textup{H}:49\textup{M}:03\textup{S} & 1\textup{H}:56\textup{M}:05\textup{S} & 2\textup{H}:49\textup{M}:23\textup{S} & 1\textup{H}:41\textup{M}:38\textup{S} \\ \hline
Train Samples / second & 1.606 & 3.169 & 2.172 & 3.620 \\ \hline
Train Samples / second / GPU & 1.606 & 1.584 & 2.172 & 1.810 \\ \hline
Train Runtime for 1 sample & 0.623 & 0.631 & 0.460 & 0.553 \\ \hline
Multi-GPU overhead & - & 1.36\% & - & 20.00\% \\ \hline
\begin{tabular}[c]{@{}c@{}}Average   GPU RAM usage\\      (MB, batch size: 1)\end{tabular} & 29633 & 35755 & \cellcolor[HTML]{9AFF99}12515 & \cellcolor[HTML]{9AFF99}12285 \\ \hline
DeepSpeed Runtime Gain & - & - & \cellcolor[HTML]{9AFF99}26.05\% & \cellcolor[HTML]{9AFF99}12.45\% \\ \hline
DeepSpeed Memory Gain & - & - & \cellcolor[HTML]{9AFF99}57.77\% & \cellcolor[HTML]{9AFF99}65.64\% \\ \hline
\end{tabular}%
}
\vskip-3pt
\end{table}

\begin{insight}
We summarize our recommendations in the following:
\begin{itemize}[leftmargin=7mm\relax]
    \item Stick with data parallelism if the model fits inside one GPU.
    \item If we could not accommodate the model inside one GPU, go with Huggingface and DeepSpeed frameworks.
\end{itemize}
\end{insight}



\section{Related Work}
\label{sec:background}

To automate the generation of program documents and facilitate the natural language code search, a transformer-based language model, called CodeBERT~\cite{codebert}, is trained on pairs of natural languages and programming languages; for example, a source code slice having few comments in natural language regarding program description followed by the code instructions. CodeBERT's architecture consists of a BERT model and two generators, one for code instructions and the other for the natural language description. CodeBERT has a few variants, such as (i) GraphCodeBERT~\cite{graph_codeBERT} that uses semantic-level data flow structure, (ii) BART~\cite{bart}, with both encoder and decoder part in its model architecture, and its extension (iii) PLBART~\cite{PLBART}. In our work, we consider only CodeBERT because it is a base model and investigate its usage in software vulnerability detection, which has not been explored so far.


A BERTBase model is used for software vulnerability detection~\cite{Bertforsecurity}. It was fined tuned with Software Assurance Reference Dataset (SARD) database of $100$K C/C++ source files and tested with 123 vulnerabilities. This work showed that the BERTBase model and BERT with RNN heads of LSTM or BiLSTM models outperform standard LSTM or BiLSTM models. The highest detection accuracy reported was 93.49\% for their dataset and model. 
In contrast to this work, we consider a large dataset and multiple model architectures like GPT-2 and MegatronBERT~\cite{shoeybi2020megatronlm}. Besides, the data input form is different; they used the source file after removing labels and comments, whereas we leverage code gadgets. 
%

Except for a few transformer-based models (aforementioned), software vulnerability detection has been investigated through (i) dynamic analysis, (ii) pattern matching, and (iii) machine learning with Convolutional Neural Networks (CNN) and Recurrent Neural Networks (RNN)-based models. 
Dynamic analysis executes a program looking for unusual or unexpected behavior. Fuzzing~\cite{fuzzing} (or Atheris~\cite{atheris} for python codes) and the taint analysis~\cite{taint} are good examples of dynamic techniques. Unfortunately, they are not efficient for a long code/program as their runtime increases exponentially with the code/program length. 

On the pattern matching side, (i) code similarity techniques, (ii) rule-based techniques, and (iii) code-property graphs are used. 
Code-similarity techniques find vulnerabilities by comparing a currently scanned code with signatures of identified vulnerable codes. Those signatures are created by hashing, selection of substrings, or abstract representation such as graphs and trees. VUDDY~\cite{vuddy} and Vulpecker~\cite{vulpecker} are examples of code-similarity techniques.
For the rule-based technique, the analyst decides on rules that identify vulnerabilities. Flawfinder~\cite{flawfinder} and Coverity~\cite{coverity} apply the rule-based method.
Unfortunately, both code-similarity and rule-based techniques produce either high false negatives or high false positives, or both. This is due to the fact that the decision about vulnerability depends solely on historical data. Consequently, new unseen vulnerabilities cannot be detected.
Vulnerability detection works in two steps for code-property graph techniques: First, for a given source code, an appropriate graph is constructed by combining its abstract syntax tree and program dependence graph. Then, the code-property graph is traversed to detect vulnerabilities~\cite{codeproperty}. Sadly, this technique still requires human intervention and supervision.
 
Machine learning automates the detection and learning process with limited or no human intervention. However, classical machine learning requires feature engineering, which is difficult, extensively laborious, error-prone, and also needs human help to some extent. Thus, a significant body of research has studied deep learning but is limited to CNN and RNN-based models~\cite{deep_representationlearning, main_paper,IRLearning}. As these models require formatted data to capture important features related to the vulnerabilities, methods such as the lexed representation of C/C++ code~\cite{deep_representationlearning}, code gadget~\cite{main_paper}, code-property graphs~\cite{vulsniper}, improved code gadget with code attention and system dependency graph~\cite{muvuldeepecker}, and a minimum intermediate representation learning~\cite{IRLearning} are proposed. 
Besides deep learning language models, some works have been done using graph neural networks in software vulnerability detection~\cite{devign}. The method is named Devign, and it captures composite programming representations (abstract syntax tree, control flow, and data flow) of source codes. 
Overall, all these methods improve one over the other, but still, the results can be improved. 

Our work focuses on the standard code gadget and its extraction rather than its improved versions to see the relevance of the approach with the transformer-based language models. Moreover, any improved versions of extraction techniques for the code representation can be easily transferred to our work by updating the code gadget. Thus, we do not use minimum intermediate representation learning and graph neural networks. However, in most cases, our results with transformer-based models and standard code gadgets are better than those with minimum intermediate representation (see Section~\ref{sec:results}). We cannot compare our results with graph neural network results due to their results in different datasets.
\section{Conclusion}
\label{sec:conclusion}

This paper studied transformer-based language models for software vulnerability detection. Firstly, it presented a systematic framework to use these models in the domain. Then, it presented the comparative performances of these models along with recurrent neural network (RNN)-based models under binary and multi-class classification tasks. For the dataset with two vulnerabilities, buffer and resource management errors, related to the API function call, the transformer-based language models outperformed BiLSTM and BiGRU in all performance metrics (\emph{e.g.}, FPR, FNR, and F1-score). More precisely, GPT-2 Large and GPT-2 XL have the best F1-score for binary and multi-class classification (global average), respectively. The overall trend for F1-score was increasing with the models' increasing size. For a separate dataset with 126 types of different vulnerabilities (related to 341 CWE IDs) falling under four broad categories -- API function call, pointer usage, array usage, and arithmetic expressions -- this paper studied the performance of BiLSTM, BiGRU, BERTBase, and GPT-2 Base. Our results in binary classification tasks demonstrated that BERTBase and GPT-2 have better F1-score, but not good than BiGRU in FNR except for the array usage category. Overall, the transformer-based language models performed well in vulnerability detection. As these language models are difficult to run due to the challenges related to GPU memory size, libraries to perform model parallelism, and installation of the dependencies to the environment, this paper analyzed the popular platforms and presented the best methods to run these models.

%
%
%
%

\section{Acknowledgement}
This material is based upon work supported by the US Army International Technology Center Indo-Pacific (ITC-IPAC) under Contract No. FA520921P0015. Any opinions, findings and conclusions, or recommendations expressed in this material are those of the author(s) and do not necessarily reflect the views of the U.S. Army ITC-IPAC. Besides, we acknowledge our vacation student Mr. Andrew Cain for his help during the preliminary stage of this work. The work of Josef Pieprzyk was supported in part by Polish National Science Center (NCN) under Grant 2018/31/B/ST6/03003.

\bibliographystyle{ACM-Reference-Format}
\bibliography{bibo}


\begin{thebibliography}{00}


\ifx \showCODEN    \undefined \def \showCODEN     #1{\unskip}     \fi
\ifx \showDOI      \undefined \def \showDOI       #1{#1}\fi
\ifx \showISBNx    \undefined \def \showISBNx     #1{\unskip}     \fi
\ifx \showISBNxiii \undefined \def \showISBNxiii  #1{\unskip}     \fi
\ifx \showISSN     \undefined \def \showISSN      #1{\unskip}     \fi
\ifx \showLCCN     \undefined \def \showLCCN      #1{\unskip}     \fi
\ifx \shownote     \undefined \def \shownote      #1{#1}          \fi
\ifx \showarticletitle \undefined \def \showarticletitle #1{#1}   \fi
\ifx \showURL      \undefined \def \showURL       {\relax}        \fi
\providecommand\bibfield[2]{#2}
\providecommand\bibinfo[2]{#2}
\providecommand\natexlab[1]{#1}
\providecommand\showeprint[2][]{arXiv:#2}

\bibitem[\protect\citeauthoryear{Ahmad, Chakraborty, Ray, and Chang}{Ahmad
  et~al\mbox{.}}{2021}]%
        {PLBART}
\bibfield{author}{\bibinfo{person}{Wasi~Uddin Ahmad}, \bibinfo{person}{Saikat
  Chakraborty}, \bibinfo{person}{Baishakhi Ray}, {and}
  \bibinfo{person}{Kai{-}Wei Chang}.} \bibinfo{year}{2021}\natexlab{}.
\newblock \showarticletitle{Unified Pre-training for Program Understanding and
  Generation}.
\newblock \bibinfo{journal}{{\em CoRR\/}}  \bibinfo{volume}{abs/2103.06333}
  (\bibinfo{year}{2021}).
\newblock
\showeprint{2103.06333}
\showURL{%
\url{https://arxiv.org/abs/2103.06333}}


\bibitem[\protect\citeauthoryear{Anaconda}{Anaconda}{}]%
        {anaconda}
\bibfield{author}{\bibinfo{person}{Anaconda}.}
\newblock \bibinfo{title}{Anaconda}.
\newblock
\newblock
\newblock
\shownote{\url{https://www.anaconda.com}.}


\bibitem[\protect\citeauthoryear{Cho, van Merrienboer, Gulcehre, Bahdanau,
  Bougares, Schwenk, and Bengio}{Cho et~al\mbox{.}}{2014}]%
        {gru}
\bibfield{author}{\bibinfo{person}{Kyunghyun Cho}, \bibinfo{person}{Bart van
  Merrienboer}, \bibinfo{person}{Caglar Gulcehre}, \bibinfo{person}{Dzmitry
  Bahdanau}, \bibinfo{person}{Fethi Bougares}, \bibinfo{person}{Holger
  Schwenk}, {and} \bibinfo{person}{Yoshua Bengio}.}
  \bibinfo{year}{2014}\natexlab{}.
\newblock \showarticletitle{Learning Phrase Representations using RNN
  Encoder-Decoder for Statistical Machine Translation}.
\newblock  (\bibinfo{year}{2014}).
\newblock
\showDOI{%
\url{https://doi.org/10.48550/ARXIV.1406.1078}}


\bibitem[\protect\citeauthoryear{Coverity}{Coverity}{}]%
        {coverity}
\bibfield{author}{\bibinfo{person}{Coverity}.}
\newblock \bibinfo{title}{Coverity}.
\newblock
\newblock
\newblock
\shownote{\url{https://scan.coverity.com/}, last accessed on 01 July 2021.}


\bibitem[\protect\citeauthoryear{Devlin, Chang, Lee, and Toutanova}{Devlin
  et~al\mbox{.}}{2019}]%
        {bert}
\bibfield{author}{\bibinfo{person}{Jacob Devlin}, \bibinfo{person}{Ming{-}Wei
  Chang}, \bibinfo{person}{Kenton Lee}, {and} \bibinfo{person}{Kristina
  Toutanova}.} \bibinfo{year}{2019}\natexlab{}.
\newblock \showarticletitle{{BERT:} Pre-training of Deep Bidirectional
  Transformers for Language Understanding}. In \bibinfo{booktitle}{{\em Proc.
  the 2019 Conference of the North American Chapter of the Association for
  Computational Linguistics: Human Language Technologies, {NAACL-HLT}}}.
  \bibinfo{pages}{4171--4186}.
\newblock
\showDOI{%
\url{https://doi.org/10.18653/v1/n19-1423}}


\bibitem[\protect\citeauthoryear{Duan, Wu, Ji, Rui, Luo, Yang, and Wu}{Duan
  et~al\mbox{.}}{2019}]%
        {vulsniper}
\bibfield{author}{\bibinfo{person}{Xu Duan}, \bibinfo{person}{Jingzheng Wu},
  \bibinfo{person}{Shouling Ji}, \bibinfo{person}{Zhiqing Rui},
  \bibinfo{person}{Tianyue Luo}, \bibinfo{person}{Mutian Yang}, {and}
  \bibinfo{person}{Yanjun Wu}.} \bibinfo{year}{2019}\natexlab{}.
\newblock \showarticletitle{VulSniper: Focus Your Attention to Shoot
  Fine-Grained Vulnerabilities}. In \bibinfo{booktitle}{{\em Proceedings of the
  28th International Joint Conference on Artificial Intelligence}} {\em
  (\bibinfo{series}{IJCAI'19})}. \bibinfo{publisher}{AAAI Press},
  \bibinfo{pages}{4665–4671}.
\newblock
\showISBNx{9780999241141}


\bibitem[\protect\citeauthoryear{Face}{Face}{}]%
        {huggingface}
\bibfield{author}{\bibinfo{person}{Hugging Face}.}
\newblock \bibinfo{title}{Transformers: State-of-the-art Machine Learning for
  JAX, PyTorch and TensorFlow}.
\newblock
\newblock
\newblock
\shownote{\url{https://huggingface.co/docs/transformers/quicktour}.}


\bibitem[\protect\citeauthoryear{Feng, Guo, Tang, Duan, Feng, Gong, Shou, Qin,
  Liu, Jiang, and Zhou}{Feng et~al\mbox{.}}{2020}]%
        {codebert}
\bibfield{author}{\bibinfo{person}{Zhangyin Feng}, \bibinfo{person}{Daya Guo},
  \bibinfo{person}{Duyu Tang}, \bibinfo{person}{Nan Duan},
  \bibinfo{person}{Xiaocheng Feng}, \bibinfo{person}{Ming Gong},
  \bibinfo{person}{Linjun Shou}, \bibinfo{person}{Bing Qin},
  \bibinfo{person}{Ting Liu}, \bibinfo{person}{Daxin Jiang}, {and}
  \bibinfo{person}{Ming Zhou}.} \bibinfo{year}{2020}\natexlab{}.
\newblock \showarticletitle{CodeBERT: A Pre-Trained Model for Programming and
  Natural Languages}. In \bibinfo{booktitle}{{\em Proc. of Findings of the
  Association for Computational Linguistics: EMNLP 2020}}.
  \bibinfo{pages}{1536–1547}.
\newblock
\showDOI{%
\url{https://doi.org/10.18653/v1/2020.findings-emnlp.139}}


\bibitem[\protect\citeauthoryear{Foundation}{Foundation}{}]%
        {horovod}
\bibfield{author}{\bibinfo{person}{The~Linux Foundation}.}
\newblock \bibinfo{title}{Horovod}.
\newblock
\newblock
\newblock
\shownote{\url{https://horovod.ai}.}


\bibitem[\protect\citeauthoryear{Gao, Biderman, Black, Golding, Hoppe, Foster,
  Phang, He, Thite, Nabeshima, Presser, and Leahy}{Gao et~al\mbox{.}}{2020}]%
        {pile}
\bibfield{author}{\bibinfo{person}{Leo Gao}, \bibinfo{person}{Stella Biderman},
  \bibinfo{person}{Sid Black}, \bibinfo{person}{Laurence Golding},
  \bibinfo{person}{Travis Hoppe}, \bibinfo{person}{Charles Foster},
  \bibinfo{person}{Jason Phang}, \bibinfo{person}{Horace He},
  \bibinfo{person}{Anish Thite}, \bibinfo{person}{Noa Nabeshima},
  \bibinfo{person}{Shawn Presser}, {and} \bibinfo{person}{Connor Leahy}.}
  \bibinfo{year}{2020}\natexlab{}.
\newblock \showarticletitle{The {P}ile: An 800GB Dataset of Diverse Text for
  Language Modeling}.
\newblock \bibinfo{journal}{{\em arXiv preprint arXiv:2101.00027\/}}
  (\bibinfo{year}{2020}).
\newblock


\bibitem[\protect\citeauthoryear{Google}{Google}{2020}]%
        {atheris}
\bibfield{author}{\bibinfo{person}{Google}.} \bibinfo{year}{2020}\natexlab{}.
\newblock \bibinfo{title}{Atheris}.
\newblock
\newblock
\newblock
\shownote{\url{https://pypi.org/project/atheris/}.}


\bibitem[\protect\citeauthoryear{Guo, Ren, Lu, Feng, Tang, Liu, Zhou, Duan,
  Svyatkovskiy, Fu, Tufano, Deng, Clement, Drain, Sundaresan, Yin, Jiang, and
  Zhou}{Guo et~al\mbox{.}}{2020}]%
        {graph_codeBERT}
\bibfield{author}{\bibinfo{person}{Daya Guo}, \bibinfo{person}{Shuo Ren},
  \bibinfo{person}{Shuai Lu}, \bibinfo{person}{Zhangyin Feng},
  \bibinfo{person}{Duyu Tang}, \bibinfo{person}{Shujie Liu},
  \bibinfo{person}{Long Zhou}, \bibinfo{person}{Nan Duan},
  \bibinfo{person}{Alexey Svyatkovskiy}, \bibinfo{person}{Shengyu Fu},
  \bibinfo{person}{Michele Tufano}, \bibinfo{person}{Shao~Kun Deng},
  \bibinfo{person}{Colin~B. Clement}, \bibinfo{person}{Dawn Drain},
  \bibinfo{person}{Neel Sundaresan}, \bibinfo{person}{Jian Yin},
  \bibinfo{person}{Daxin Jiang}, {and} \bibinfo{person}{Ming Zhou}.}
  \bibinfo{year}{2020}\natexlab{}.
\newblock \showarticletitle{GraphCodeBERT: Pre-training Code Representations
  with Data Flow}.
\newblock \bibinfo{journal}{{\em CoRR\/}}  \bibinfo{volume}{abs/2009.08366}
  (\bibinfo{year}{2020}).
\newblock
\showeprint{2009.08366}
\showURL{%
\url{https://arxiv.org/abs/2009.08366}}


\bibitem[\protect\citeauthoryear{Hasheminezhad, Shirzad, Wu, Diehl, Schulz, and
  Kaiser}{Hasheminezhad et~al\mbox{.}}{2020}]%
        {horovod1}
\bibfield{author}{\bibinfo{person}{B. Hasheminezhad}, \bibinfo{person}{S.
  Shirzad}, \bibinfo{person}{N. Wu}, \bibinfo{person}{P. Diehl},
  \bibinfo{person}{H. Schulz}, {and} \bibinfo{person}{H. Kaiser}.}
  \bibinfo{year}{2020}\natexlab{}.
\newblock \showarticletitle{Towards a Scalable and Distributed Infrastructure
  for Deep Learning Applications}. In \bibinfo{booktitle}{{\em 2020 IEEE/ACM
  Fifth Workshop on Deep Learning on Supercomputers (DLS)}}.
  \bibinfo{publisher}{IEEE Computer Society}, \bibinfo{address}{Los Alamitos,
  CA, USA}, \bibinfo{pages}{20--30}.
\newblock
\showDOI{%
\url{https://doi.org/10.1109/DLS51937.2020.00008}}


\bibitem[\protect\citeauthoryear{Hochreiter and Schmidhuber}{Hochreiter and
  Schmidhuber}{1997}]%
        {lstm}
\bibfield{author}{\bibinfo{person}{Sepp Hochreiter} {and}
  \bibinfo{person}{Jurgen Schmidhuber}.} \bibinfo{year}{1997}\natexlab{}.
\newblock \showarticletitle{Long short-term memory}.
\newblock \bibinfo{journal}{{\em Neural Computation\/}} \bibinfo{volume}{9},
  \bibinfo{number}{8} (\bibinfo{year}{1997}), \bibinfo{pages}{1735--1780}.
\newblock
\showURL{%
\url{https://direct.mit.edu/neco/article-abstract/9/8/1735/6109/Long-Short-Term-Memory?redirectedFrom=fulltext}}


\bibitem[\protect\citeauthoryear{Huggingface.co}{Huggingface.co}{}]%
        {huggingface_token}
\bibfield{author}{\bibinfo{person}{Huggingface.co}.}
\newblock \bibinfo{title}{Tokenizer summary}.
\newblock
\newblock
\newblock
\shownote{\url{https://huggingface.co/transformers/v3.0.2/tokenizer_summary.html}.}


\bibitem[\protect\citeauthoryear{Husain, Wu, Gazit, Allamanis, and
  Brockschmidt}{Husain et~al\mbox{.}}{2019}]%
        {codeserachnet}
\bibfield{author}{\bibinfo{person}{Hamel Husain}, \bibinfo{person}{Ho{-}Hsiang
  Wu}, \bibinfo{person}{Tiferet Gazit}, \bibinfo{person}{Miltiadis Allamanis},
  {and} \bibinfo{person}{Marc Brockschmidt}.} \bibinfo{year}{2019}\natexlab{}.
\newblock \showarticletitle{CodeSearchNet Challenge: Evaluating the State of
  Semantic Code Search}.
\newblock \bibinfo{journal}{{\em CoRR\/}}  \bibinfo{volume}{abs/1909.09436}
  (\bibinfo{year}{2019}).
\newblock
\showeprint{1909.09436}
\showURL{%
\url{http://arxiv.org/abs/1909.09436}}


\bibitem[\protect\citeauthoryear{Kim, Woo, Lee, and Oh}{Kim
  et~al\mbox{.}}{2017}]%
        {vuddy}
\bibfield{author}{\bibinfo{person}{Seulbae Kim}, \bibinfo{person}{Seunghoon
  Woo}, \bibinfo{person}{Heejo Lee}, {and} \bibinfo{person}{Hakjoo Oh}.}
  \bibinfo{year}{2017}\natexlab{}.
\newblock \showarticletitle{VUDDY: A Scalable Approach for Vulnerable Code
  Clone Discovery}. In \bibinfo{booktitle}{{\em 2017 IEEE Symposium on Security
  and Privacy (SP)}}. \bibinfo{pages}{595--614}.
\newblock
\showDOI{%
\url{https://doi.org/10.1109/SP.2017.62}}


\bibitem[\protect\citeauthoryear{Lewis, Liu, Goyal, Ghazvininejad, Mohamed,
  Levy, Stoyanov, and Zettlemoyer}{Lewis et~al\mbox{.}}{2019}]%
        {bart}
\bibfield{author}{\bibinfo{person}{Mike Lewis}, \bibinfo{person}{Yinhan Liu},
  \bibinfo{person}{Naman Goyal}, \bibinfo{person}{Marjan Ghazvininejad},
  \bibinfo{person}{Abdelrahman Mohamed}, \bibinfo{person}{Omer Levy},
  \bibinfo{person}{Veselin Stoyanov}, {and} \bibinfo{person}{Luke
  Zettlemoyer}.} \bibinfo{year}{2019}\natexlab{}.
\newblock \showarticletitle{{BART:} Denoising Sequence-to-Sequence Pre-training
  for Natural Language Generation, Translation, and Comprehension}.
\newblock \bibinfo{journal}{{\em CoRR\/}}  \bibinfo{volume}{abs/1910.13461}
  (\bibinfo{year}{2019}).
\newblock
\showeprint{1910.13461}
\showURL{%
\url{http://arxiv.org/abs/1910.13461}}


\bibitem[\protect\citeauthoryear{Li, Wang, Xin, Yang, and Chen}{Li
  et~al\mbox{.}}{2020}]%
        {IRLearning}
\bibfield{author}{\bibinfo{person}{Xin Li}, \bibinfo{person}{Lu Wang},
  \bibinfo{person}{Yang Xin}, \bibinfo{person}{Yixian Yang}, {and}
  \bibinfo{person}{Yuling Chen}.} \bibinfo{year}{2020}\natexlab{}.
\newblock \showarticletitle{Automated Vulnerability Detection in Source Code
  Using Minimum Intermediate Representation Learning}.
\newblock \bibinfo{journal}{{\em Applied Sciences\/}} \bibinfo{volume}{10},
  \bibinfo{number}{5} (\bibinfo{year}{2020}).
\newblock
\showISSN{2076-3417}
\showDOI{%
\url{https://doi.org/10.3390/app10051692}}


\bibitem[\protect\citeauthoryear{Li, Zou, Xu, Chen, Zhu, and Jin}{Li
  et~al\mbox{.}}{2021}]%
        {vuldeelocator}
\bibfield{author}{\bibinfo{person}{Zhen Li}, \bibinfo{person}{Deqing Zou},
  \bibinfo{person}{Shouhuai Xu}, \bibinfo{person}{Zhaoxuan Chen},
  \bibinfo{person}{Yawei Zhu}, {and} \bibinfo{person}{Hai Jin}.}
  \bibinfo{year}{2021}\natexlab{}.
\newblock \showarticletitle{VulDeeLocator: A Deep Learning-based Fine-grained
  Vulnerability Detector}.
\newblock \bibinfo{journal}{{\em IEEE Transactions on Dependable and Secure
  Computing\/}} (\bibinfo{year}{2021}), \bibinfo{pages}{1--1}.
\newblock
\showDOI{%
\url{https://doi.org/10.1109/TDSC.2021.3076142}}


\bibitem[\protect\citeauthoryear{Li, Zou, Xu, Jin, Qi, and Hu}{Li
  et~al\mbox{.}}{2016}]%
        {vulpecker}
\bibfield{author}{\bibinfo{person}{Zhen Li}, \bibinfo{person}{Deqing Zou},
  \bibinfo{person}{Shouhuai Xu}, \bibinfo{person}{Hai Jin},
  \bibinfo{person}{Hanchao Qi}, {and} \bibinfo{person}{Jie Hu}.}
  \bibinfo{year}{2016}\natexlab{}.
\newblock \showarticletitle{VulPecker: an automated vulnerability detection
  system based on code similarity analysis}. In \bibinfo{booktitle}{{\em Proc.
  of the 32nd Annual Conference on Computer Security Applications}}.
  \bibinfo{pages}{201--213}.
\newblock
\showDOI{%
\url{https://doi.org/10.1145/2991079.2991102}}


\bibitem[\protect\citeauthoryear{Li, Zou, Xu, Jin, Zhu, and Chen}{Li
  et~al\mbox{.}}{2021}]%
        {sysevr}
\bibfield{author}{\bibinfo{person}{Zhen Li}, \bibinfo{person}{Deqing Zou},
  \bibinfo{person}{Shouhuai Xu}, \bibinfo{person}{Hai Jin},
  \bibinfo{person}{Yawei Zhu}, {and} \bibinfo{person}{Zhaoxuan Chen}.}
  \bibinfo{year}{2021}\natexlab{}.
\newblock \showarticletitle{SySeVR: A framework for using deep learning to
  detect software vulnerabilities}.
\newblock \bibinfo{journal}{{\em IEEE Trans. Dependable Sec. Comput\/}}
  (\bibinfo{year}{2021}).
\newblock
\showDOI{%
\url{https://doi.org/abs/1807.06756}}


\bibitem[\protect\citeauthoryear{Li, Zou, Xu, Ou, Jin, Wang, Deng, and
  Zhong}{Li et~al\mbox{.}}{2018}]%
        {main_paper}
\bibfield{author}{\bibinfo{person}{Zhen Li}, \bibinfo{person}{Deqing Zou},
  \bibinfo{person}{Shouhuai Xu}, \bibinfo{person}{Xinyu Ou},
  \bibinfo{person}{Hai Jin}, \bibinfo{person}{Sujuan Wang},
  \bibinfo{person}{Zhijun Deng}, {and} \bibinfo{person}{Yuyi Zhong}.}
  \bibinfo{year}{2018}\natexlab{}.
\newblock \showarticletitle{VulDeePecker: A Deep Learning-Based System for
  Vulnerability Detection}. In \bibinfo{booktitle}{{\em Proc. Network and
  Distributed System Security Symposium (NDSS)}}. \bibinfo{publisher}{Internet
  Society}.
\newblock
\showDOI{%
\url{https://doi.org/10.14722/ndss.2018.23158}}


\bibitem[\protect\citeauthoryear{Liu, Ott, Goyal, Du, Joshi, Chen, Levy, Lewis,
  Zettlemoyer, and Stoyanov}{Liu et~al\mbox{.}}{2019}]%
        {roberta}
\bibfield{author}{\bibinfo{person}{Yinhan Liu}, \bibinfo{person}{Myle Ott},
  \bibinfo{person}{Naman Goyal}, \bibinfo{person}{Jingfei Du},
  \bibinfo{person}{Mandar Joshi}, \bibinfo{person}{Danqi Chen},
  \bibinfo{person}{Omer Levy}, \bibinfo{person}{Mike Lewis},
  \bibinfo{person}{Luke Zettlemoyer}, {and} \bibinfo{person}{Veselin
  Stoyanov}.} \bibinfo{year}{2019}\natexlab{}.
\newblock \showarticletitle{RoBERTa: {A} Robustly Optimized {BERT} Pretraining
  Approach}.
\newblock \bibinfo{journal}{{\em CoRR\/}}  \bibinfo{volume}{abs/1907.11692}
  (\bibinfo{year}{2019}).
\newblock
\showeprint[arxiv]{1907.11692}
\showURL{%
\url{http://arxiv.org/abs/1907.11692}}


\bibitem[\protect\citeauthoryear{Microsoft}{Microsoft}{2022}]%
        {deepspeed}
\bibfield{author}{\bibinfo{person}{Microsoft}.}
  \bibinfo{year}{2022}\natexlab{}.
\newblock \bibinfo{title}{DeepSpeed}.
\newblock
\newblock
\newblock
\shownote{\url{https://www.deepspeed.ai}.}


\bibitem[\protect\citeauthoryear{Newsome and Song}{Newsome and Song}{2005}]%
        {taint}
\bibfield{author}{\bibinfo{person}{James Newsome} {and} \bibinfo{person}{Dawn
  Song}.} \bibinfo{year}{2005}\natexlab{}.
\newblock \showarticletitle{Dynamic Taint Analysis for Automatic Detection,
  Analysis, and Signature Generation of Exploits on Commodity Software}. In
  \bibinfo{booktitle}{{\em Proc. NDSS}}.
\newblock


\bibitem[\protect\citeauthoryear{Radford, Narasimhan, Salimans, and
  Sutskever}{Radford et~al\mbox{.}}{2018}]%
        {gpt}
\bibfield{author}{\bibinfo{person}{Alec Radford}, \bibinfo{person}{Karthik
  Narasimhan}, \bibinfo{person}{Tim Salimans}, {and} \bibinfo{person}{Ilya
  Sutskever}.} \bibinfo{year}{2018}\natexlab{}.
\newblock \bibinfo{title}{Improving Language Understanding by Generative
  Pre-Training}.
\newblock
\newblock
\newblock
\shownote{\url{https://cdn.openai.com/research-covers/language-unsupervised/language\_understanding\_paper.pdf}.}


\bibitem[\protect\citeauthoryear{Radford, Wu, Child, Luan, Amodei, and
  Sutskever}{Radford et~al\mbox{.}}{2019}]%
        {gpt2}
\bibfield{author}{\bibinfo{person}{Alec Radford}, \bibinfo{person}{Jeff Wu},
  \bibinfo{person}{Rewon Child}, \bibinfo{person}{David Luan},
  \bibinfo{person}{Dario Amodei}, {and} \bibinfo{person}{Ilya Sutskever}.}
  \bibinfo{year}{2019}\natexlab{}.
\newblock \showarticletitle{Language Models are Unsupervised Multitask
  Learners}.
\newblock  (\bibinfo{year}{2019}).
\newblock
\newblock
\shownote{\url{https://cdn.openai.com/better-language-models/language\_models\_are\_unsupervised\_multitask\_learners.pdf}.}


\bibitem[\protect\citeauthoryear{Rogers, Kovaleva, and Rumshisky}{Rogers
  et~al\mbox{.}}{2020}]%
        {how_bert_works}
\bibfield{author}{\bibinfo{person}{Anna Rogers}, \bibinfo{person}{Olga
  Kovaleva}, {and} \bibinfo{person}{Anna Rumshisky}.}
  \bibinfo{year}{2020}\natexlab{}.
\newblock \showarticletitle{A Primer in BERTology: What We Know About How BERT
  Works}.
\newblock \bibinfo{journal}{{\em Trans. of the Association for Computational
  Linguistics\/}}  \bibinfo{volume}{8} (\bibinfo{year}{2020}),
  \bibinfo{pages}{842–866}.
\newblock
\showDOI{%
\url{https://doi.org/10.1162/tacl_a_00349}}


\bibitem[\protect\citeauthoryear{Russell, Kim, Hamilton, Lazovich, Harer,
  Ozdemir, Ellingwood, and McConley}{Russell et~al\mbox{.}}{2018}]%
        {deep_representationlearning}
\bibfield{author}{\bibinfo{person}{Rebecca~L. Russell}, \bibinfo{person}{Louis
  Kim}, \bibinfo{person}{Lei~H. Hamilton}, \bibinfo{person}{Tomo Lazovich},
  \bibinfo{person}{Jacob~A. Harer}, \bibinfo{person}{Onur Ozdemir},
  \bibinfo{person}{Paul~M. Ellingwood}, {and} \bibinfo{person}{Marc~W.
  McConley}.} \bibinfo{year}{2018}\natexlab{}.
\newblock \showarticletitle{Automated Vulnerability Detection in Source Code
  Using Deep Representation Learning}. In \bibinfo{booktitle}{{\em Proc.
  ICMLA}}. \bibinfo{pages}{757-- 762}.
\newblock


\bibitem[\protect\citeauthoryear{Sanh, Debut, Chaumond, and Wolf}{Sanh
  et~al\mbox{.}}{2019}]%
        {distilbert}
\bibfield{author}{\bibinfo{person}{Victor Sanh}, \bibinfo{person}{Lysandre
  Debut}, \bibinfo{person}{Julien Chaumond}, {and} \bibinfo{person}{Thomas
  Wolf}.} \bibinfo{year}{2019}\natexlab{}.
\newblock \showarticletitle{DistilBERT, a distilled version of {BERT:} smaller,
  faster, cheaper and lighter}. In \bibinfo{booktitle}{{\em Proc. ${EMC}^2$:
  5th Edition Co-located with NeurIPS’19}}. \bibinfo{pages}{1--5}.
\newblock


\bibitem[\protect\citeauthoryear{Shoeybi, Patwary, Puri, LeGresley, Casper, and
  Catanzaro}{Shoeybi et~al\mbox{.}}{2020}]%
        {shoeybi2020megatronlm}
\bibfield{author}{\bibinfo{person}{Mohammad Shoeybi}, \bibinfo{person}{Mostofa
  Patwary}, \bibinfo{person}{Raul Puri}, \bibinfo{person}{Patrick LeGresley},
  \bibinfo{person}{Jared Casper}, {and} \bibinfo{person}{Bryan Catanzaro}.}
  \bibinfo{year}{2020}\natexlab{}.
\newblock \bibinfo{title}{Megatron-LM: Training Multi-Billion Parameter
  Language Models Using Model Parallelism}.
\newblock
\newblock
\showeprint[arxiv]{cs.CL/1909.08053}


\bibitem[\protect\citeauthoryear{Sutton, Greene, and Amini}{Sutton
  et~al\mbox{.}}{2007}]%
        {fuzzing}
\bibfield{author}{\bibinfo{person}{Michael Sutton}, \bibinfo{person}{Adam
  Greene}, {and} \bibinfo{person}{Pedram Amini}.}
  \bibinfo{year}{2007}\natexlab{}.
\newblock \showarticletitle{Fuzzing: Brute Force Vulnerability Discovery}.
\newblock \bibinfo{journal}{{\em Pearson Education\/}} (\bibinfo{year}{2007}).
\newblock
\showDOI{%
\url{https://doi.org/books?hl=en\&lr=\&id=DPAwwn7QDy8C\&oi=fnd\&pg=PT21\&ots=4yt9E59Owq\&sig=-Ik4SyRTD9YTvmMnYcpKQMH2jz4\&redir\_esc=y\#v=onepage\&q\&f=false}}


\bibitem[\protect\citeauthoryear{TIOBE}{TIOBE}{}]%
        {useofC}
\bibfield{author}{\bibinfo{person}{the software quality~company TIOBE}.}
\newblock \bibinfo{title}{TIOBE Index for May 2022}.
\newblock
\newblock
\newblock
\shownote{\url{https://www.tiobe.com/tiobe-index/}.}


\bibitem[\protect\citeauthoryear{Vaswani, Shazeer, Parmar, Uszkoreit, Jones,
  Gomez, Łukasz Kaiser, and Polosukhin}{Vaswani et~al\mbox{.}}{2017}]%
        {attention}
\bibfield{author}{\bibinfo{person}{Ashish Vaswani}, \bibinfo{person}{Noam
  Shazeer}, \bibinfo{person}{Niki Parmar}, \bibinfo{person}{Jakob Uszkoreit},
  \bibinfo{person}{Llion Jones}, \bibinfo{person}{Aidan~N Gomez},
  \bibinfo{person}{Łukasz Kaiser}, {and} \bibinfo{person}{Illia Polosukhin}.}
  \bibinfo{year}{2017}\natexlab{}.
\newblock \showarticletitle{Attention is all you need}. In
  \bibinfo{booktitle}{{\em Proc. Advances in neural information processing
  systems}}, Vol.~\bibinfo{volume}{30}. \bibinfo{publisher}{Curran Associates,
  Inc.}, \bibinfo{pages}{5998–6008}.
\newblock


\bibitem[\protect\citeauthoryear{VULDB}{VULDB}{a}]%
        {macos}
\bibfield{author}{\bibinfo{person}{VULDB}.}
\newblock \bibinfo{title}{Apple macOS USD File buffer overflow}.
\newblock
\newblock
\newblock
\shownote{\url{https://vuldb.com/?id.163591}.}


\bibitem[\protect\citeauthoryear{VULDB}{VULDB}{b}]%
        {whatsapp}
\bibfield{author}{\bibinfo{person}{VULDB}.}
\newblock \bibinfo{title}{Facebook WhatsApp on Android Video Stream buffer
  overflow}.
\newblock
\newblock
\newblock
\shownote{\url{https://vuldb.com/?id.160672}.}


\bibitem[\protect\citeauthoryear{VULDB}{VULDB}{c}]%
        {nvidia}
\bibfield{author}{\bibinfo{person}{VULDB}.}
\newblock \bibinfo{title}{NVIDIA Shield TV up to 8.2.1 NVDEC buffer overflow}.
\newblock
\newblock
\newblock
\shownote{\url{https://vuldb.com/?id.168508}.}


\bibitem[\protect\citeauthoryear{VulDeePecker}{VulDeePecker}{}]%
        {vuldeepecker_data}
\bibfield{author}{\bibinfo{person}{VulDeePecker}.}
\newblock \bibinfo{title}{Database of "VulDeePecker: A Deep Learning-Based
  System for Vulnerability Detection"}.
\newblock
\newblock
\newblock
\shownote{\url{https://github.com/CGCL-codes/VulDeePecker}.}


\bibitem[\protect\citeauthoryear{Wheeler}{Wheeler}{2018}]%
        {flawfinder}
\bibfield{author}{\bibinfo{person}{David~A. Wheeler}.}
  \bibinfo{year}{2018}\natexlab{}.
\newblock \bibinfo{title}{Flawfinder}.
\newblock
\newblock
\newblock
\shownote{\url{https://dwheeler.com/flawfinder/}.}


\bibitem[\protect\citeauthoryear{Yamaguchi, Golde, Arp, and Rieck}{Yamaguchi
  et~al\mbox{.}}{2014}]%
        {codeproperty}
\bibfield{author}{\bibinfo{person}{Fabian Yamaguchi}, \bibinfo{person}{Nico
  Golde}, \bibinfo{person}{Daniel Arp}, {and} \bibinfo{person}{Konrad Rieck}.}
  \bibinfo{year}{2014}\natexlab{}.
\newblock \showarticletitle{Modeling and discovering vulnerabilities with code
  property graphs}. In \bibinfo{booktitle}{{\em Proc. IEEE Symposium on
  Security and Privacy}}. \bibinfo{pages}{590--604}.
\newblock


\bibitem[\protect\citeauthoryear{Zhou, Liu, Siow, Du, and Liu}{Zhou
  et~al\mbox{.}}{2019}]%
        {devign}
\bibfield{author}{\bibinfo{person}{Yaqin Zhou}, \bibinfo{person}{Shangqing
  Liu}, \bibinfo{person}{Jingkai Siow}, \bibinfo{person}{Xiaoning Du}, {and}
  \bibinfo{person}{Yang Liu}.} \bibinfo{year}{2019}\natexlab{}.
\newblock \showarticletitle{Devign: Effective Vulnerability Identification by
  Learning Comprehensive Program Semantics via Graph Neural Networks}. In
  \bibinfo{booktitle}{{\em Proc. NeurIPS}}.
\newblock


\bibitem[\protect\citeauthoryear{Zhu, Kiros, Zemel, Salakhutdinov, Urtasun,
  Torralba, and Fidler}{Zhu et~al\mbox{.}}{2015}]%
        {bookcorpusdataset}
\bibfield{author}{\bibinfo{person}{Yukun Zhu}, \bibinfo{person}{Ryan Kiros},
  \bibinfo{person}{Richard~S. Zemel}, \bibinfo{person}{Ruslan Salakhutdinov},
  \bibinfo{person}{Raquel Urtasun}, \bibinfo{person}{Antonio Torralba}, {and}
  \bibinfo{person}{Sanja Fidler}.} \bibinfo{year}{2015}\natexlab{}.
\newblock \showarticletitle{Aligning Books and Movies: Towards Story-like
  Visual Explanations by Watching Movies and Reading Books}.
\newblock \bibinfo{journal}{{\em CoRR\/}}  \bibinfo{volume}{abs/1506.06724}
  (\bibinfo{year}{2015}).
\newblock
\showeprint{1506.06724}
\showURL{%
\url{http://arxiv.org/abs/1506.06724}}


\bibitem[\protect\citeauthoryear{Ziems and Wu}{Ziems and Wu}{2021}]%
        {Bertforsecurity}
\bibfield{author}{\bibinfo{person}{Noah Ziems} {and} \bibinfo{person}{Shaoen
  Wu}.} \bibinfo{year}{2021}\natexlab{}.
\newblock \bibinfo{title}{Security Vulnerability Detection Using Deep Learning
  Natural Language Processing}.
\newblock
\newblock
\showeprint[arxiv]{cs.CR/2105.02388}


\bibitem[\protect\citeauthoryear{Zou, Wang, Xu, Li, and Jin}{Zou
  et~al\mbox{.}}{2019}]%
        {muvuldeepecker}
\bibfield{author}{\bibinfo{person}{Deqing Zou}, \bibinfo{person}{Sujuan Wang},
  \bibinfo{person}{Shouhuai Xu}, \bibinfo{person}{Zhen Li}, {and}
  \bibinfo{person}{Hai Jin}.} \bibinfo{year}{2019}\natexlab{}.
\newblock \showarticletitle{µVulDeePecker: A deep learning-based system for
  multiclass vulnerability detection}.
\newblock \bibinfo{journal}{{\em IEEE Trans. Dependable Sec. Comput\/}}
  (\bibinfo{year}{2019}).
\newblock
\showDOI{%
\url{https://doi.org/10.1109/TDSC.2019.2942930}}


\end{thebibliography}

\appendix
\balance   

\section{Datasets}

\subsection{VulDeePecker Data}
\label{appendix:vul_data}
\begin{figure}[t]
	\centering
		\subfigure[Vulnerable with Buffer Error]{
			\includegraphics[width=0.45\columnwidth]{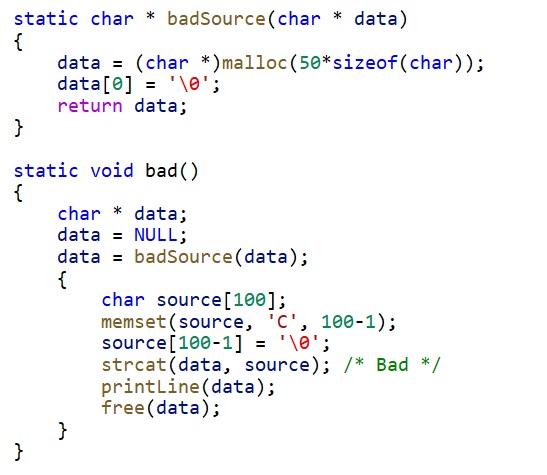}
		}
		\hspace{2mm}
		\subfigure[Non-vulnerable]{
			\includegraphics[width=0.45\columnwidth]{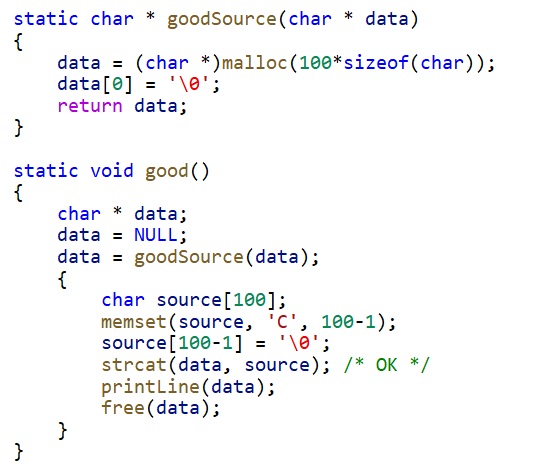}
		}
		\caption{An example of a program with Buffer Error and its corrected version.}
		\label{fig:buffer_error}
    \vspace{10pt}
	\centering
		\subfigure[Vulnerable with RME Error]{
			\includegraphics[width=0.4\columnwidth]{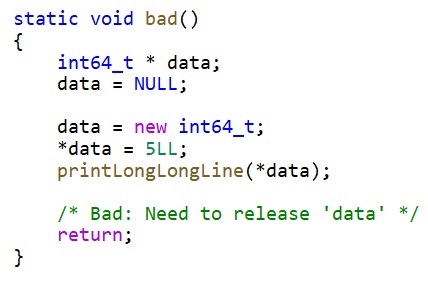}
		}
		\hspace{10mm}
		\subfigure[Non-vulnerable]{
			\includegraphics[width=0.4\columnwidth]{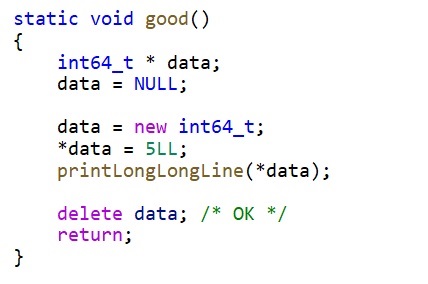}
		}
		\caption{An example of a program with Resource Management Error (RME) and its corrected version.}
		\label{fig:RME_error}
		\vskip5pt
\end{figure}

We consider the dataset published by CGCL/SCTS/BDTS Lab1 of Huazhong University of Science and Technology~\cite{vuldeepecker_data}. We call this dataset \emph{VulDeePecker dataset} because it was released as a part of their work that proposed a deep learning-based system for vulnerability detection called VulDeePecker~\cite{main_paper}. The dataset contains two common types of vulnerabilities collected from (i) syntactic and academic security flaws and (ii) popular open-source projects, including Linux kernel, Thunderbird, Wireshark, Apache HTTP Server, Xen, OpenSSL, and VLC media player, mostly available on the National Vulnerability Database (NVD) and the NIST Software Assurance Reference Dataset (SARD). The vulnerabilities are the following:
\begin{itemize}[leftmargin=3mm]
    \item \textbf{CWE-119 Buffer Error (BE):} BE covers buffer vulnerabilities caused by direct addressing of memory location without proper validation of a referenced memory buffer. Refer to Figure~\ref{fig:buffer_error} for an example. 
    \item \textbf{CWE-399 Resource Management Error (RME):} RME includes resource management vulnerabilities induced by improper handling of resources, which may lead to a variety of errors such as resource exhaustion, memory leakage, channel, and path exceptions. Refer to Figure~\ref{fig:RME_error} for an example.   
\end{itemize}

\subsection{SeVC Data}
\label{appendix:sysevr_data}
We consider the Semantics-based Vulnerability Candidate (SeVC) dataset having 126 types of different vulnerabilities~\cite{sysevr}. This dataset is collected from 1591 open-source C/C++ programs from the National Vulnerability Database (NVD) and 14000 programs from the Software Assurance Reference Dataset (SARD). Moreover, it has 56395 and 364232 vulnerable and clean samples. The SeVC dataset is divided into four major categories based on the cause of the vulnerability in the following way:
\begin{enumerate}
    \item Library/API Function Call: This category has vulnerabilities related to library/API function calls.
    \item Array Usage: This category has vulnerabilities related to arrays, like address transfer as a function parameter and improper array element access.
    \item Pointer Usage: This category has vulnerabilities related to pointers, like improper use in pointer arithmetic and wrong referencing.
    \item Arithmetic Expression: This category has vulnerabilities related to arithmetic expressions, like integer overflow. 
\end{enumerate}

\section{Models}
\label{appendix:models}
\subsection{Bidirectional Long Short Term Memory}
\begin{figure}[]
\centering
   \includegraphics[trim=0.2cm 0cm 0.2cm 0cm, clip=true,width=0.65\linewidth]{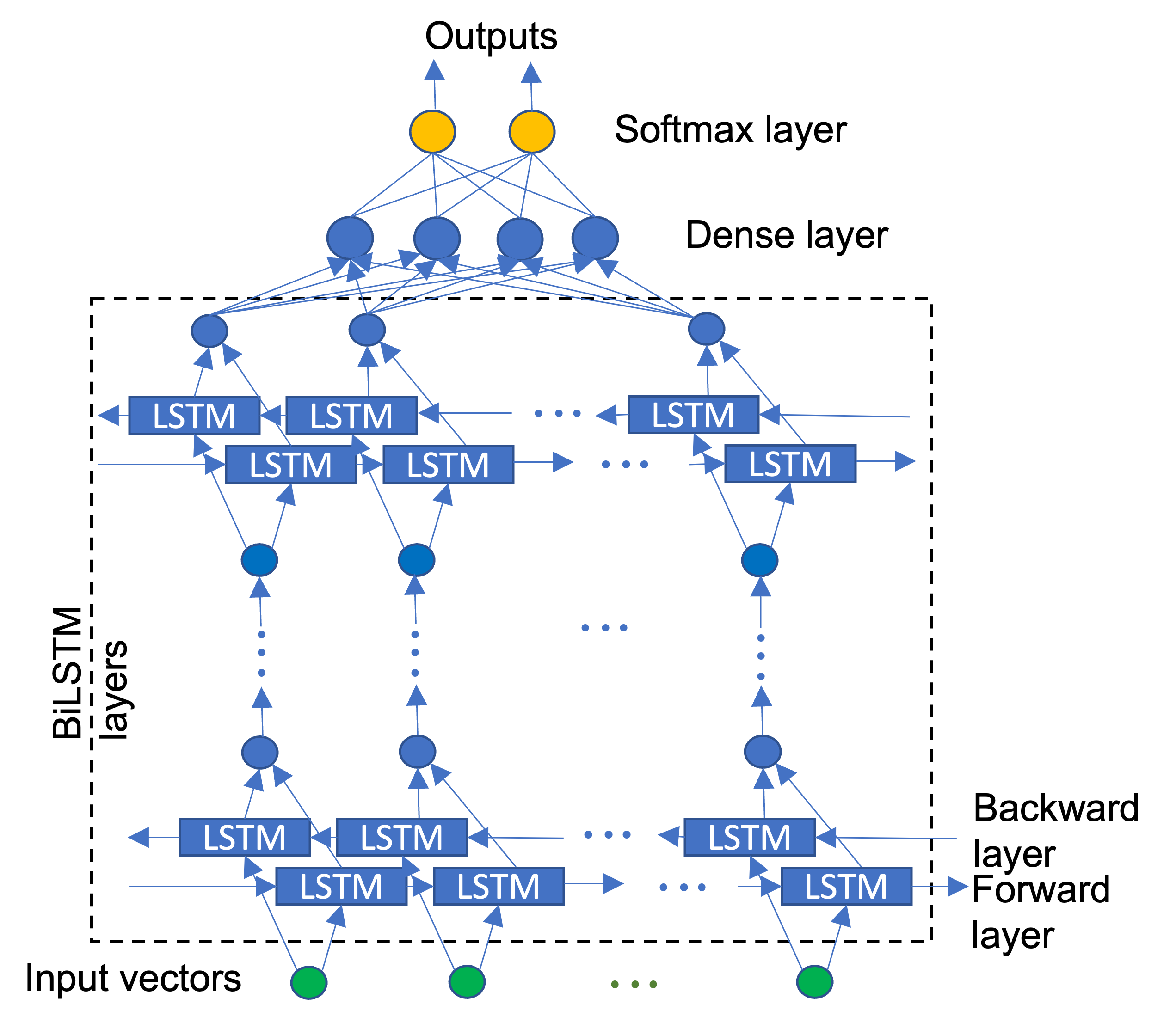}\
   \caption{BiLSTM architecture.}
\label{fig:bilstm}
\end{figure}

Bidirectional Long Short Term Memory (BiLSTM) are recurrent neural networks. It has two long short-term memory (LSTMs)~\cite{lstm} -- one LSTM takes the input in the forward direction and the other in the backward direction -- which enables BiLSTM to learn long-term dependencies in the input sequence efficiently. BiLSTM architecture is depicted in Figure~\ref{fig:bilstm}. The LSTM block in the BiLSTM has the cell state regulated by three gates, namely forget gate, update gate, and output gate. Each LSTM cell has three inputs: (i) memory component $C_{t-1}$ and activation component $a_{t-1}$ are the two inputs from the previous cell, and (ii) word embedding $x_t$ at time $t$ of input data. Computations are carried out as presented in the following:
\begin{align*}
    &\textup{Forget gate: } f_t =  \sigma (W_f \cdot [a_{t-1}, x_t] + b_f),\\
    &\textup{Update gate: } i_t = \sigma (W_u \cdot [a_{t-1}, x_t] + b_u),\\
    &\textup{Output gate: } o_t = \sigma (W_o \cdot [a_{t-1}, x_t] + b_o),\\
\end{align*}
where $b$ is the bias, $W$ is the learning parameter (weight matrix), and $\sigma$ is the Sigmoid function. The memory component $C_t$ is calculated as:
\begin{align*}
    C_t = f_t * C_{t-1} + i_t * \tilde{C}_t,
\end{align*}
where $\tilde{C}_t = tanh (W_c[a_{t-1}, x_t] + b_c$, and `*` is elementwise multiplication. The activation component for the next cell is calculated as:
\begin{align*}
    a_t = o_t * \textup{tanh} (C_t).
\end{align*}
\vskip-7pt
%

\subsection{Bidirectional Gated Recurrent Unit}
Bidirectional Gated Recurrent Unit (BiGRU)~\cite{gru} is a recurrent neural network with gated recurrent units (GRUs). A GRU has a gating mechanism, and it is similar to LSTM but has a forget gate, fewer parameters, and no output gate. The architecture of BiGRU is the same as in Figure~\ref{fig:bilstm} where each LSTM cell is replaced by a GRU cell. Moreover, each GRU cell has two inputs; memory component $C_{t-1}$ from the previous cell and word embedding $x_t$ at time $t$ of input data. Computations are carried out as follows:
\begin{align*}
    &\textup{Update gate: } i_t = \sigma (W_u \cdot x_t + U_u \cdot C_{t-1} + b_u),\\
    &\textup{Reset gate: } r_t = \sigma (W_r \cdot x_t + U_r \cdot C_{t-1} + b_r),\\
    &\textup{Current memory content: } \tilde{C}_t = \phi (W_h \cdot x_t + U_h(r_t \odot C_{t-1}) + b_h),\\
    &\textup{Final memory content: } {C}_t = i_t \odot C_{t-1}  + (1 - i_t) \odot \tilde{C}_t,\\
\end{align*}
where $b$ is the bias, $W$ and $U$ are the learning parameter (weight matrix), $\sigma$ is the Sigmoid function, $\phi$ is a hyperbolic tangent function, and $\odot$ is the Hadamard (element-wise) product.

\subsection{Transformers}
\begin{figure}[t]
\centering
   \includegraphics[trim=1cm 1cm 0.5cm 1cm, clip=true,width=1\linewidth]{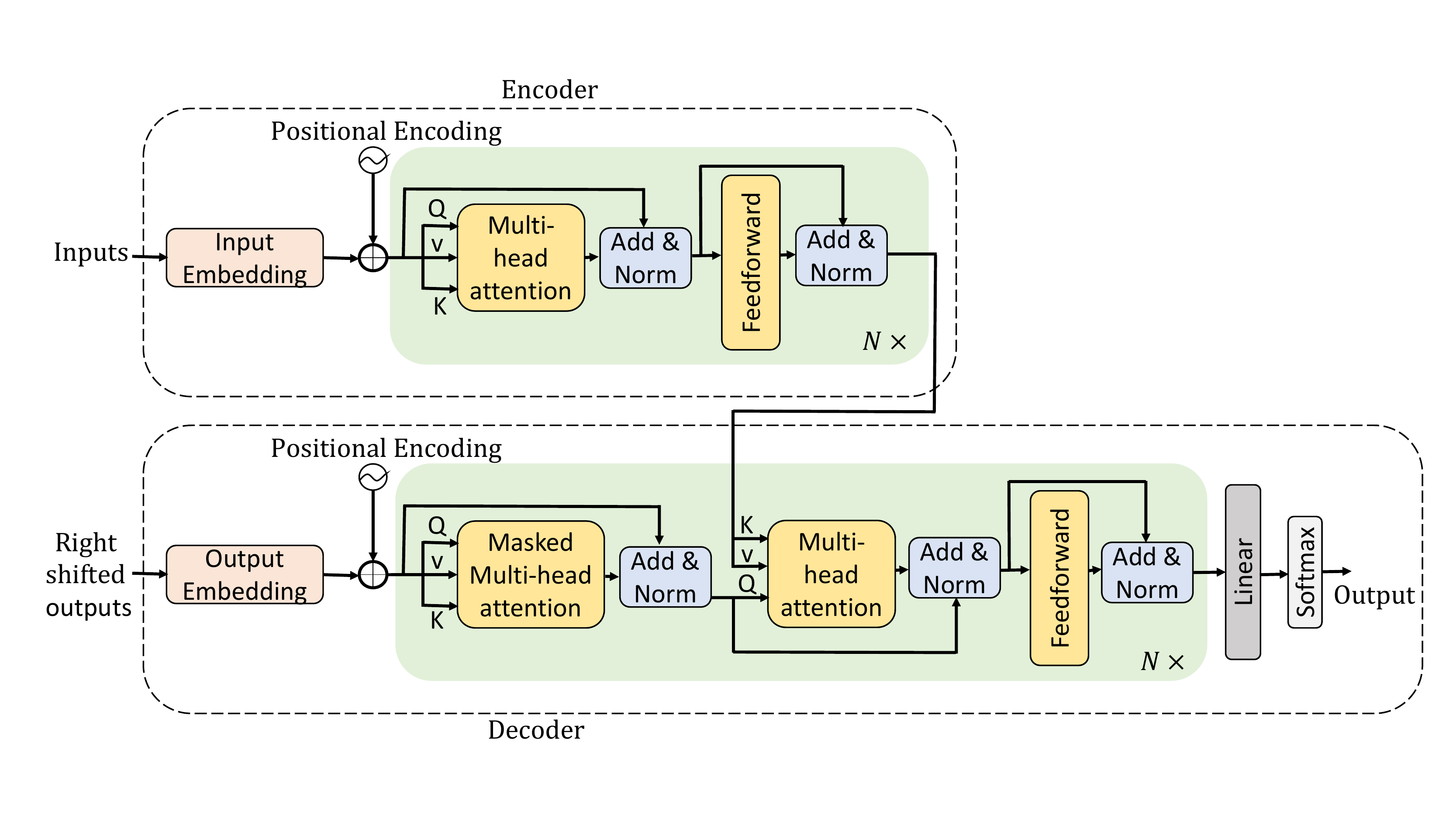}\
   \caption{Transformer architecture.}
\label{fig:tranformer}
\vskip8pt
\end{figure}

Transformers are deep neural network-based models based on an attention mechanism that differentially weights the importance and context of each token in the input sentence~\cite{attention}. Refer to Figure~\ref{fig:tranformer} for an illustration of transformer architecture. Its network architecture has encoder and decoder blocks, and the core elements in these blocks are positional encoding, multi-head attention, and fully connected layers. 

\begin{figure}[t]
\centering
   \includegraphics[trim=8cm 5cm 8cm 4.5cm, clip=true, width=0.8\linewidth]{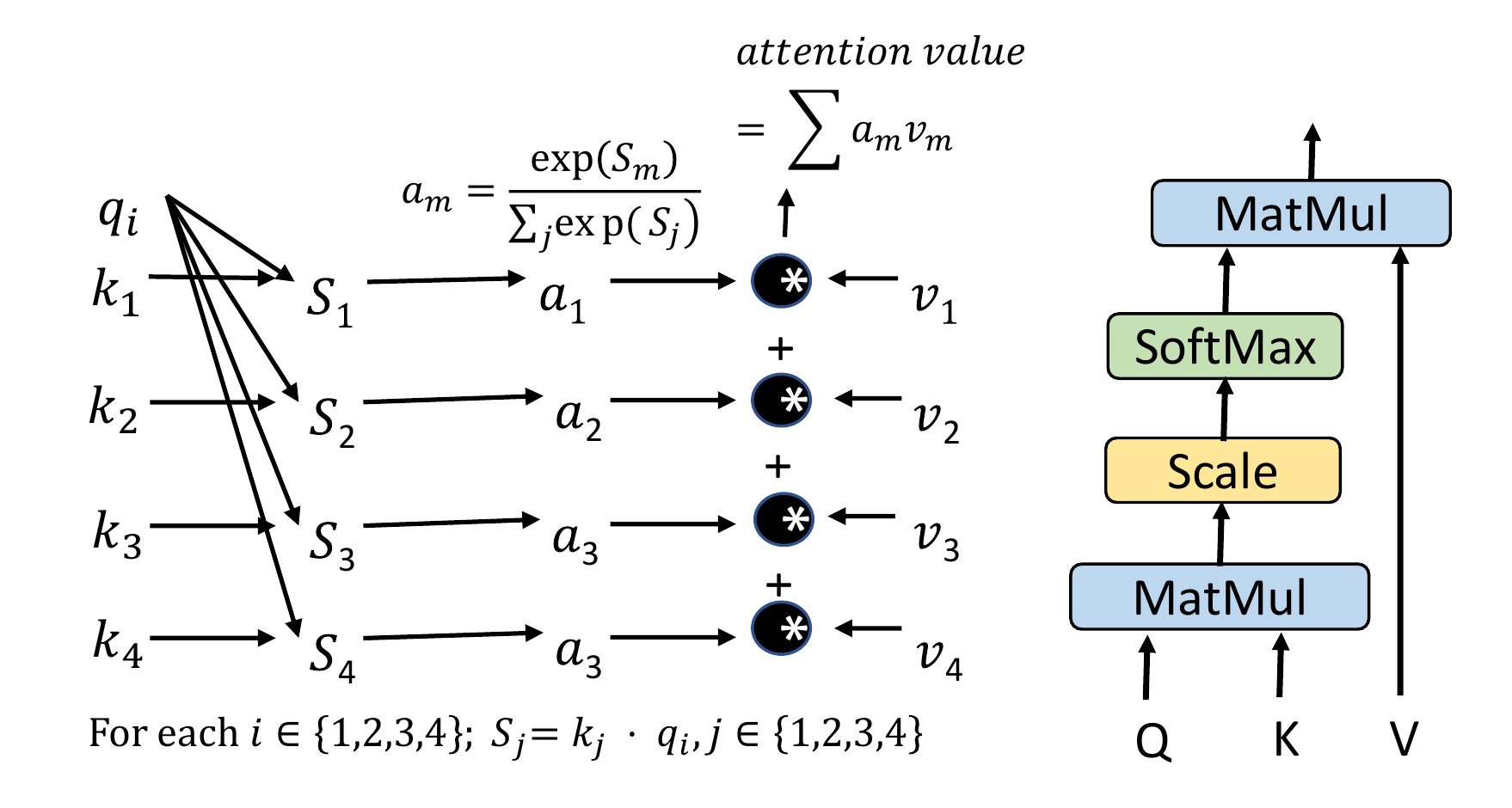}
   \caption{Illustration of operations in self-attention; scaled dot-product attention with four keys and four values associated with four tokens in an input sequence (left figure) and its system block (right figure).}
\label{fig:self_attention}
\end{figure}

\begin{figure}[t]
\centering
   \includegraphics[trim=11cm 6cm 9cm 5cm, clip=true,width=0.5\linewidth]{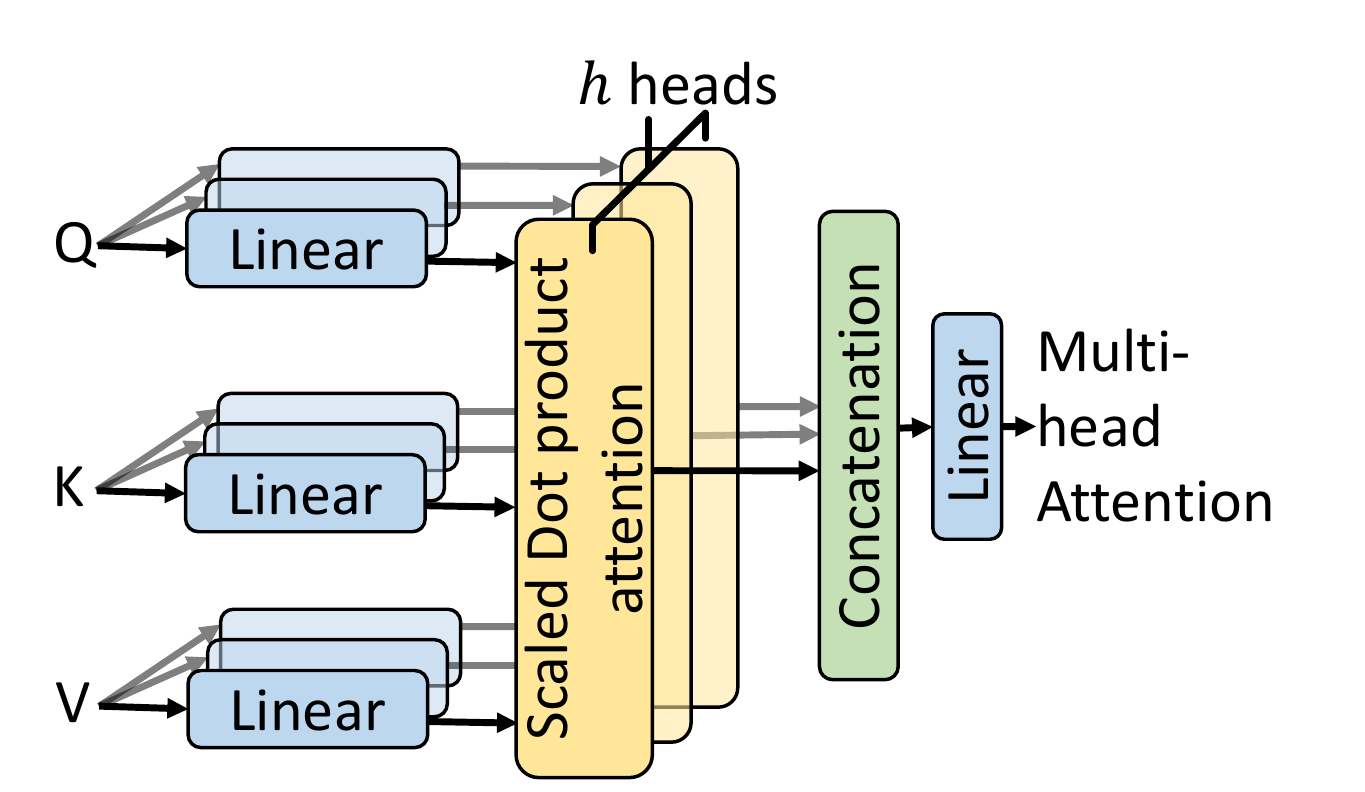}\
   \caption{Multi-head attention mechanism.}
\label{fig:multihead_attention}
\end{figure}

Positional encoding provides the relative or exact position of the tokens (of data points) in the input sequence, and it is applied to the embeddings, which convert the input tokens (encoder block) and output tokens (decoder block) to vectors of dimension $d_{\textup{model}}$. Sinusoidal functions can be used for the positional encodings, for example, $sin(pos/10000^{{2i}/d_{\textup{model}}})$ for even positions, and $cos(pos/10000^{{2i}/d_{\textup{model}}})$ for odd positions, where $i$ and $pos$ are the dimension and position index, respectively.     
An attention mechanism is applied to represent the input sequence by relating each word/token to related words/tokens in the input sequence. This is called \emph{self-attention}. One of its examples is scaled dot-product attention, whose calculation is illustrated in Figure~\ref{fig:self_attention}. Firstly, each output vector of the positional encoding is converted to three vectors; a query $q$, a key $k$, and a value $v$. The conversion is performed by multiplying the output vector by matrices $W^Q$, $W^K$, and $W^V$, respectively. These matrices are continuously updated during training for better projections. Secondly, the dot products of query (a vector $q$) with all keys (all vectors $k$) of the input sequence generates scores $S_i$, $i\in \{1,2,3,\dots\}$. The scores are normalized, and then softmax is calculated. This provides the weight of other parts of the input sequence while encoding a word at a specific position. Finally, each value vector is multiplied by their corresponding softmax scores and summed together to yield the attention value of the token. The following equation represents all these operations in matrix form:
\begin{equation}
    \textup{Attention}(Q,K,V) = \textup{softmax}(\frac{QK^T}{\sqrt{d_k}})V,
\end{equation}
where $d_k$ is the key vector's dimension, and $Q$, $K$ and $V$ are matrices packed with all (multi-head) queries, (multi-head) keys and (multi-head) values, respectively.   

The output of the attention function is further improved by collectively attending to the information from different representation subspaces. This is called \emph{multi-head attention}, and performed as follows: (i) The queries, values, and keys are linearly projected to $h$ times, (ii) the attention function is calculated in each projection in parallel, and (iii) all the outputs of the attention functions are concatenated and projected through the linear layer as shown in Figure~\ref{fig:multihead_attention}. Its representation in matrix form is the following:  
\begin{equation}
    \textup{MultiHead}(Q,K,V) = \textup{Concat}(\textup{head}_1, \dotsc, \textup{head}_h) W^O,
\end{equation}
where $\textup{head}_i = \textup{Attention}(QW_i^Q, KW_i^K, VW_i^V)$, and $W^O$ is a weight matrix.

In transformer architecture, unlike in encoder block, the decoder block has two layers of multi-head attention. The first one is masked multi-head attention that enables target sequences, \emph{i.e.}, its right-shifted previous outputs, paying attention to itself, and ignoring the future decoder's input. Next is the multi-head attention that enables the target sequence to pay attention to the input sequences. Consequently, the attention scores of each target sequence, considering the attention scores of the input sequence (coming from the encoder block), are generated, and these are transformed from their embedding space to probability space by the final fully connected layer and softmax layer.   

Now we present various transformer-based models that are considered in this paper in the following paragraphs.

\paragraph{Bidirectional Encoder Representations from Transformers} 
Bidirectional Encoder Representations from Transformers (BERT) is a transformer~\cite{attention} whose architecture is solely based only on its encoder blocks~\cite{bert}. It scans the entire surrounding context of its input all at once, processes it through encoder blocks, and collectively fuses the left and right context in all of its layers to learn syntactic information and acquires some semantic and world knowledge~\cite{how_bert_works}. It is a state-of-art language model.
In this project, we have utilised two BERT models pre-trained on lower-cased English datasets such as Wikipedia (2500M words) and BooksCorpus dataset (800M words)~\cite{bookcorpusdataset}. 
The BERT Base model has 110M parameters with \emph{12 layers, 768 hidden size, and 12 attention heads}.
The BERT Large model has 340M parameters with \emph{24 layers, 1024 hidden size, 16 attention heads}.

\paragraph{DistilBERT}
DistilBERT~\cite{distilbert} is a smaller, faster, cheaper, and lighter version of BERT while retaining the close performance (e.g., 95\%) of the original BERT's performance. The size of DistilBERT is 40\% smaller than BERT, and it is obtained by leveraging the knowledge distillation technique in the pre-training phase. Knowledge distillation  is a compression technique where a smaller model, called student, is trained such that it reproduces the learning of the larger model, called teacher. DistillBERT has 66M parameters with \emph{6 layers, 768 hidden size, and 12 attention heads}. 

\paragraph{RoBERTa}
Robustly optimized BERT approach (RoBERTa)~\cite{roberta} updates the key hyper-parameters during BERT's pre-training towards better optimization, performance, and robustness across NLP tasks. The updates include (1) longer model training time with more data and large mini-batch, (2) training on a longer sequence, (3) removing the next sentence prediction (an objective method applied for the BERT's pre-training to predict if the two segments are following each other or belong to different documents to create a semantic inference), and (4) dynamic masking to the training data. In this paper, we consider the RoBERTa with the BERT Base architecture, and it has \emph{12 layers, 768 hidden size, and 12 attention heads}.

\paragraph{CodeBERT} 
CodeBERT~\cite{codebert} is a bimodal pre-trained transformer model for both programming and natural languages. It is trained on bimodal data (code \& documents)~\cite{codeserachnet}, with codes from Python, Java, JavaScript, Ruby, Go, and PHP. Its architecture follows BERT~\cite{bert} and RoBERTa~\cite{roberta}. During pre-training, its architecture has two generators, namely NL-generator and Code-generator, and one NL-code discriminator. NL-generator and Code-generator generate plausible tokens for masked positions based on surrounding contexts for natural language input and code input, respectively. NL-Code discriminator is trained on the tokens sampled from NL-generator and Code-generator. Its architecture is the same as RoBERTa, and it is the targeted model used for the fine-tuning purpose. Moreover, it has 125M parameters with \emph{12 layers, 768 hidden size, and 12 attention heads}.

\paragraph{Generative Pre-trained Transformer}
The architecture of Generative Pre-trained Transformer (GPT) is based on decoder blocks of transformer and masked self-attention~\cite{gpt}. GPT learns long-range dependencies between words and sentence and world knowledge through the generative pre-training, which is unsupervised learning, and this learning are transferred to specific task through fine-tuning, which is supervised learning. In contrast to BERT, GPT outputs one token at a time, and that is added to its input sequence for the next round. Consequently, each token in the input sentence has a context of the previous words only. Thus, GPT is called an auto-regressive model. GPT outperforms available models that are based on recursive neural networks, convolutional neural networks, and LSTMs~\cite{gpt}. Moreover, the GPT model is a powerful predictor of the next token in a sequence, so it is popular in text generation tasks.   
GPT models have been improved by increasing their model parameters and rigorous pre-training on a large corpus of English text datasets, called WebText~\cite{gpt2}. More precisely, improved GPT models consider task conditioning that enables the model to produce task-specific outputs for the same input. In this paper, we consider GPT-2 models of various sizes: (1) GPT-2 Base, which has 117M parameters with \emph{12 layers, 768 hidden size, and 12 attention heads}, (2) GPT-2 Large, which has 774M parameters with \emph{36 layers, 1280 hidden size, and 20 attention heads}, (3) GPT-2 XL, which has 1.5B parameters with \emph{48 layers, 1600 hidden size, and 25 attention heads}, and (4) GPT-J, which has 6B parameters with \emph{28 layers, 4096 hidden size, and 16 attention heads}, pre-trained on the dataset called Pile having a diverse text data~\cite{pile}.

\paragraph{Megatron-LM} 
Megatron-LMs are transformer-based models developed by NVIDIA. They are generated by enabling intra-layer model-parallelism on the architecture level of the existing language models, such as BERT and GPT-2~\cite{shoeybi2020megatronlm}. The model parallelism includes two-dimensional model parallelism: tensor model parallelism and pipeline model parallelism. The tensor model parallelism splits a single transformer across multiple GPUs, while the pipeline model parallelism splits transformer layers across multiple GPUs. Thus, these models efficiently utilize multiple GPU environments to train large models that can not be fitted inside one GPU. 
The resulting models do not require changes in compilers and libraries, and they enable an efficient way to scale up their model parameters to billions (e.g., 8.3B) for an increased performance~\cite{shoeybi2020megatronlm}. Moreover, Megatron's BERT version has rearranged normalization layers and residual connections to allow the direct flow of gradients through the network without going through the normalization layers. This enables increasing in its performance with the increase in the model's size.
In this paper, we consider Megatron versions of BERT and GPT-2 models provided by Nvidia; (1) MegatronBERT having 345M parameters with \emph{24 layers, 1024 hidden size, and 16 attention heads}, and (2) Megatron-GPT2 having 345M parameters with \emph{24 layers, 1024 hidden size, and 16 attention heads}. These models are pre-trained on text data sourced from Wikipedia, RealNews, OpenWebText, and CC-Stories~\cite{shoeybi2020megatronlm}.

\begin{table*}[hbt!]
\centering
\caption{Test performance of various models for the binary classification on clean VulDeePecker dataset.}
\resizebox{\textwidth}{!}{%
\begin{tabular}{|c|c|c|ccc|ccc|ccc|ccc|ccc|ccc|}
\hline
\multirow{2}{*}{\textbf{\begin{tabular}[c]{@{}c@{}}Dataset and\\ Vulnerability\end{tabular}}}                 & \multirow{2}{*}{\textbf{Metrics}} & \multirow{2}{*}{\textbf{\begin{tabular}[c]{@{}c@{}}VulDeePecker\\      Original\end{tabular}}} & \multicolumn{3}{c|}{\textbf{BiLSTM}}                                  & \multicolumn{3}{c|}{\textbf{BiGRU}}                                   & \multicolumn{3}{c|}{\textbf{BERT Base}}                               & \multicolumn{3}{c|}{\textbf{GPT-2 Base}}                              & \multicolumn{3}{c|}{\textbf{CodeBERT}}                                & \multicolumn{3}{c|}{\textbf{DistilBERT}}                              \\ \cline{4-21} 
                                                                                                              &                                   &                                                                                                & \multicolumn{1}{c|}{Fold 1}  & \multicolumn{1}{c|}{Fold 2}  & Fold 3  & \multicolumn{1}{c|}{Fold 1}  & \multicolumn{1}{c|}{Fold 2}  & Fold 3  & \multicolumn{1}{c|}{Fold 1}  & \multicolumn{1}{c|}{Fold 2}  & Fold 3  & \multicolumn{1}{c|}{Fold 1}  & \multicolumn{1}{c|}{Fold 2}  & Fold 3  & \multicolumn{1}{c|}{Fold 1}  & \multicolumn{1}{c|}{Fold 2}  & Fold 3  & \multicolumn{1}{c|}{Fold 1}  & \multicolumn{1}{c|}{Fold 2}  & Fold 3  \\ \hline
\multirow{5}{*}{\begin{tabular}[c]{@{}c@{}}Group 1, \\ Buffer Error \\ (BE)\end{tabular}}                     & FPR                               & 2.90\%                                                                                         & \multicolumn{1}{c|}{33.25\%} & \multicolumn{1}{c|}{36.36\%} & 31.98\% & \multicolumn{1}{c|}{18.41\%} & \multicolumn{1}{c|}{11.64\%} & 15.52\% & \multicolumn{1}{c|}{4.80\%}  & \multicolumn{1}{c|}{3.16\%}  & 4.20\%  & \multicolumn{1}{c|}{4.11\%}  & \multicolumn{1}{c|}{3.82\%}  & 4.68\%  & \multicolumn{1}{c|}{2.89\%}  & \multicolumn{1}{c|}{2.45\%}  & 3.58\%  & \multicolumn{1}{c|}{3.70\%}  & \multicolumn{1}{c|}{3.99\%}  & 3.87\%  \\ \cline{2-21} 
                                                                                                              & FNR                               & 18.00\%                                                                                        & \multicolumn{1}{c|}{17.61\%} & \multicolumn{1}{c|}{10.85\%} & 17.37\% & \multicolumn{1}{c|}{31.54\%} & \multicolumn{1}{c|}{39.34\%} & 35.58\% & \multicolumn{1}{c|}{6.90\%}  & \multicolumn{1}{c|}{5.71\%}  & 6.95\%  & \multicolumn{1}{c|}{6.11\%}  & \multicolumn{1}{c|}{6.98\%}  & 6.23\%  & \multicolumn{1}{c|}{4.47\%}  & \multicolumn{1}{c|}{4.70\%}  & 5.37\%  & \multicolumn{1}{c|}{6.77\%}  & \multicolumn{1}{c|}{6.15\%}  & 7.34\%  \\ \cline{2-21} 
                                                                                                              & Precision                         & 82.00\%                                                                                        & \multicolumn{1}{c|}{71.25\%} & \multicolumn{1}{c|}{71.03\%} & 72.10\% & \multicolumn{1}{c|}{69.70\%} & \multicolumn{1}{c|}{77.35\%} & 72.07\% & \multicolumn{1}{c|}{92.31\%} & \multicolumn{1}{c|}{95.13\%} & 93.24\% & \multicolumn{1}{c|}{93.40\%} & \multicolumn{1}{c|}{94.09\%} & 92.56\% & \multicolumn{1}{c|}{95.34\%} & \multicolumn{1}{c|}{96.22\%} & 94.26\% & \multicolumn{1}{c|}{93.97\%} & \multicolumn{1}{c|}{93.90\%} & 93.70\% \\ \cline{2-21} 
                                                                                                              & Recall                            & 91.70\%                                                                                        & \multicolumn{1}{c|}{82.39\%} & \multicolumn{1}{c|}{89.15\%} & 82.63\% & \multicolumn{1}{c|}{68.46\%} & \multicolumn{1}{c|}{60.66\%} & 64.42\% & \multicolumn{1}{c|}{93.10\%} & \multicolumn{1}{c|}{94.29\%} & 93.05\% & \multicolumn{1}{c|}{93.89\%} & \multicolumn{1}{c|}{93.02\%} & 93.77\% & \multicolumn{1}{c|}{95.53\%} & \multicolumn{1}{c|}{95.30\%} & 94.63\% & \multicolumn{1}{c|}{93.23\%} & \multicolumn{1}{c|}{93.85\%} & 92.66\% \\ \cline{2-21} 
                                                                                                              & F1-score                          & 86.60\%                                                                                        & \multicolumn{1}{c|}{76.42\%} & \multicolumn{1}{c|}{79.07\%} & 77.01\% & \multicolumn{1}{c|}{69.08\%} & \multicolumn{1}{c|}{67.99\%} & 68.03\% & \multicolumn{1}{c|}{92.71\%} & \multicolumn{1}{c|}{94.71\%} & 93.15\% & \multicolumn{1}{c|}{93.64\%} & \multicolumn{1}{c|}{93.55\%} & 93.16\% & \multicolumn{1}{c|}{95.44\%} & \multicolumn{1}{c|}{95.76\%} & 94.44\% & \multicolumn{1}{c|}{93.60\%} & \multicolumn{1}{c|}{93.87\%} & 93.18\% \\ \hline
\multirow{5}{*}{\begin{tabular}[c]{@{}c@{}}Group 2, \\ Resource \\ Management \\ Error \\ (RME)\end{tabular}} & FPR                               & 2.80\%                                                                                         & \multicolumn{1}{c|}{16.79\%} & \multicolumn{1}{c|}{18.18\%} & 13.33\% & \multicolumn{1}{c|}{4.13\%}  & \multicolumn{1}{c|}{3.35\%}  & 5.73\%  & \multicolumn{1}{c|}{3.05\%}  & \multicolumn{1}{c|}{4.04\%}  & 2.86\%  & \multicolumn{1}{c|}{4.33\%}  & \multicolumn{1}{c|}{3.75\%}  & 3.36\%  & \multicolumn{1}{c|}{3.24\%}  & \multicolumn{1}{c|}{3.45\%}  & 2.57\%  & \multicolumn{1}{c|}{4.82\%}  & \multicolumn{1}{c|}{5.03\%}  & 3.36\%  \\ \cline{2-21} 
                                                                                                              & FNR                               & 4.70\%                                                                                         & \multicolumn{1}{c|}{11.94\%} & \multicolumn{1}{c|}{9.46\%}  & 16.48\% & \multicolumn{1}{c|}{10.45\%} & \multicolumn{1}{c|}{11.87\%} & 8.70\%  & \multicolumn{1}{c|}{6.34\%}  & \multicolumn{1}{c|}{4.82\%}  & 6.30\%  & \multicolumn{1}{c|}{5.04\%}  & \multicolumn{1}{c|}{3.90\%}  & 6.11\%  & \multicolumn{1}{c|}{5.60\%}  & \multicolumn{1}{c|}{3.53\%}  & 5.00\%  & \multicolumn{1}{c|}{6.53\%}  & \multicolumn{1}{c|}{6.68\%}  & 8.15\%  \\ \cline{2-21} 
                                                                                                              & Precision                         & 95.30\%                                                                                        & \multicolumn{1}{c|}{83.99\%} & \multicolumn{1}{c|}{83.28\%} & 86.23\% & \multicolumn{1}{c|}{91.95\%} & \multicolumn{1}{c|}{93.32\%} & 89.47\% & \multicolumn{1}{c|}{94.18\%} & \multicolumn{1}{c|}{92.60\%} & 94.58\% & \multicolumn{1}{c|}{92.04\%} & \multicolumn{1}{c|}{93.17\%} & 93.72\% & \multicolumn{1}{c|}{93.88\%} & \multicolumn{1}{c|}{93.69\%} & 95.18\% & \multicolumn{1}{c|}{91.09\%} & \multicolumn{1}{c|}{90.79\%} & 93.58\% \\ \cline{2-21} 
                                                                                                              & Recall                            & 94.60\%                                                                                        & \multicolumn{1}{c|}{88.06\%} & \multicolumn{1}{c|}{90.54\%} & 83.52\% & \multicolumn{1}{c|}{89.55\%} & \multicolumn{1}{c|}{88.13\%} & 91.30\% & \multicolumn{1}{c|}{93.66\%} & \multicolumn{1}{c|}{95.18\%} & 93.70\% & \multicolumn{1}{c|}{94.96\%} & \multicolumn{1}{c|}{96.10\%} & 93.89\% & \multicolumn{1}{c|}{94.40\%} & \multicolumn{1}{c|}{96.47\%} & 95.00\% & \multicolumn{1}{c|}{93.47\%} & \multicolumn{1}{c|}{93.32\%} & 91.85\% \\ \cline{2-21} 
                                                                                                              & F1-score                          & 95.00\%                                                                                        & \multicolumn{1}{c|}{85.97\%} & \multicolumn{1}{c|}{86.76\%} & 84.85\% & \multicolumn{1}{c|}{90.74\%} & \multicolumn{1}{c|}{90.65\%} & 90.38\% & \multicolumn{1}{c|}{93.92\%} & \multicolumn{1}{c|}{93.87\%} & 94.14\% & \multicolumn{1}{c|}{93.48\%} & \multicolumn{1}{c|}{94.61\%} & 93.80\% & \multicolumn{1}{c|}{94.14\%} & \multicolumn{1}{c|}{95.06\%} & 95.09\% & \multicolumn{1}{c|}{92.27\%} & \multicolumn{1}{c|}{92.04\%} & 92.71\% \\ \hline
\end{tabular}
}
\label{tab:3fold1}
\end{table*}
 
 
\begin{table*}[hbt!]
\vskip5pt
\centering
\caption{Test performance of various models for the binary classification on clean VulDeePecker dataset.}
\resizebox{\textwidth}{!}{%
\begin{tabular}{|c|c|ccc|ccc|ccc|ccc|ccc|ccc|}
\hline
\multirow{2}{*}{\textbf{\begin{tabular}[c]{@{}c@{}}Dataset and\\ Vulnerability\end{tabular}}}                 & \multirow{2}{*}{\textbf{Metrics}} & \multicolumn{3}{c|}{\textbf{RoBERTa}}                                 & \multicolumn{3}{c|}{\textbf{GPT-2 Large}}                             & \multicolumn{3}{c|}{\textbf{GPT-2 XL}}                                & \multicolumn{3}{c|}{\textbf{MegatronBERT}}                            & \multicolumn{3}{c|}{\textbf{MegatronGPT2}}                            & \multicolumn{3}{c|}{\textbf{GPT-J}}                                   \\ \cline{3-20} 
                                                                                                              &                                   & \multicolumn{1}{c|}{Fold 1}  & \multicolumn{1}{c|}{Fold 2}  & Fold 3  & \multicolumn{1}{c|}{Fold 1}  & \multicolumn{1}{c|}{Fold 2}  & Fold 3  & \multicolumn{1}{c|}{Fold 1}  & \multicolumn{1}{c|}{Fold 2}  & Fold 3  & \multicolumn{1}{c|}{Fold 1}  & \multicolumn{1}{c|}{Fold 2}  & Fold 3  & \multicolumn{1}{c|}{Fold 1}  & \multicolumn{1}{c|}{Fold 2}  & Fold 3  & \multicolumn{1}{c|}{Fold 1}  & \multicolumn{1}{c|}{Fold 2}  & Fold 3  \\ \hline
\multirow{5}{*}{\begin{tabular}[c]{@{}c@{}}Group 1, \\ Buffer Error \\ (BE)\end{tabular}}                     & FPR                               & \multicolumn{1}{c|}{3.98\%}  & \multicolumn{1}{c|}{4.41\%}  & 5.05\%  & \multicolumn{1}{c|}{2.72\%}  & \multicolumn{1}{c|}{2.12\%}  & 3.18\%  & \multicolumn{1}{c|}{2.76\%}  & \multicolumn{1}{c|}{2.33\%}  & 2.89\%  & \multicolumn{1}{c|}{3.41\%}  & \multicolumn{1}{c|}{2.83\%}  & 3.50\%  & \multicolumn{1}{c|}{2.97\%}  & \multicolumn{1}{c|}{2.58\%}  & 2.89\%  & \multicolumn{1}{c|}{3.66\%}  & \multicolumn{1}{c|}{2.49\%}  & 2.08\%  \\ \cline{2-20} 
                                                                                                              & FNR                               & \multicolumn{1}{c|}{5.65\%}  & \multicolumn{1}{c|}{8.25\%}  & 5.77\%  & \multicolumn{1}{c|}{4.66\%}  & \multicolumn{1}{c|}{4.63\%}  & 4.85\%  & \multicolumn{1}{c|}{4.99\%}  & \multicolumn{1}{c|}{4.51\%}  & 5.31\%  & \multicolumn{1}{c|}{4.86\%}  & \multicolumn{1}{c|}{4.82\%}  & 6.03\%  & \multicolumn{1}{c|}{5.39\%}  & \multicolumn{1}{c|}{5.08\%}  & 6.36\%  & \multicolumn{1}{c|}{3.61\%}  & \multicolumn{1}{c|}{5.33\%}  & 8.32\%  \\ \cline{2-20} 
                                                                                                              & Precision                         & \multicolumn{1}{c|}{93.61\%} & \multicolumn{1}{c|}{93.17\%} & 92.06\% & \multicolumn{1}{c|}{95.59\%} & \multicolumn{1}{c|}{96.72\%} & 94.90\% & \multicolumn{1}{c|}{95.51\%} & \multicolumn{1}{c|}{96.41\%} & 95.32\% & \multicolumn{1}{c|}{94.52\%} & \multicolumn{1}{c|}{95.66\%} & 94.34\% & \multicolumn{1}{c|}{95.18\%} & \multicolumn{1}{c|}{96.02\%} & 95.27\% & \multicolumn{1}{c|}{94.22\%} & \multicolumn{1}{c|}{96.13\%} & 96.48\% \\ \cline{2-20} 
                                                                                                              & Recall                            & \multicolumn{1}{c|}{94.35\%} & \multicolumn{1}{c|}{91.75\%} & 94.23\% & \multicolumn{1}{c|}{95.34\%} & \multicolumn{1}{c|}{95.37\%} & 95.15\% & \multicolumn{1}{c|}{95.01\%} & \multicolumn{1}{c|}{95.49\%} & 94.69\% & \multicolumn{1}{c|}{95.14\%} & \multicolumn{1}{c|}{95.18\%} & 93.97\% & \multicolumn{1}{c|}{94.61\%} & \multicolumn{1}{c|}{94.92\%} & 93.64\% & \multicolumn{1}{c|}{96.39\%} & \multicolumn{1}{c|}{94.67\%} & 91.68\% \\ \cline{2-20} 
                                                                                                              & F1-score                          & \multicolumn{1}{c|}{93.98\%} & \multicolumn{1}{c|}{92.46\%} & 93.13\% & \multicolumn{1}{c|}{95.46\%} & \multicolumn{1}{c|}{96.04\%} & 95.03\% & \multicolumn{1}{c|}{95.26\%} & \multicolumn{1}{c|}{95.95\%} & 95.00\% & \multicolumn{1}{c|}{94.83\%} & \multicolumn{1}{c|}{95.42\%} & 94.16\% & \multicolumn{1}{c|}{94.89\%} & \multicolumn{1}{c|}{95.47\%} & 94.45\% & \multicolumn{1}{c|}{95.29\%} & \multicolumn{1}{c|}{95.40\%} & 94.02\% \\ \hline
\multirow{5}{*}{\begin{tabular}[c]{@{}c@{}}Group 2, \\ Resource \\ Management \\ Error \\ (RME)\end{tabular}} & FPR                               & \multicolumn{1}{c|}{3.34\%}  & \multicolumn{1}{c|}{3.16\%}  & 2.27\%  & \multicolumn{1}{c|}{2.16\%}  & \multicolumn{1}{c|}{1.18\%}  & 1.78\%  & \multicolumn{1}{c|}{2.16\%}  & \multicolumn{1}{c|}{1.58\%}  & 1.58\%  & \multicolumn{1}{c|}{3.15\%}  & \multicolumn{1}{c|}{2.27\%}  & 1.78\%  & \multicolumn{1}{c|}{2.75\%}  & \multicolumn{1}{c|}{1.97\%}  & 2.76\%  & \multicolumn{1}{c|}{2.16\%}  & \multicolumn{1}{c|}{1.08\%}  & 3.26\%  \\ \cline{2-20} 
                                                                                                              & FNR                               & \multicolumn{1}{c|}{5.60\%}  & \multicolumn{1}{c|}{4.45\%}  & 5.56\%  & \multicolumn{1}{c|}{3.36\%}  & \multicolumn{1}{c|}{2.78\%}  & 3.15\%  & \multicolumn{1}{c|}{2.80\%}  & \multicolumn{1}{c|}{3.15\%}  & 3.89\%  & \multicolumn{1}{c|}{2.80\%}  & \multicolumn{1}{c|}{4.08\%}  & 3.70\%  & \multicolumn{1}{c|}{2.43\%}  & \multicolumn{1}{c|}{3.34\%}  & 3.33\%  & \multicolumn{1}{c|}{3.73\%}  & \multicolumn{1}{c|}{5.19\%}  & 2.96\%  \\ \cline{2-20} 
                                                                                                              & Precision                         & \multicolumn{1}{c|}{93.70\%} & \multicolumn{1}{c|}{94.15\%} & 95.68\% & \multicolumn{1}{c|}{95.93\%} & \multicolumn{1}{c|}{97.76\%} & 96.67\% & \multicolumn{1}{c|}{95.95\%} & \multicolumn{1}{c|}{97.03\%} & 97.01\% & \multicolumn{1}{c|}{94.21\%} & \multicolumn{1}{c|}{95.74\%} & 96.65\% & \multicolumn{1}{c|}{94.92\%} & \multicolumn{1}{c|}{96.30\%} & 94.91\% & \multicolumn{1}{c|}{95.91\%} & \multicolumn{1}{c|}{97.89\%} & 94.08\% \\ \cline{2-20} 
                                                                                                              & Recall                            & \multicolumn{1}{c|}{94.40\%} & \multicolumn{1}{c|}{95.55\%} & 94.44\% & \multicolumn{1}{c|}{96.64\%} & \multicolumn{1}{c|}{97.22\%} & 96.85\% & \multicolumn{1}{c|}{97.20\%} & \multicolumn{1}{c|}{96.85\%} & 96.11\% & \multicolumn{1}{c|}{97.20\%} & \multicolumn{1}{c|}{95.92\%} & 96.30\% & \multicolumn{1}{c|}{97.57\%} & \multicolumn{1}{c|}{96.66\%} & 96.67\% & \multicolumn{1}{c|}{96.27\%} & \multicolumn{1}{c|}{94.81\%} & 97.04\% \\ \cline{2-20} 
                                                                                                              & F1-score                          & \multicolumn{1}{c|}{94.05\%} & \multicolumn{1}{c|}{94.84\%} & 95.06\% & \multicolumn{1}{c|}{96.28\%} & \multicolumn{1}{c|}{97.49\%} & 96.76\% & \multicolumn{1}{c|}{96.57\%} & \multicolumn{1}{c|}{96.94\%} & 96.56\% & \multicolumn{1}{c|}{95.68\%} & \multicolumn{1}{c|}{95.83\%} & 96.47\% & \multicolumn{1}{c|}{96.23\%} & \multicolumn{1}{c|}{96.48\%} & 95.78\% & \multicolumn{1}{c|}{96.09\%} & \multicolumn{1}{c|}{96.32\%} & 95.53\% \\ \hline
\end{tabular}
}
\label{tab:3fold2}
\end{table*}

\begin{table*}[hbt!]
\vskip5pt
\centering
\caption{Test performance of various models for the multi-class classification on clean VulDeePecker dataset.}
\resizebox{\textwidth}{!}{%
\begin{tabular}{|c|c|c|ccc|ccc|ccc|ccc|ccc|ccc|}
\hline
\multirow{2}{*}{\textbf{\begin{tabular}[c]{@{}c@{}}Dataset and \\ Vulnerability\end{tabular}}}               & \multirow{2}{*}{\textbf{Metrics}} & \multirow{2}{*}{\textbf{\begin{tabular}[c]{@{}c@{}}VulDeePecker\\      Original\end{tabular}}} & \multicolumn{3}{c|}{\textbf{BiLSTM}}                                  & \multicolumn{3}{c|}{\textbf{BiGRU}}                                   & \multicolumn{3}{c|}{\textbf{BERTBase}}                                & \multicolumn{3}{c|}{\textbf{GPT-2 Base}}                              & \multicolumn{3}{c|}{\textbf{CodeBERT}}                                & \multicolumn{3}{c|}{\textbf{DistilBERT}}                              \\ \cline{4-21} 
                                                                                                             &                                   &                                                                                                & \multicolumn{1}{c|}{Fold1}   & \multicolumn{1}{c|}{Fold 2}  & Fold 3  & \multicolumn{1}{c|}{Fold1}   & \multicolumn{1}{c|}{Fold 2}  & Fold 3  & \multicolumn{1}{c|}{Fold1}   & \multicolumn{1}{c|}{Fold 2}  & Fold 3  & \multicolumn{1}{c|}{Fold1}   & \multicolumn{1}{c|}{Fold 2}  & Fold 3  & \multicolumn{1}{c|}{Fold1}   & \multicolumn{1}{c|}{Fold 2}  & Fold 3  & \multicolumn{1}{c|}{Fold1}   & \multicolumn{1}{c|}{Fold 2}  & Fold 3  \\ \hline
\multirow{5}{*}{\begin{tabular}[c]{@{}c@{}}Group 3, \\ Buffer Error \\ (BE)\end{tabular}}                    & FPR                               & 2.90\%                                                                                         & \multicolumn{1}{c|}{21.19\%} & \multicolumn{1}{c|}{20.80\%} & 21.89\% & \multicolumn{1}{c|}{11.96\%} & \multicolumn{1}{c|}{4.97\%}  & 4.16\%  & \multicolumn{1}{c|}{2.23\%}  & \multicolumn{1}{c|}{2.59\%}  & 2.30\%  & \multicolumn{1}{c|}{2.58\%}  & \multicolumn{1}{c|}{3.30\%}  & 2.98\%  & \multicolumn{1}{c|}{1.65\%}  & \multicolumn{1}{c|}{2.08\%}  & 2.00\%  & \multicolumn{1}{c|}{2.00\%}  & \multicolumn{1}{c|}{2.69\%}  & 2.58\%  \\ \cline{2-21} 
                                                                                                             & FNR                               & 18.00\%                                                                                        & \multicolumn{1}{c|}{12.89\%} & \multicolumn{1}{c|}{15.49\%} & 13.45\% & \multicolumn{1}{c|}{26.18\%} & \multicolumn{1}{c|}{43.56\%} & 46.19\% & \multicolumn{1}{c|}{5.20\%}  & \multicolumn{1}{c|}{4.83\%}  & 4.53\%  & \multicolumn{1}{c|}{5.39\%}  & \multicolumn{1}{c|}{5.65\%}  & 5.18\%  & \multicolumn{1}{c|}{5.46\%}  & \multicolumn{1}{c|}{5.02\%}  & 4.00\%  & \multicolumn{1}{c|}{6.97\%}  & \multicolumn{1}{c|}{5.33\%}  & 5.18\%  \\ \cline{2-21} 
                                                                                                             & Precision                         & 82.00\%                                                                                        & \multicolumn{1}{c|}{60.99\%} & \multicolumn{1}{c|}{61.88\%} & 60.15\% & \multicolumn{1}{c|}{70.13\%} & \multicolumn{1}{c|}{81.94\%} & 83.16\% & \multicolumn{1}{c|}{94.18\%} & \multicolumn{1}{c|}{93.63\%} & 94.05\% & \multicolumn{1}{c|}{93.32\%} & \multicolumn{1}{c|}{91.96\%} & 92.39\% & \multicolumn{1}{c|}{95.61\%} & \multicolumn{1}{c|}{94.80\%} & 94.82\% & \multicolumn{1}{c|}{94.65\%} & \multicolumn{1}{c|}{93.36\%} & 93.35\% \\ \cline{2-21} 
                                                                                                             & Recall                            & 91.70\%                                                                                        & \multicolumn{1}{c|}{87.11\%} & \multicolumn{1}{c|}{84.51\%} & 86.55\% & \multicolumn{1}{c|}{73.82\%} & \multicolumn{1}{c|}{56.44\%} & 53.81\% & \multicolumn{1}{c|}{94.80\%} & \multicolumn{1}{c|}{95.17\%} & 95.47\% & \multicolumn{1}{c|}{94.61\%} & \multicolumn{1}{c|}{94.35\%} & 94.82\% & \multicolumn{1}{c|}{94.54\%} & \multicolumn{1}{c|}{94.98\%} & 96.00\% & \multicolumn{1}{c|}{93.03\%} & \multicolumn{1}{c|}{94.67\%} & 94.82\% \\ \cline{2-21} 
                                                                                                             & F1-score                          & 86.60\%                                                                                        & \multicolumn{1}{c|}{71.74\%} & \multicolumn{1}{c|}{71.44\%} & 70.97\% & \multicolumn{1}{c|}{71.92\%} & \multicolumn{1}{c|}{66.84\%} & 65.34\% & \multicolumn{1}{c|}{94.49\%} & \multicolumn{1}{c|}{94.40\%} & 94.76\% & \multicolumn{1}{c|}{93.96\%} & \multicolumn{1}{c|}{93.14\%} & 93.59\% & \multicolumn{1}{c|}{95.07\%} & \multicolumn{1}{c|}{94.89\%} & 95.40\% & \multicolumn{1}{c|}{93.83\%} & \multicolumn{1}{c|}{94.01\%} & 94.08\% \\ \hline
\multirow{5}{*}{\begin{tabular}[c]{@{}c@{}}Group 3,\\ Resource \\ Management \\ Error \\ (RME)\end{tabular}} & FPR                               & 2.80\%                                                                                         & \multicolumn{1}{c|}{3.20\%}  & \multicolumn{1}{c|}{3.18\%}  & 3.84\%  & \multicolumn{1}{c|}{0.80\%}  & \multicolumn{1}{c|}{0.48\%}  & 0.69\%  & \multicolumn{1}{c|}{0.78\%}  & \multicolumn{1}{c|}{0.56\%}  & 0.65\%  & \multicolumn{1}{c|}{1.02\%}  & \multicolumn{1}{c|}{0.97\%}  & 1.07\%  & \multicolumn{1}{c|}{0.62\%}  & \multicolumn{1}{c|}{0.64\%}  & 0.65\%  & \multicolumn{1}{c|}{0.72\%}  & \multicolumn{1}{c|}{0.68\%}  & 0.79\%  \\ \cline{2-21} 
                                                                                                             & FNR                               & 4.70\%                                                                                         & \multicolumn{1}{c|}{11.65\%} & \multicolumn{1}{c|}{12.59\%} & 7.82\%  & \multicolumn{1}{c|}{5.55\%}  & \multicolumn{1}{c|}{8.21\%}  & 8.70\%  & \multicolumn{1}{c|}{4.25\%}  & \multicolumn{1}{c|}{4.93\%}  & 3.02\%  & \multicolumn{1}{c|}{5.18\%}  & \multicolumn{1}{c|}{5.47\%}  & 3.37\%  & \multicolumn{1}{c|}{4.07\%}  & \multicolumn{1}{c|}{4.74\%}  & 3.55\%  & \multicolumn{1}{c|}{4.44\%}  & \multicolumn{1}{c|}{5.84\%}  & 3.37\%  \\ \cline{2-21} 
                                                                                                             & Precision                         & 95.30\%                                                                                        & \multicolumn{1}{c|}{75.04\%} & \multicolumn{1}{c|}{75.20\%} & 73.20\% & \multicolumn{1}{c|}{92.74\%} & \multicolumn{1}{c|}{95.45\%} & 93.80\% & \multicolumn{1}{c|}{93.00\%} & \multicolumn{1}{c|}{94.90\%} & 94.46\% & \multicolumn{1}{c|}{90.96\%} & \multicolumn{1}{c|}{91.52\%} & 91.12\% & \multicolumn{1}{c|}{94.36\%} & \multicolumn{1}{c|}{94.22\%} & 94.43\% & \multicolumn{1}{c|}{93.49\%} & \multicolumn{1}{c|}{93.82\%} & 93.31\% \\ \cline{2-21} 
                                                                                                             & Recall                            & 94.60\%                                                                                        & \multicolumn{1}{c|}{88.35\%} & \multicolumn{1}{c|}{87.41\%} & 92.18\% & \multicolumn{1}{c|}{94.45\%} & \multicolumn{1}{c|}{91.79\%} & 91.30\% & \multicolumn{1}{c|}{95.75\%} & \multicolumn{1}{c|}{95.07\%} & 96.98\% & \multicolumn{1}{c|}{94.82\%} & \multicolumn{1}{c|}{94.53\%} & 96.63\% & \multicolumn{1}{c|}{95.93\%} & \multicolumn{1}{c|}{95.26\%} & 96.45\% & \multicolumn{1}{c|}{95.56\%} & \multicolumn{1}{c|}{94.16\%} & 96.63\% \\ \cline{2-21} 
                                                                                                             & F1-score                          & 95.00\%                                                                                        & \multicolumn{1}{c|}{81.15\%} & \multicolumn{1}{c|}{80.84\%} & 81.60\% & \multicolumn{1}{c|}{93.59\%} & \multicolumn{1}{c|}{93.58\%} & 92.53\% & \multicolumn{1}{c|}{94.35\%} & \multicolumn{1}{c|}{94.99\%} & 95.71\% & \multicolumn{1}{c|}{92.85\%} & \multicolumn{1}{c|}{93.00\%} & 93.79\% & \multicolumn{1}{c|}{95.14\%} & \multicolumn{1}{c|}{94.74\%} & 95.43\% & \multicolumn{1}{c|}{94.52\%} & \multicolumn{1}{c|}{93.99\%} & 94.94\% \\ \hline
\multirow{3}{*}{\begin{tabular}[c]{@{}c@{}}Group 3, \\ BE + RME \\ (Global Avg.)\end{tabular}}               & Precision                         & N/A                                                                                            & \multicolumn{1}{c|}{64.17\%} & \multicolumn{1}{c|}{64.92\%} & 63.34\% & \multicolumn{1}{c|}{75.92\%} & \multicolumn{1}{c|}{86.35\%} & 86.96\% & \multicolumn{1}{c|}{93.87\%} & \multicolumn{1}{c|}{93.95\%} & 94.16\% & \multicolumn{1}{c|}{92.68\%} & \multicolumn{1}{c|}{91.84\%} & 92.04\% & \multicolumn{1}{c|}{95.28\%} & \multicolumn{1}{c|}{94.65\%} & 94.71\% & \multicolumn{1}{c|}{94.33\%} & \multicolumn{1}{c|}{93.48\%} & 93.34\% \\ \cline{2-21} 
                                                                                                             & Recall                            & N/A                                                                                            & \multicolumn{1}{c|}{87.43\%} & \multicolumn{1}{c|}{85.26\%} & 88.07\% & \multicolumn{1}{c|}{79.23\%} & \multicolumn{1}{c|}{65.57\%} & 63.92\% & \multicolumn{1}{c|}{95.05\%} & \multicolumn{1}{c|}{95.15\%} & 95.88\% & \multicolumn{1}{c|}{94.66\%} & \multicolumn{1}{c|}{94.39\%} & 95.30\% & \multicolumn{1}{c|}{94.91\%} & \multicolumn{1}{c|}{95.05\%} & 96.12\% & \multicolumn{1}{c|}{93.69\%} & \multicolumn{1}{c|}{94.54\%} & 95.30\% \\ \cline{2-21} 
                                                                                                             & F1-score                          & N/A                                                                                            & \multicolumn{1}{c|}{74.02\%} & \multicolumn{1}{c|}{73.71\%} & 73.68\% & \multicolumn{1}{c|}{77.54\%} & \multicolumn{1}{c|}{74.54\%} & 73.68\% & \multicolumn{1}{c|}{94.46\%} & \multicolumn{1}{c|}{94.55\%} & 95.01\% & \multicolumn{1}{c|}{93.66\%} & \multicolumn{1}{c|}{93.10\%} & 93.64\% & \multicolumn{1}{c|}{95.09\%} & \multicolumn{1}{c|}{94.85\%} & 95.41\% & \multicolumn{1}{c|}{94.01\%} & \multicolumn{1}{c|}{94.00\%} & 94.31\% \\ \hline
\multirow{3}{*}{\begin{tabular}[c]{@{}c@{}}Group 3, \\ BE + RME \\ (Macro Avg.)\end{tabular}}                & Precision                         & 88.65\%                                                                                        & \multicolumn{1}{c|}{68.01\%} & \multicolumn{1}{c|}{68.54\%} & 66.67\% & \multicolumn{1}{c|}{81.43\%} & \multicolumn{1}{c|}{88.69\%} & 88.48\% & \multicolumn{1}{c|}{93.59\%} & \multicolumn{1}{c|}{94.26\%} & 94.26\% & \multicolumn{1}{c|}{92.14\%} & \multicolumn{1}{c|}{91.74\%} & 91.76\% & \multicolumn{1}{c|}{94.99\%} & \multicolumn{1}{c|}{94.51\%} & 94.63\% & \multicolumn{1}{c|}{94.07\%} & \multicolumn{1}{c|}{93.59\%} & 93.33\% \\ \cline{2-21} 
                                                                                                             & Recall                            & 93.15\%                                                                                        & \multicolumn{1}{c|}{87.73\%} & \multicolumn{1}{c|}{85.96\%} & 89.37\% & \multicolumn{1}{c|}{84.14\%} & \multicolumn{1}{c|}{74.12\%} & 72.55\% & \multicolumn{1}{c|}{95.28\%} & \multicolumn{1}{c|}{95.12\%} & 96.23\% & \multicolumn{1}{c|}{94.71\%} & \multicolumn{1}{c|}{94.44\%} & 95.72\% & \multicolumn{1}{c|}{95.24\%} & \multicolumn{1}{c|}{95.12\%} & 96.22\% & \multicolumn{1}{c|}{94.30\%} & \multicolumn{1}{c|}{94.41\%} & 95.72\% \\ \cline{2-21} 
                                                                                                             & F1-score                          & 90.80\%                                                                                        & \multicolumn{1}{c|}{76.62\%} & \multicolumn{1}{c|}{76.27\%} & 76.37\% & \multicolumn{1}{c|}{82.76\%} & \multicolumn{1}{c|}{80.75\%} & 79.73\% & \multicolumn{1}{c|}{94.43\%} & \multicolumn{1}{c|}{94.69\%} & 95.23\% & \multicolumn{1}{c|}{93.41\%} & \multicolumn{1}{c|}{93.07\%} & 93.70\% & \multicolumn{1}{c|}{95.11\%} & \multicolumn{1}{c|}{94.82\%} & 95.42\% & \multicolumn{1}{c|}{94.18\%} & \multicolumn{1}{c|}{94.00\%} & 94.51\% \\ \hline
\end{tabular}
}
\label{tab:3fold3}
\end{table*}

\begin{table*}[hbt!]
\vskip5pt
\centering
\caption{Test performance of various models for the multi-class classification on clean VulDeePecker dataset.}
\resizebox{\textwidth}{!}{%
\begin{tabular}{|c|c|ccc|ccc|ccc|ccc|ccc|ccc|}
\hline
\multirow{2}{*}{\textbf{\begin{tabular}[c]{@{}c@{}}Dataset and\\ Vulnerability\end{tabular}}}              & \multirow{2}{*}{\textbf{Metrics}} & \multicolumn{3}{c|}{\textbf{RoBERTa}}                                 & \multicolumn{3}{c|}{\textbf{GPT-2 Large}}                             & \multicolumn{3}{c|}{\textbf{GPT-2 XL}}                                & \multicolumn{3}{c|}{\textbf{MegatronBERT}}                            & \multicolumn{3}{c|}{\textbf{MegatronGPT-2}}                           & \multicolumn{3}{c|}{\textbf{GPT-J}}                                   \\ \cline{3-20} 
                                                                                                           &                                   & \multicolumn{1}{c|}{Fold 1}  & \multicolumn{1}{c|}{Fold 2}  & Fold 3  & \multicolumn{1}{c|}{Fold 1}  & \multicolumn{1}{c|}{Fold 2}  & Fold 3  & \multicolumn{1}{c|}{Fold 1}  & \multicolumn{1}{c|}{Fold 2}  & Fold 3  & \multicolumn{1}{c|}{Fold 1}  & \multicolumn{1}{c|}{Fold 2}  & Fold 3  & \multicolumn{1}{c|}{Fold 1}  & \multicolumn{1}{c|}{Fold 2}  & Fold 3  & \multicolumn{1}{c|}{Fold 1}  & \multicolumn{1}{c|}{Fold 2}  & Fold 3  \\ \hline
\multirow{5}{*}{\begin{tabular}[c]{@{}c@{}}Group 3, \\ Buffer Error\\ (BE)\end{tabular}}                   & FPR                               & \multicolumn{1}{c|}{2.15\%}  & \multicolumn{1}{c|}{2.54\%}  & 2.53\%  & \multicolumn{1}{c|}{1.50\%}  & \multicolumn{1}{c|}{1.78\%}  & 1.53\%  & \multicolumn{1}{c|}{1.50\%}  & \multicolumn{1}{c|}{1.50\%}  & 1.40\%  & \multicolumn{1}{c|}{1.90\%}  & \multicolumn{1}{c|}{1.78\%}  & 1.80\%  & \multicolumn{1}{c|}{1.80\%}  & \multicolumn{1}{c|}{2.31\%}  & 1.98\%  & \multicolumn{1}{c|}{1.60\%}  & \multicolumn{1}{c|}{1.47\%}  & 1.25\%  \\ \cline{2-20} 
                                                                                                           & FNR                               & \multicolumn{1}{c|}{5.66\%}  & \multicolumn{1}{c|}{6.54\%}  & 4.86\%  & \multicolumn{1}{c|}{5.13\%}  & \multicolumn{1}{c|}{4.38\%}  & 5.38\%  & \multicolumn{1}{c|}{4.80\%}  & \multicolumn{1}{c|}{4.57\%}  & 4.92\%  & \multicolumn{1}{c|}{5.20\%}  & \multicolumn{1}{c|}{4.76\%}  & 3.87\%  & \multicolumn{1}{c|}{5.33\%}  & \multicolumn{1}{c|}{5.52\%}  & 4.40\%  & \multicolumn{1}{c|}{5.00\%}  & \multicolumn{1}{c|}{4.83\%}  & 5.84\%  \\ \cline{2-20} 
                                                                                                           & Precision                         & \multicolumn{1}{c|}{94.34\%} & \multicolumn{1}{c|}{93.64\%} & 93.49\% & \multicolumn{1}{c|}{96.01\%} & \multicolumn{1}{c|}{95.56\%} & 95.94\% & \multicolumn{1}{c|}{96.02\%} & \multicolumn{1}{c|}{96.22\%} & 96.28\% & \multicolumn{1}{c|}{94.99\%} & \multicolumn{1}{c|}{95.54\%} & 95.32\% & \multicolumn{1}{c|}{95.23\%} & \multicolumn{1}{c|}{94.24\%} & 94.86\% & \multicolumn{1}{c|}{95.76\%} & \multicolumn{1}{c|}{96.27\%} & 96.63\% \\ \cline{2-20} 
                                                                                                           & Recall                            & \multicolumn{1}{c|}{94.34\%} & \multicolumn{1}{c|}{93.46\%} & 95.14\% & \multicolumn{1}{c|}{94.87\%} & \multicolumn{1}{c|}{95.62\%} & 94.62\% & \multicolumn{1}{c|}{95.20\%} & \multicolumn{1}{c|}{95.43\%} & 95.08\% & \multicolumn{1}{c|}{94.80\%} & \multicolumn{1}{c|}{95.24\%} & 96.13\% & \multicolumn{1}{c|}{94.67\%} & \multicolumn{1}{c|}{94.48\%} & 95.60\% & \multicolumn{1}{c|}{95.00\%} & \multicolumn{1}{c|}{95.17\%} & 94.16\% \\ \cline{2-20} 
                                                                                                           & F1-score                          & \multicolumn{1}{c|}{94.34\%} & \multicolumn{1}{c|}{93.55\%} & 94.31\% & \multicolumn{1}{c|}{95.43\%} & \multicolumn{1}{c|}{95.59\%} & 95.28\% & \multicolumn{1}{c|}{95.61\%} & \multicolumn{1}{c|}{95.82\%} & 95.68\% & \multicolumn{1}{c|}{94.90\%} & \multicolumn{1}{c|}{95.39\%} & 95.72\% & \multicolumn{1}{c|}{94.95\%} & \multicolumn{1}{c|}{94.36\%} & 95.23\% & \multicolumn{1}{c|}{95.38\%} & \multicolumn{1}{c|}{95.72\%} & 95.38\% \\ \hline
\multirow{5}{*}{\begin{tabular}[c]{@{}c@{}}Group 3, \\ Resource\\ Management\\ Error\\ (RME)\end{tabular}} & FPR                               & \multicolumn{1}{c|}{0.88\%}  & \multicolumn{1}{c|}{0.62\%}  & 0.75\%  & \multicolumn{1}{c|}{0.40\%}  & \multicolumn{1}{c|}{0.36\%}  & 0.50\%  & \multicolumn{1}{c|}{0.44\%}  & \multicolumn{1}{c|}{0.38\%}  & 0.42\%  & \multicolumn{1}{c|}{0.46\%}  & \multicolumn{1}{c|}{0.42\%}  & 0.63\%  & \multicolumn{1}{c|}{0.58\%}  & \multicolumn{1}{c|}{0.46\%}  & 0.65\%  & \multicolumn{1}{c|}{0.22\%}  & \multicolumn{1}{c|}{0.22\%}  & 0.30\%  \\ \cline{2-20} 
                                                                                                           & FNR                               & \multicolumn{1}{c|}{4.99\%}  & \multicolumn{1}{c|}{4.01\%}  & 3.20\%  & \multicolumn{1}{c|}{3.70\%}  & \multicolumn{1}{c|}{2.55\%}  & 2.31\%  & \multicolumn{1}{c|}{3.70\%}  & \multicolumn{1}{c|}{3.28\%}  & 2.31\%  & \multicolumn{1}{c|}{4.25\%}  & \multicolumn{1}{c|}{4.01\%}  & 2.66\%  & \multicolumn{1}{c|}{2.77\%}  & \multicolumn{1}{c|}{4.01\%}  & 3.02\%  & \multicolumn{1}{c|}{5.18\%}  & \multicolumn{1}{c|}{7.30\%}  & 5.15\%  \\ \cline{2-20} 
                                                                                                           & Precision                         & \multicolumn{1}{c|}{92.11\%} & \multicolumn{1}{c|}{94.43\%} & 93.64\% & \multicolumn{1}{c|}{96.30\%} & \multicolumn{1}{c|}{96.74\%} & 95.65\% & \multicolumn{1}{c|}{95.95\%} & \multicolumn{1}{c|}{96.54\%} & 96.32\% & \multicolumn{1}{c|}{95.75\%} & \multicolumn{1}{c|}{96.16\%} & 94.65\% & \multicolumn{1}{c|}{94.77\%} & \multicolumn{1}{c|}{95.81\%} & 94.46\% & \multicolumn{1}{c|}{97.90\%} & \multicolumn{1}{c|}{97.88\%} & 97.27\% \\ \cline{2-20} 
                                                                                                           & Recall                            & \multicolumn{1}{c|}{95.01\%} & \multicolumn{1}{c|}{95.99\%} & 96.80\% & \multicolumn{1}{c|}{96.30\%} & \multicolumn{1}{c|}{97.45\%} & 97.69\% & \multicolumn{1}{c|}{96.30\%} & \multicolumn{1}{c|}{96.72\%} & 97.69\% & \multicolumn{1}{c|}{95.75\%} & \multicolumn{1}{c|}{95.99\%} & 97.34\% & \multicolumn{1}{c|}{97.23\%} & \multicolumn{1}{c|}{95.99\%} & 96.98\% & \multicolumn{1}{c|}{94.82\%} & \multicolumn{1}{c|}{92.70\%} & 94.85\% \\ \cline{2-20} 
                                                                                                           & F1-score                          & \multicolumn{1}{c|}{93.54\%} & \multicolumn{1}{c|}{95.20\%} & 95.20\% & \multicolumn{1}{c|}{96.30\%} & \multicolumn{1}{c|}{97.09\%} & 96.66\% & \multicolumn{1}{c|}{96.13\%} & \multicolumn{1}{c|}{96.63\%} & 97.00\% & \multicolumn{1}{c|}{95.75\%} & \multicolumn{1}{c|}{96.07\%} & 95.97\% & \multicolumn{1}{c|}{95.99\%} & \multicolumn{1}{c|}{95.90\%} & 95.71\% & \multicolumn{1}{c|}{96.34\%} & \multicolumn{1}{c|}{95.22\%} & 96.04\% \\ \hline
\multirow{3}{*}{\begin{tabular}[c]{@{}c@{}}Group 3, \\ BE + RME\\ (Global Avg.)\end{tabular}}              & Precision                         & \multicolumn{1}{c|}{93.74\%} & \multicolumn{1}{c|}{93.85\%} & 93.53\% & \multicolumn{1}{c|}{96.08\%} & \multicolumn{1}{c|}{95.86\%} & 95.86\% & \multicolumn{1}{c|}{96.00\%} & \multicolumn{1}{c|}{96.31\%} & 96.29\% & \multicolumn{1}{c|}{95.19\%} & \multicolumn{1}{c|}{95.70\%} & 95.13\% & \multicolumn{1}{c|}{95.11\%} & \multicolumn{1}{c|}{94.64\%} & 94.75\% & \multicolumn{1}{c|}{96.31\%} & \multicolumn{1}{c|}{96.68\%} & 96.80\% \\ \cline{2-20} 
                                                                                                           & Recall                            & \multicolumn{1}{c|}{94.52\%} & \multicolumn{1}{c|}{94.11\%} & 95.59\% & \multicolumn{1}{c|}{95.25\%} & \multicolumn{1}{c|}{96.09\%} & 95.45\% & \multicolumn{1}{c|}{95.49\%} & \multicolumn{1}{c|}{95.76\%} & 95.78\% & \multicolumn{1}{c|}{95.05\%} & \multicolumn{1}{c|}{95.43\%} & 96.45\% & \multicolumn{1}{c|}{95.34\%} & \multicolumn{1}{c|}{94.87\%} & 95.98\% & \multicolumn{1}{c|}{94.95\%} & \multicolumn{1}{c|}{94.54\%} & 94.35\% \\ \cline{2-20} 
                                                                                                           & F1-score                          & \multicolumn{1}{c|}{94.13\%} & \multicolumn{1}{c|}{93.98\%} & 94.55\% & \multicolumn{1}{c|}{95.66\%} & \multicolumn{1}{c|}{95.98\%} & 95.65\% & \multicolumn{1}{c|}{95.74\%} & \multicolumn{1}{c|}{96.03\%} & 96.04\% & \multicolumn{1}{c|}{95.12\%} & \multicolumn{1}{c|}{95.57\%} & 95.79\% & \multicolumn{1}{c|}{95.23\%} & \multicolumn{1}{c|}{94.75\%} & 95.36\% & \multicolumn{1}{c|}{95.63\%} & \multicolumn{1}{c|}{95.59\%} & 95.56\% \\ \hline
\multirow{3}{*}{\begin{tabular}[c]{@{}c@{}}Group 3, \\ BE + RME\\ (Macro Avg.)\end{tabular}}               & Precision                         & \multicolumn{1}{c|}{93.23\%} & \multicolumn{1}{c|}{94.04\%} & 93.57\% & \multicolumn{1}{c|}{96.15\%} & \multicolumn{1}{c|}{96.15\%} & 95.80\% & \multicolumn{1}{c|}{95.98\%} & \multicolumn{1}{c|}{96.38\%} & 96.30\% & \multicolumn{1}{c|}{95.37\%} & \multicolumn{1}{c|}{95.85\%} & 94.98\% & \multicolumn{1}{c|}{95.00\%} & \multicolumn{1}{c|}{95.02\%} & 94.66\% & \multicolumn{1}{c|}{96.83\%} & \multicolumn{1}{c|}{97.08\%} & 96.95\% \\ \cline{2-20} 
                                                                                                           & Recall                            & \multicolumn{1}{c|}{94.68\%} & \multicolumn{1}{c|}{94.72\%} & 95.97\% & \multicolumn{1}{c|}{95.59\%} & \multicolumn{1}{c|}{96.53\%} & 96.16\% & \multicolumn{1}{c|}{95.75\%} & \multicolumn{1}{c|}{96.07\%} & 96.38\% & \multicolumn{1}{c|}{95.28\%} & \multicolumn{1}{c|}{95.61\%} & 96.73\% & \multicolumn{1}{c|}{95.95\%} & \multicolumn{1}{c|}{95.23\%} & 96.29\% & \multicolumn{1}{c|}{94.91\%} & \multicolumn{1}{c|}{93.94\%} & 94.50\% \\ \cline{2-20} 
                                                                                                           & F1-score                          & \multicolumn{1}{c|}{93.95\%} & \multicolumn{1}{c|}{94.38\%} & 94.75\% & \multicolumn{1}{c|}{95.87\%} & \multicolumn{1}{c|}{96.34\%} & 95.98\% & \multicolumn{1}{c|}{95.87\%} & \multicolumn{1}{c|}{96.23\%} & 96.34\% & \multicolumn{1}{c|}{95.32\%} & \multicolumn{1}{c|}{95.73\%} & 95.85\% & \multicolumn{1}{c|}{95.47\%} & \multicolumn{1}{c|}{95.13\%} & 95.47\% & \multicolumn{1}{c|}{95.86\%} & \multicolumn{1}{c|}{95.48\%} & 95.71\% \\ \hline
\end{tabular}
}
\label{tab:3fold4}
\end{table*}

\section{Performance metrics}
\label{appendix:performance_metrics}
Before defining the evaluation metrics presented in this paper, we define the following terms, where the positive class samples refer to vulnerable code gadgets and the negative class samples refer to non-vulnerable code gadgets.

\begin{itemize}
    \item True Positive (TP) is the number of the positive class samples that are correctly classified.
    \item False Positive (FP) is the number of the negative class samples that are misclassified as the positive class.
    \item True Negative (TN) is the number of the negative class samples that are correctly classified.
    \item False Negative (FN) is the number of the positive class samples that are misclassified as the negative class.
\end{itemize}

\noindent
\textbf{False Positive Rate}: False Positive Rate (FPR) specifies the proportion of the negative class (i.e., non-vulnerable code gadgets) misclassified as a positive class (i.e., vulnerable code gadgets) and calculated as
\begin{equation*}
    \text{FPR} = \frac{\text{FP}}{\text{FP}+ \text{TN}}.
\end{equation*}

\noindent
\textbf{False Negative Rate}: False Negative Rate (FNR) specifies the proportion of the positive class (i.e., vulnerable code gadgets) misclassified as a negative class (i.e., non-vulnerable code gadgets) and calculated as
\begin{equation*}
    \text{FNR} = \frac{\text{FN}}{\text{FN}+ \text{TP}}.
\end{equation*}

\noindent
\textbf{Precision}: Precision specifies the classifier's resistance to misclassifying negative class samples to positive class and calculated as
\begin{equation*}
    \text{Precision} = \frac{\text{TP}}{\text{TP}+ \text{FP}}.
\end{equation*}

\noindent
\textbf{Recall}: Recall specifies the classifier's ability to correctly classify a positive class, and is calculated as
\begin{equation*}
    \text{Recall} = \frac{\text{TP}}{\text{TP}+ \text{FN}}.
\end{equation*}

\noindent
\textbf{F1-score}: F1-score considers FP and FN together, and it is a harmonic mean of Precision and Recall. It is calculated as
\begin{equation*}
    \text{F1-score} = 2\times \frac{\text{Precision}\times \text{Recall}}{\text{Precision}+ \text{Recall}}.
\end{equation*}

\section{Three Folds Results}
\label{appendix:threefold}
The 3-fold individual results for the VulDeePecker dataset are presented in Table~\ref{tab:3fold1}, Table~\ref{tab:3fold2}, Table~\ref{tab:3fold3}, and Table~\ref{tab:3fold4}.

\end{document}